\documentclass[structabstract]{aa}  
\usepackage{graphicx}
\usepackage{natbib}
\bibpunct{(}{)}{;}{a}{}{,} % to follow the A&A style
\usepackage{lscape}
\usepackage{longtable}
\usepackage{lscape}
\usepackage{gensymb}
\usepackage{color}
\usepackage{epstopdf}
\usepackage{rotating}
\usepackage{numprint}
\usepackage[percent]{overpic}
\usepackage{txfonts}
\usepackage{multirow}

\begin{document}

\title{New low surface brightness dwarf galaxies in the Centaurus group}

   \author{Oliver M\"uller\inst{1}
          \and
          Helmut Jerjen\inst{2}
                                        \and
          Bruno Binggeli\inst{1}
          }

   \institute{Departement Physik, Universit\"at Basel, Klingelbergstr. 82, CH-4056 Basel, Switzerland\\
         \email{oliver89.mueller@unibas.ch; bruno.binggeli@unibas.ch}
         \and
         Research School of Astronomy and Astrophysics, Australian National University, Canberra, ACT 2611, Australia\\
         \email{helmut.jerjen@anu.edu.au}
             %\thanks{thanks text here}
             }

   \date{Received 13 May 2016; accepted 24 August 2016}

 \abstract
  {The distribution of satellite galaxies around the Milky Way and Andromeda and their correlation in phase space pose a major challenge to the standard $\Lambda$CDM model of structure formation. Other nearby groups of galaxies are now being scrutinized to test for the ubiquity of the phenomenon. }
   {We conducted an extensive CCD {imaging survey} for faint, unresolved dwarf galaxies of very low surface brightness in the whole Centaurus group region, encompassing the Cen\,A and M\,83 subgroups lying at a distance of roughly 4 and 5 Mpc, respectively. The aim is to significantly increase the sample of known Centaurus group members down to a fainter level of completeness, serving as a basis for future studies of the 3D structure of the group.}
   {Following our previous survey of 60 square degrees covering the M\,83 subgroup, we extended and completed our survey of the Centaurus group region by imaging another 500 square degrees area in the $g$ and $r$ bands with the wide-field Dark Energy Survey camera at the 4 m Blanco telescope at CTIO. {The surface brightness limit reached for unresolved dwarf galaxies is $\mu_r \approx 29 $ mag arcsec$^{-2}$. The faintest suspected Centaurus members found have $m_r \approx$ 19.5 mag or $M_r \approx -8.8$ mag at the mean distance of the group}. 
        The images were enhanced using different filtering techniques.}
   {We found 41 new dwarf galaxy candidates, which together with the previously discovered 16 dwarf candidates in the M\,83 subgroup amounts to almost a doubling of the number of known galaxies in the Centaurus complex, if the candidates are confirmed. We carried out surface photometry in $g$ and $r$, and report the photometric parameters derived therefrom, for all new candidates as well as previously known members in the surveyed area. The photometric properties of the candidates, when compared to those of Local Group dwarfs and previously known Centaurus dwarfs, suggest membership in the Centaurus group. The sky distribution of the new objects is generally following a common envelope around the Cen\,A and M\,83 subgroups. How the new dwarfs are connected to the intriguing double-planar feature recently reported must await distance information for the candidates.}
  % conclusions heading (optional) &  leave it empty if necessary 
 {}


   \maketitle
%
%________________________________________________________________

\section{Introduction}
In addition to their traditional role as dark matter (DM) tracers by their internal dynamics \citep{2013pss5.book.1039W}, 
faint dwarf galaxies are a very powerful testbed for DM and structure formation models by their mere abundance 
and spatial distribution. There is the long-standing missing satellite problem \citep[e.g.,][]{1993MNRAS.264..201K,1999ApJ...522...82K,1999ApJ...524L..19M} and the too big to fail problem \citep{2011MNRAS.415L..40B}, both of 
which might be attributable to an incomplete understanding of baryonic physics \citep[e.g.,][]{2007ApJ...670..313S,2016arXiv160205957W}. 
However, a major challenge for the standard picture of structure formation with DM is now posed by the highly 
asymmetric features found in the distributions of dwarf galaxies in the Local Group {\citep{2005A&A...431..517K}}. There is the
vast polar structure \citep[VPOS;][]{2015MNRAS.453.1047P,2012MNRAS.423.1109P}, which is a thin (rms 
height $\approx$ 30 kpc) highly inclined, corotating substructure of faint satellite galaxies, young globular 
clusters, and streams, spreading in Galactocentric distance between 10 and 250\,kpc. Following an earlier 
suggestion by \cite{2006AJ....131.1405K}, a similar feature was found in the Andromeda galaxy surroundings \citep{2007MNRAS.374.1125M,2013Natur.493...62I}, called the Great Plane of Andromeda (GPoA). Moreover, there are two 
galaxy planes (diameters of 1-2 Mpc) that contain all but one of the 15 nonsatellite galaxies in the 
Local Group \citep{2013MNRAS.435.1928P}. Such planar structures on galactic and intergalactic scales are difficult 
to accommodate in a standard $\Lambda$CDM scenario, where extreme satellite planes are found in $<0.1\%$ 
of simulated systems \citep[e.g.,][]{2014MNRAS.442.2362P}. 
Still, the most conservative estimate from cosmological simulations including the look-elsewhere effect, but ignoring observational 
uncertainties, finds the frequency of two prominent satellite structures in the Local Group  to be 
$\sim 1$\,per cent \citep{2015MNRAS.452.3838C}. These controversial results demonstrate the need for more 
observational data to scrutinize $\Lambda$CDM predictions and assess the degree of conflict with that model.

If the relative sparseness and asymmetric distributions of low-mass dwarf galaxies are a common phenomenon in the local universe,
a major revision of our view of structure formation would be necessary. Recently, \cite{2015ApJ...802L..25T} reported evidence of a 
double-planar structure in the nearby Centaurus group of galaxies, based on hitherto known (i.e.,\,still fairly massive) galaxy 
members of the group. This result is encouraging, as it means that systematic studies of the spatial distribution of fainter 
dwarf galaxies in nearby groups can provide important observational constraints for further testing structure formation models. 
In a first step, deep and wide-field imaging is required to detect dwarf galaxy members of nearby galaxy groups with faint 
luminosity and surface brightness levels. Present-day technologies allow a dwarf galaxy census of other nearby groups 
down to $M_V \approx -10$, equivalent to Local Group dwarfs like Sculptor,  Sextans, and Tucana, clearly surpassing the achievements 
of the Sloan Digital Sky Survey \citep[SDSS;][]{2014ApJS..211...17A,2000AJ....120.1579Y} {with respect to the detection of {unresolved} dwarf candidates}.

Several international teams have taken up the effort to conduct dedicated imaging surveys of other nearby galaxy groups in
the search for faint and ultra-faint dwarf galaxies in the northern hemisphere; see, for example,\ \cite{2009AJ....137.3009C,2013AJ....146..126C}
for the M\,81 group (14 confirmed new members over 65 deg$^2$), and \cite{2014ApJ...787L..37M} {and \cite{2016A&A...588A..89J}} for the M\,101 group
(8 dwarf candidates over 7 deg$^2$). In the southern hemisphere, the deep but spatially limited Panoramic Imaging Survey of Centaurus and Sculptor (PISCeS)
 of NGC\,253 in the Sculptor group and NGC\,5128 (Cen\,A) in the Centaurus group \citep{2014ApJ...793L...7S, 2014ApJ...795L..35C,2015arXiv151205366C} 
revealed 9 extremely faint dwarf galaxies ($25.0<\mu_{r,0}<27.3$, $-13<M_V<-7.2$) in the vicinity ($\sim11$\,deg$^2$) of Cen\,A. Group memberships of these dwarfs
have been confirmed with the tip of the red giant branch (TRGB) method.

In the same spirit we conducted a large-scale survey of the Centaurus Group using the Dark Energy Camera (DECam) 
at the 4 m Blanco telescope at CTIO. Our survey has a photometric surface brightness limit that is slightly less sensitive than 
PISCeS, but a 50 times larger footprint. The survey covers a region of $\approx$ 550 deg$^2$, thus providing {complete} 
CCD coverage of this southern galaxy group, for the first time, going significantly deeper than  with the SDSS {in the north outside the Local Group}. {Owing to its greater depth, PISCeS  is a search for resolved dwarf objects, while our survey is able to detect only unresolved dwarf members of the Centaurus group.} This paper is the second report on our DECam survey of the Centaurus group region. 
We refer to the Centaurus group as the whole complex and the two main concentrations as Cen\,A and M\,83 subgroup, 
respectively. The Cen\,A  subgroup is dominated by the massive peculiar galaxy Cen\,A  (=NGC5128) at a mean distance of 3.8 Mpc and the M\,83 subgroup by the giant spiral M\,83 (=NGC 5236) at a mean distance of 4.9 Mpc \citep{2004AJ....127.2031K,2013AJ....145..101K, 2015ApJ...802L..25T, 2015AJ....149..171T}. In our first paper \citep[][hereafter MJB15]{2015A&A...583A..79M} we 
reported the discovery of 16 new dwarf galaxy candidates from our survey of the M\,83 subgroup, covering an area of 60 deg$^2$ based on the images taken with DECam.
One of the new dwarfs, dw1335-29, has already been confirmed as group member based on  HST archival data 
\citep{2016AAS...22713625C}. 

In this paper we present our extended DECam survey of the entire Centaurus group and report 
on the discovery of another 41 new dwarf candidates in addition to the 16  
dwarf galaxy candidates reported in MJB15. Even if we assume that a few objects will turn out to be background galaxies,
this sample essentially doubles the number of known galaxies in the Centaurus group.

The paper is organized as follows. In Sect.\,2 we give the details of the DECam observations. 
Sect.\,3 describes our search strategy for, and detection of, faint diffuse dwarf galaxy candidates in the 
survey footprint. In Sect.\,4 we present the results from the surface photometry analysis conducted for the new 
candidates and the known Centaurus group members. Finally, a first assessment of the dwarf galaxy distribution 
and a critical discussion of our findings are given in Sect.\,5, followed by our conclusions in Sect.\,6.   

\section{Observations and photometric calibration}
\begin{figure}[t]
\hspace*{-0.7cm}
\includegraphics[width=10.5cm]{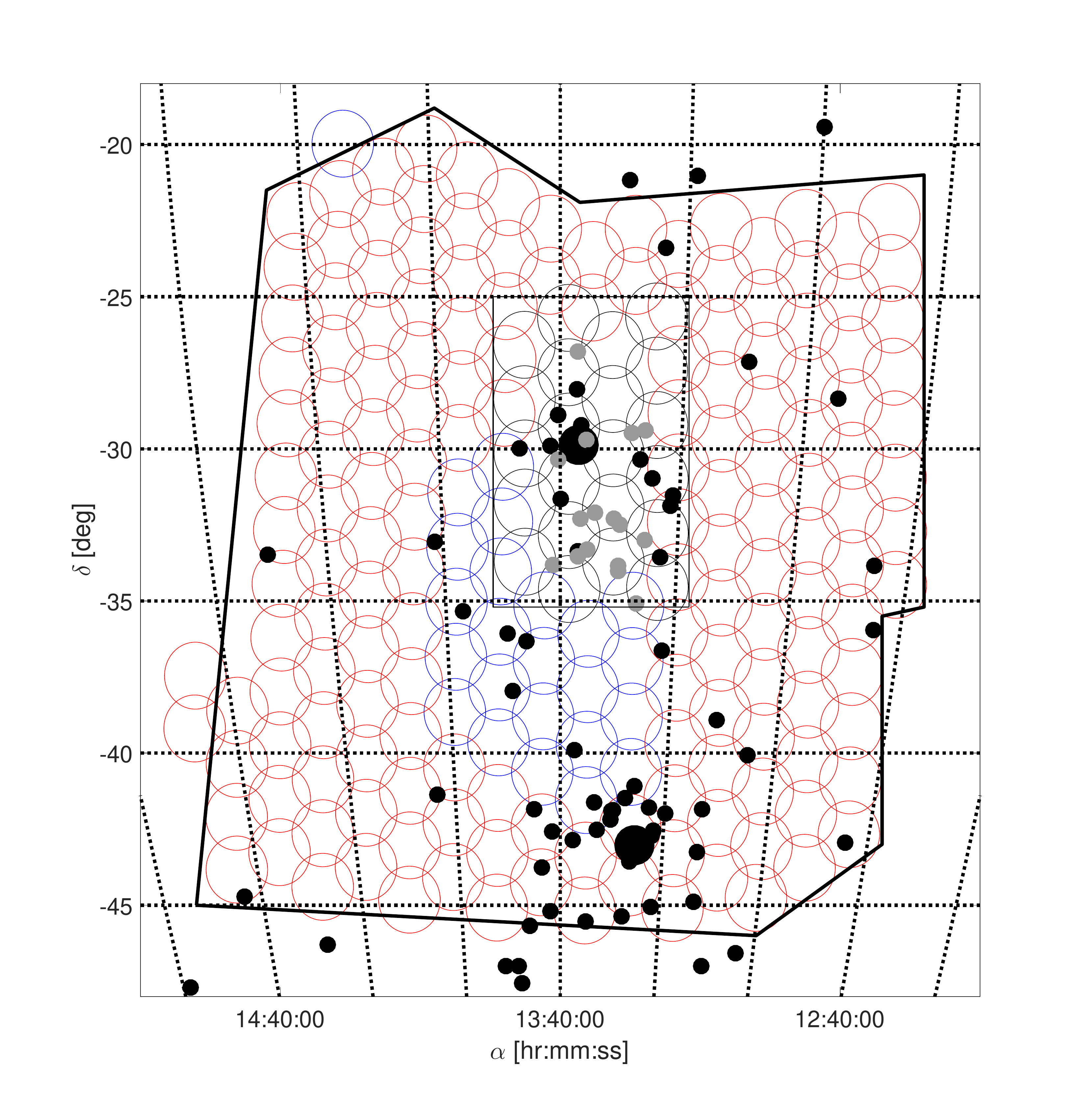}
 \caption{ Surveyed area of $\approx$550 square degrees in the Centaurus group. The individual DECam fields are represented 
 by circles. The colors indicate the three different sets of data; the fields around M83 were discussed in the MJB15 study and 
 are shown as black circles. The blue and red fields were observed in 2014 and 2015, respectively, and analyzed for the 
 present paper. The small black dots are the known dwarfs in the Centaurus group listed in the Local Volume 
 Catalog \citep{2004AJ....127.2031K,2013AJ....145..101K} complemented by the recently discovered nine dwarfs 
 of \cite{2014ApJ...795L..35C,2015arXiv151205366C}. The larger black dots are the two dominant group galaxies 
 M\,83 (13h37m00.9s, $-$29d51m56s) and Cen\,A (13h25m27.6s, $-$43d01m09s).   }
 \label{fieldImage}
\end{figure}

We obtained images in the $g$ and $r$ bands over two observing runs on 2014 July 17--19 and 
2015 June 4--9 using the Dark Energy Camera at the 4 m Blanco telescope at Cerro Tololo Inter-American Observatory (CTIO) as 
part of the observing proposals 2014A-0624 and 2015A-0616 (both PI: H.~Jerjen).
With an array of 62 2k$\times$4k CCD detectors the DECam has a 3 square degree field of view and a pixel scale of 
$0\farcs27$. In 2014 we obtained a complete data set for 24 fields under dark time conditions (blue circles in Fig.\ref{fieldImage}).
Exposure times were $3\times40$\,sec in both bands. During the 2015 observing run two exposures were taken 
in each band for a total of 163 fields (red circles in Fig.\ref{fieldImage}). 
To fill the inter-chip gaps, we dithered diagonally by half of a CCD chip. 
The measured median seeing was $1\farcs0$. As we were observing under 
waning moon conditions, we strategically collected the $r$-band images in the 
first four nights with exposure times between $2\times120$ and $2\times210$\,sec 
and the $g$-band images in the last four nights with exposure times between $2\times100$ and $2\times170$\,sec, 
depending on the sky brightness and the angular distance of the target field from the moon.

The images were fully reduced and stacked using the DECam community pipeline \citep{2014ASPC..485..379V}. 
Fig.\,\ref{fieldImage} shows the survey footprint superimposed on the distribution of the known galaxies in the Centaurus group. The circles correspond to the 
individual DECam fields while the colors indicate the different data sets. Black circles indicate the 22 DECam fields of MJB15.  

To determine the photometric zero points and color terms for each DECam field, we matched the instrumental 
magnitudes of typically 100-200 stars in each field with their corresponding photometric data from the AAVSO 
Photometric All-Sky Survey (APASS) catalog \citep{2014CoSka..43..518H} using the DAOPHOT package 
\citep{1987PASP...99..191S} in IRAF and fitted the following two equations:
$$m_g = m_{g,instr} + Z_g + c_{g} \cdot (m_{g,instr}-m_{r,instr}) - k_{g}X$$
$$m_r = m_{r,instr} + Z_p+ c_{r} \cdot (m_{g,instr} - m_{r,instr}) - k_{r}X,$$
where $Z_g$ and $Z_r$ are the photometric zero points, $c_g$ and $c_r$ are
the color terms, $k_g$ and $k_r$ are the atmospheric extinction coefficients, and $X$ is the 
mean airmass. The most recent extinction values $k_r=0.10$ and $k_g=0.20$ for CTIO were
kindly provided by the Dark Energy Survey team. The airmass $X$ was given in the header 
of each exposure.

To allow a direct comparison with available photometry for Local Group dwarf galaxies in the literature, 
we converted our $gr$ photometry (see section 4) into the $V$ band using the transformation equation by \cite{SloanConv} as follows:
\begin{eqnarray}
V = g - 0.5784\cdot(g - r) - 0.0038.
\end{eqnarray}
This formula can be used for the total magnitudes and surface brightness parameters of the galaxy. 
To further compare our results from the S\'ersic profile fitting with the $B$-band results for Local Group and 
Virgo cluster dwarf galaxies, we also converted the literature values from the $B$ to $r$ band using the equation \citep{SloanConv},
\begin{eqnarray}
r= B-1.3130\cdot(g-r)-0.2271,
\end{eqnarray}
where we adopted a color index of $(g-r)= 0.6$ suitable for early-type dwarf galaxies \citep{2008AJ....135..380L}.
The entire survey area (this paper and MJB15) is subdivided into three different data sets (see Fig.\,\ref{fieldImage}). As we mentioned before, the exposure times for the fields obtained
in the 2015 run (red circles in Fig.\,\ref{fieldImage}) were adjusted to compensate for the sky brightness variation due to lunar illumination 
to achieve approximately equal photometric depth across the survey area. To test photometric uniformity we sampled the faintest stars 
in different regions and measured their apparent magnitudes. The variance is in the range of 0.3\,mag. 
Overlapping DECam fields from different data sets were also compared for their detection quality. We find no significant difference.  

\begin{figure}[ht]
\hspace*{-0.6cm}
\includegraphics[width=10.5cm]{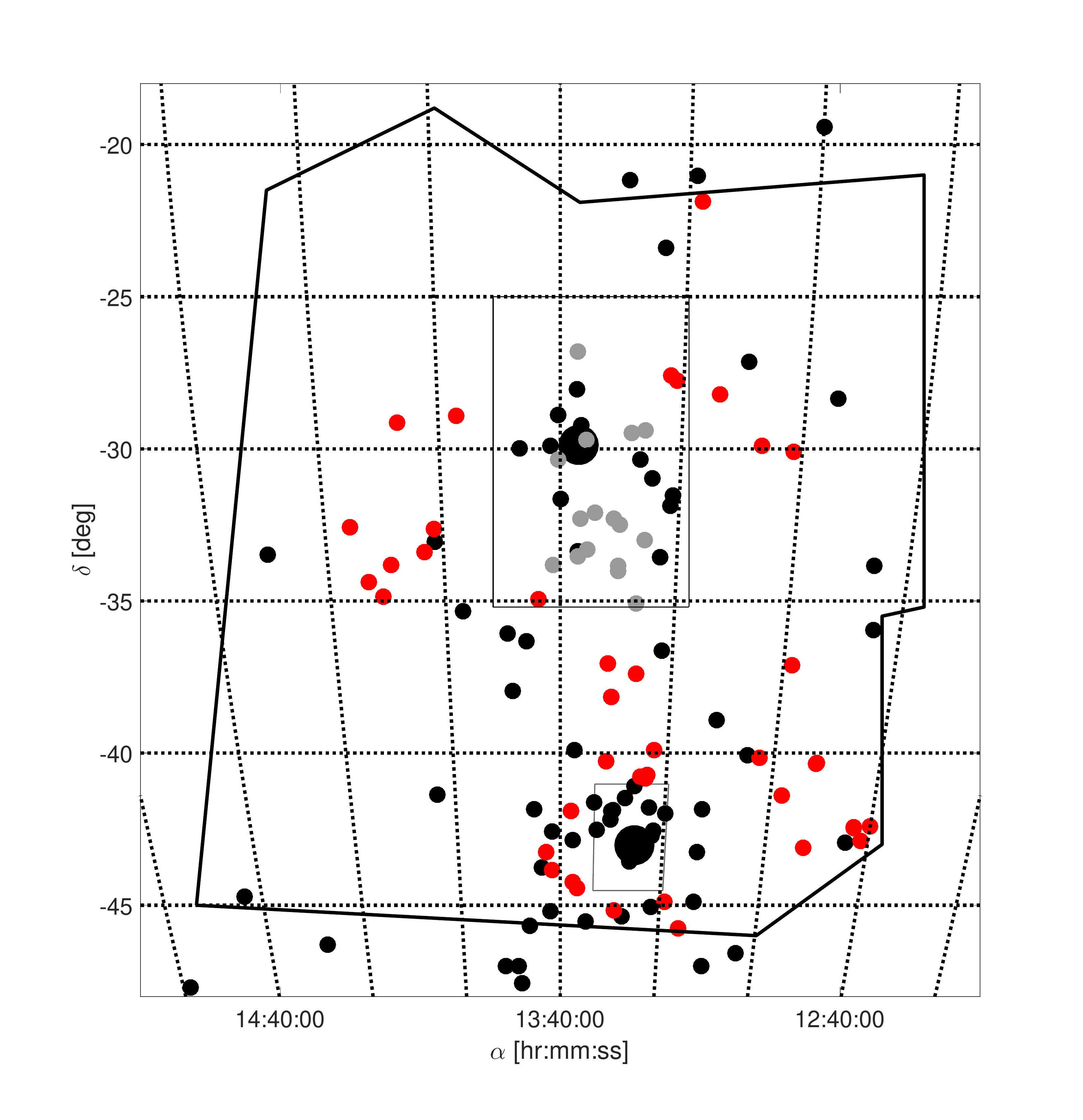}
\caption{Same as Fig.\,\ref{fieldImage} but without the DECam pointings, showing the 41 newly detected dwarf galaxy candidates as red dots. 
The 16 dwarf candidates we previously reported in the vicinity of M\,83 (MJB15; survey footprint shown as large rectangle) are indicated as small gray circles.
The survey area of the \cite{2014ApJ...795L..35C,2015arXiv151205366C} study around Cen\,A is approximated with the small rectangle.}
  \label{fieldDepth}
\end{figure}

% discovery table
\begin{table}
\setlength{\tabcolsep}{3pt}
\caption{Names and coordinates of the 41 new dwarf galaxy candidates.
}
\label{table:1}
\begin{tabular}{lccll}
\hline\hline
& $\alpha$ & $\delta$ & \\ 
Name & (J2000) & (J2000) & Type & Notes\\ 
\hline \\[-2mm]
dw1240-42 & 12:40:02 & $-$ 42:24:44 &dSph&  \\ 
dw1241-32 & 12:41:27 & $-$ 42:53:45 &dSph &\\ 
dw1243-42 & 12:43:13 & $-$ 42:27:48 & dSph&pair: dw1243-42b\\ 
dw1243-42b & 12:43:11 & $-$ 42:26:37 &dIrr &\\ 
dw1251-40 & 12:51:56 & $-$ 40:19:53&dSph &pair: dw1252-40\\ 
dw1252-40 & 12:52:01 & $-$ 40:21:55 &dSph &bg? \\ 
dw1252-43 & 12:52:25 & $-$ 43:05:58 &dSph & \\ 
dw1257-41 & 12:57:45 & $-$ 41:22:52 &dSph & \\ 
dw1258-37 & 12:58:29 & $-$ 37:07:21 & dSph& \\ 
dw1301-30 & 13:01:28 & $-$ 30:06:43 & dSph&\\ 
dw1302-40 & 13:02:49 & $-$ 40:08:35 & dSph& \\ 
dw1306-29 & 13:06:48 & $-$ 29:53:30 &dSph &bg? \\ 
dw1314-28 & 13:14:02 & $-$ 28:12:12 &dIrr/dSph &bg?\\ 
dw1315-45 & 13:15:56 & $-$ 45:45:02 &dIrr &\\ 
dw1318-21 & 13:18:04 & $-$ 21:53:06 &dSph &bg? \\ 
dw1318-44 & 13:18:58 & $-$ 44:53:41 & dSph& \\ 
dw1321-27 &  13:21:08& $-$ 27:44:56 & dSph&\\ 
dw1322-27 & 13:22:06 & $-$ 27:34:45 &dIrr/dSph,N &bg? \\ 
dw1322-39 & 13:22:32 & $-$ 39:54:20 & dIrr &\\ 
dw1323-40 & 13:24:53 & $-$ 40:45:41 & dSph &\\ 
dw1323-40b & 13:23:55 & $-$ 40:50:09 & dSph &\\ 
dw1323-40c & 13:23:37& $-$ 40:43:17& dSph &\\ 
dw1326-37 & 13:26:22 & $-$ 37:23:08 &dIrr? &bg? \\ 
dw1329-45 & 13:29:10 & $-$ 45:10:31 &dSph & \\ 
dw1330-38 &  13:30:41& $-$ 38:10:03 &cirrus? & \\ 
dw1331-37 & 13:31:32 & $-$ 37:03:29 & dSph& \\ 
dw1331-40 & 13:31:26 & $-$ 40:15:47 & cirrus?& \\ 
dw1336-44 & 13:36:44  & $-$ 44:26:50 &dIrr &\\ 
dw1337-41 & 13:37:55 & $-$  41:54:11&cirrus? & \\ 
dw1337-44 & 13:37:34  & $-$ 44:13:07 &dIrr? &\\ 
dw1341-43 & 13:41:37  & $-$ 43:51:17 & dSph&\\ 
dw1342-43 & 13:42:44 & $-$ 43:15:19 & dIrr? &\\ 
dw1343-34 & 13:43:49 & $-$  34:56:07&cirrus? & \\ 
dw1357-28 & 13:57:00 & $-$ 28:55:15 & dSph &\\ 
dw1401-32 & 14:01:25 & $-$ 32:37:46 &  dSph&\\ 
dw1403-33 & 14:03:18 & $-$ 33:24:14 & dSph &\\ 
dw1406-29 & 14:06:41 & $-$ 29:08:10 & dSph &\\ 
dw1409-33 & 14:09:03 & $-$ 33:49:40 & dSph &\\ 
dw1410-34 & 14:10:47 & $-$ 34:52:07 & dIrr &\\ 
dw1413-34 & 14:13:08 & $-$ 34:23:33 & dSph &\\ 
dw1415-32 & 14:15:41 & $-$ 32:34:21 & dIrr? &\\ 
\\
%dw &  & $-$  &  \\ 
%dw &  & $-$  &  \\ 
%dw &  & $-$  &  \\ 
\hline\hline
\end{tabular}
\tablefoot{
The morphological type of the galaxies listed here is a first guess. Deeper imaging is needed to identify their real morphology. Question marks indicate an uncertainty of the classification. If a candidate turns out to be a background galaxy, its morphological type needs to be revised.
}
\end{table}
 %______________________________________________________________

\section{Search and detection of new dwarf candidates}
Finding new dwarf galaxy candidates in the Centaurus group required the 
search for unresolved, low surface brightness objects in the DECam images. 
With the relative short exposure times we cannot resolve galaxies at the distance 
of the Centaurus group into RGB stars. A quick estimate of the TRBG magnitude 
shows that we miss the RGB tip by a few tenths of a magnitude. For this estimate 
we took the stellar population of the Sculptor dwarf galaxy as a reference with $M_I = - 4.1$ and $V-I = 1.5$ for its TRGB \citep{2007MNRAS.380.1255R}. 
This translates into $M_r$ $\approx$ $-2.8$ with an assumed color index of $V-r$ $\approx$ 0.2. We assume a mean 
Galactic extinction of $A_r = 0.15$ for our survey field; see Table 2 for the exact extinction values for all the galaxies 
in the survey. At the distance of 4.9 Mpc (M\,83)  this gives an expected apparent magnitude of $m_r = 25.8$ for the RGB tip. At the distance of 3.8 Mpc (Cen\,A) the expected apparent magnitude is $m_r = 25.3$. The faintest stars detectable in the survey data have a magnitude $\approx25$, and thus
we can expect to see the brightest stars in galaxies only if they have a shorter distance than Cen\,A. Faint individual 
stars are indeed visible in some of the galaxy candidates, but in most of the cases we miss the tip. Therefore this is a search 
for unresolved stellar systems.\\

All $gr$ band images available for an individual DECam field were co-added 
using the SWarp program \citep{2002ASPC..281..228B}. The SWarp program subtracts the background of every frame, resamples 
them onto a common coordinate system, stacks them, and puts the combined image into a single file (hereafter deep image). The frames were 
combined using the weighted co-addition algorithm.

Gray-level manipulation was applied on the deep images to enhance the contrast. 
We carefully estimated the local background $RMS$ noise and chose a range 
of $2\times RMS$ below and above the estimated sky background level. This is the regime where we expect 
the low surface brightness dwarfs to be most prominent. In a first step, the deep images were visually inspected. 
Then different filtering techniques like the Gaussian convolution and the ring median filter \citep{1995PASP..107..496S} 
were applied to enhance the presence of any low-surface brightness features. This strategy 
can potentially lead to losing high surface brightness objects, such as~bright background galaxies or Blue Compact Dwarfs (BCD)
in the Centaurus group. We refer to section 3 in MJB15 for more details about the search strategy.

\begin{figure}[ht]
\includegraphics[width=9cm]{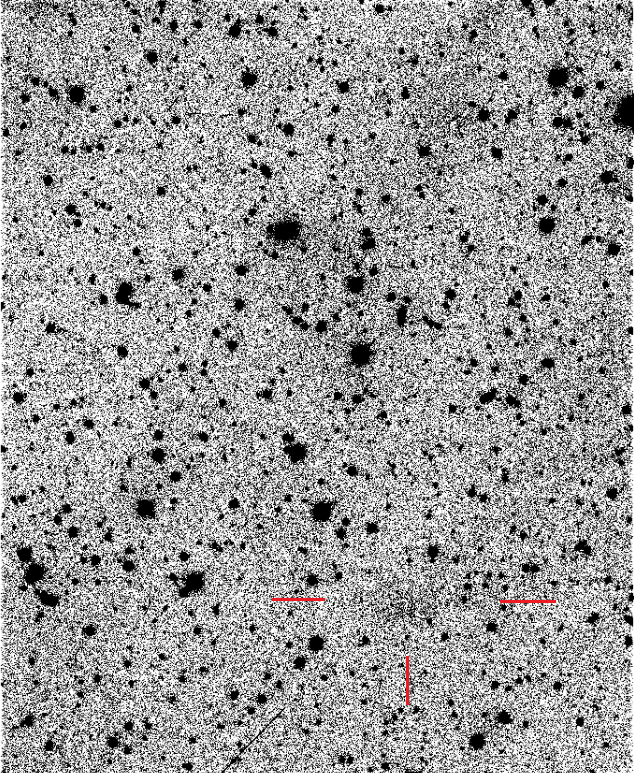}
  \caption{Example of a low surface brightness feature (indicated with red lines) close to Galactic cirrus. 
  The object was dismissed as dwarf galaxy candidate
  for reasons explained in the text.  }
  \label{cirrus}
\end{figure}

The region of the Centaurus group is at low Galactic latitudes and thus has a relatively high level of contamination 
from foreground stars and Galactic cirrus. Although cirrus can sometimes resemble low surface brightness dwarf 
galaxies in shape and size, it is often possible to distinguish them morphologically. When a low surface brightness 
object was detected in or near a structure of Galactic nebula (cirrus) it was dismissed as a dwarf galaxy candidate
(see Fig.~\ref{cirrus} for an example). As part of this decision process real dwarf galaxies could have been accidentally rejected.

\begin{figure*}
\centering
\includegraphics[width=3.6cm]{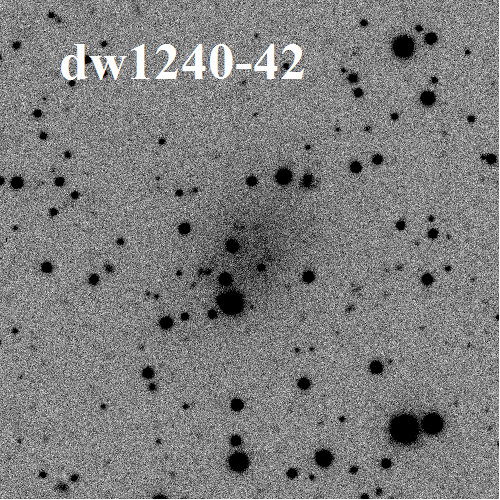}
\includegraphics[width=3.6cm]{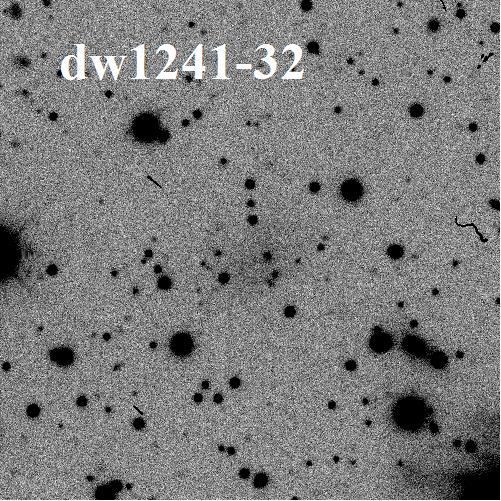}
\includegraphics[width=3.6cm]{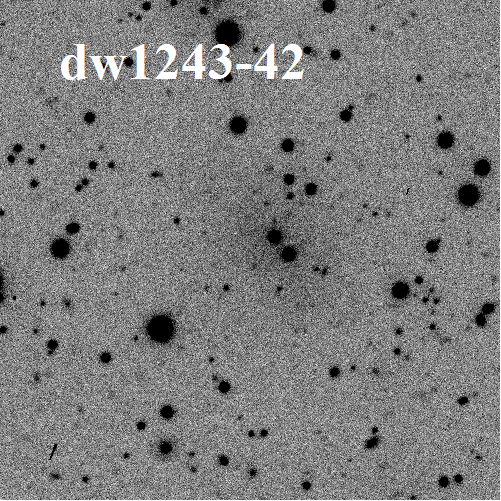}
\includegraphics[width=3.6cm]{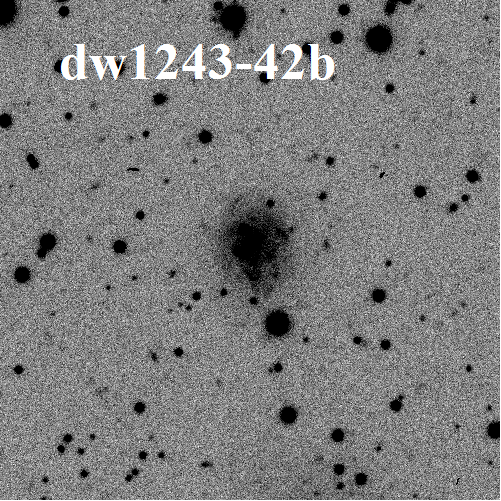}
\includegraphics[width=3.6cm]{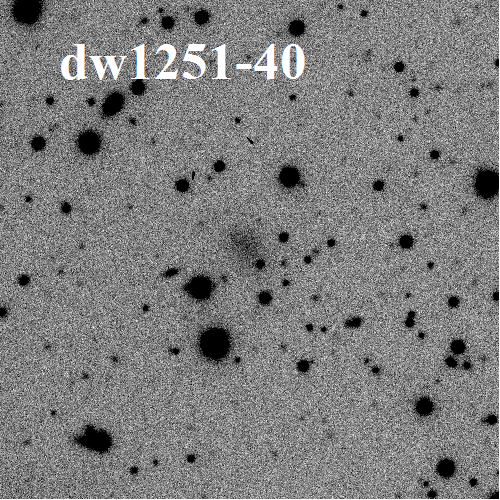}\\
\includegraphics[width=3.6cm]{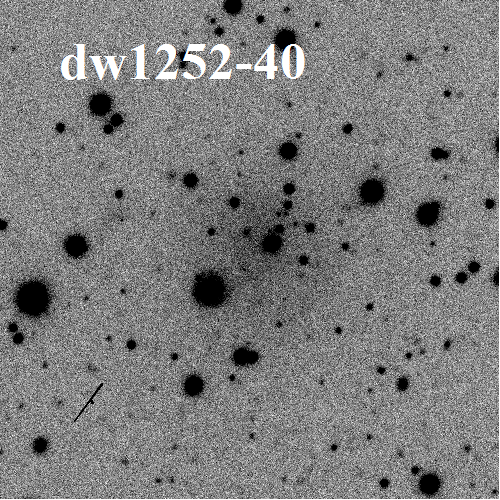}
\includegraphics[width=3.6cm]{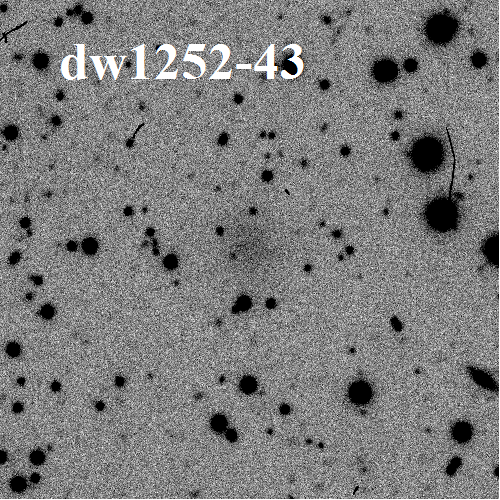}
\includegraphics[width=3.6cm]{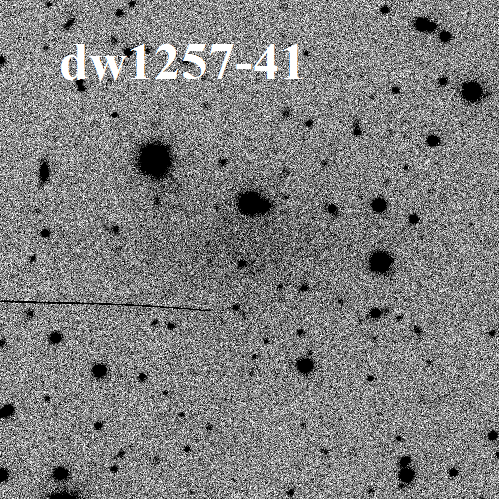}
\includegraphics[width=3.6cm]{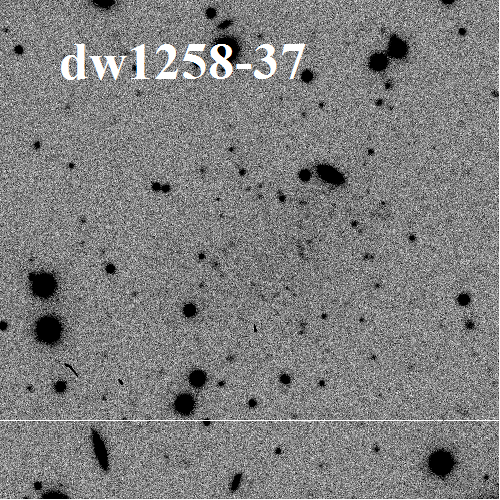}
\includegraphics[width=3.6cm]{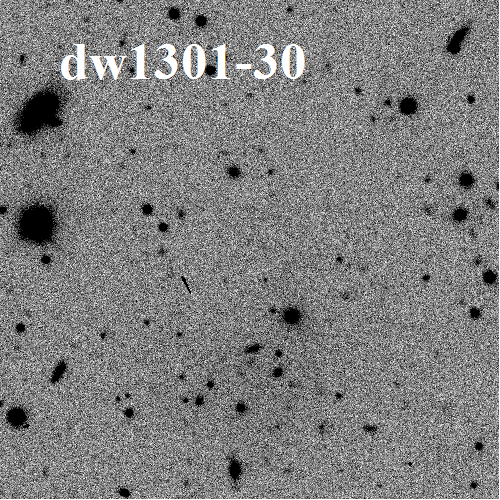}\\
\includegraphics[width=3.6cm]{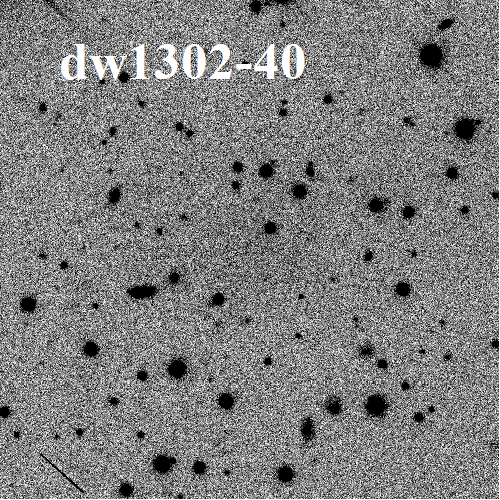}
\includegraphics[width=3.6cm]{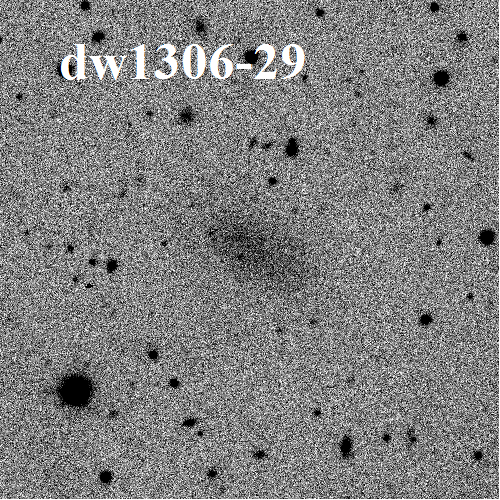}
\includegraphics[width=3.6cm]{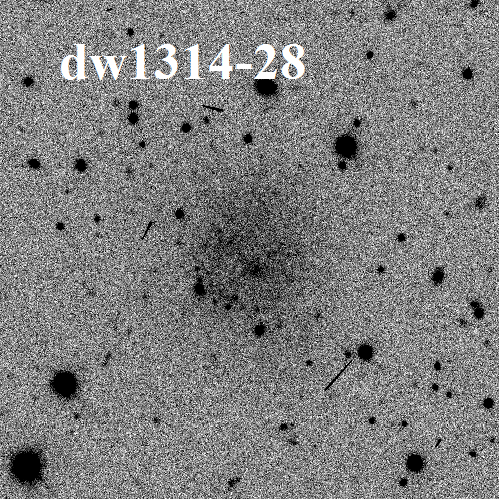}
\includegraphics[width=3.6cm]{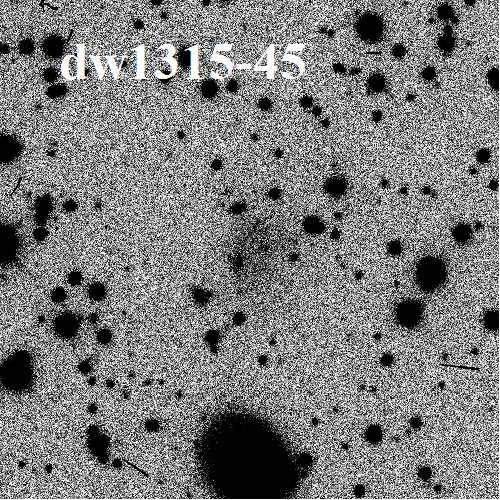}
\includegraphics[width=3.6cm]{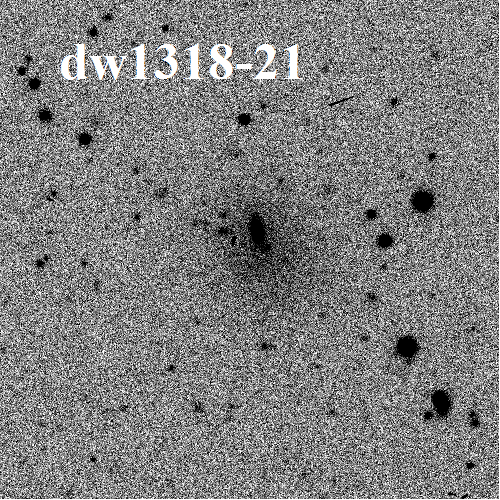}\\
\includegraphics[width=3.6cm]{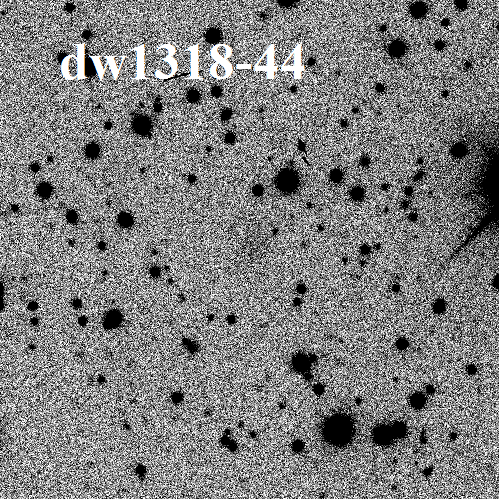}
\includegraphics[width=3.6cm]{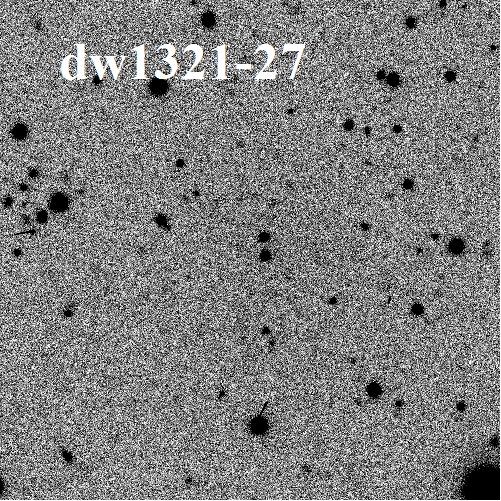}
\includegraphics[width=3.6cm]{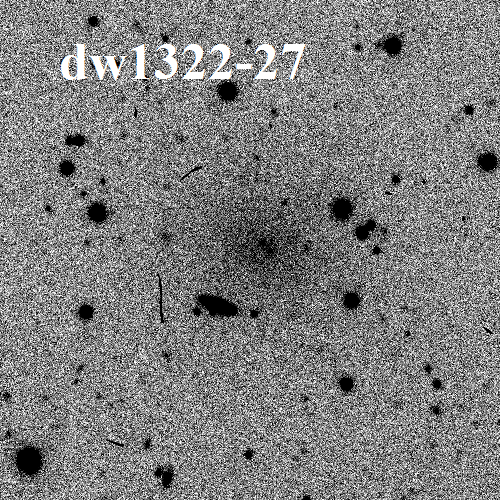}
\includegraphics[width=3.6cm]{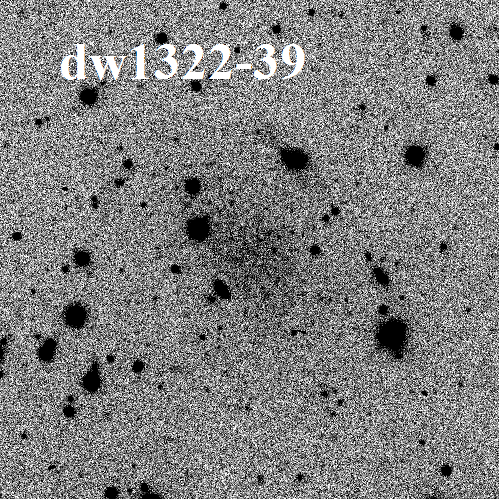}
\includegraphics[width=3.6cm]{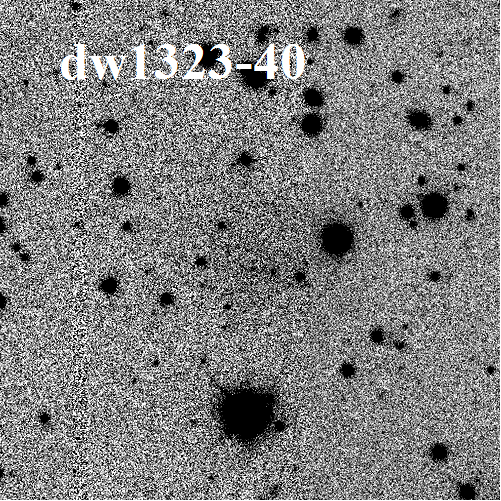}\\
\includegraphics[width=3.6cm]{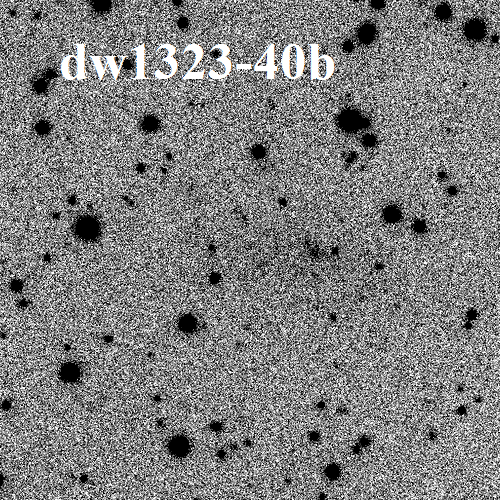}
\caption{Gallery showing DECam $r$-band images of the new Centaurus group dwarf galaxy candidates. 
One side of an image is 2.25\,arcmin or 3.0\,kpc at 4.5\,Mpc. North is to the top, east to the {left}.}
\label{sample0}
\end{figure*}

\begin{figure*}
\centering
\includegraphics[width=3.6cm]{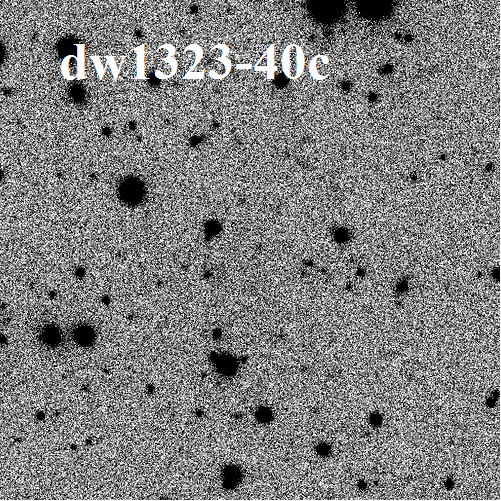}
\includegraphics[width=3.6cm]{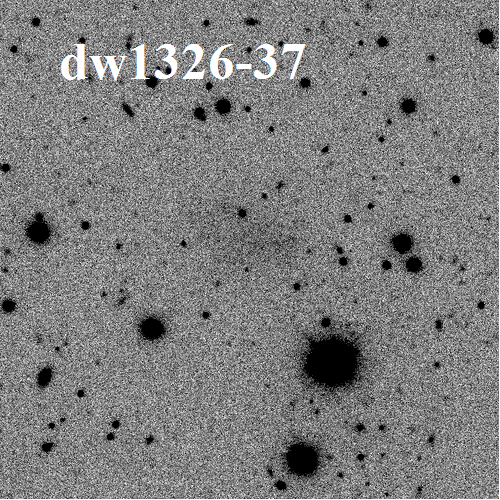}
\includegraphics[width=3.6cm]{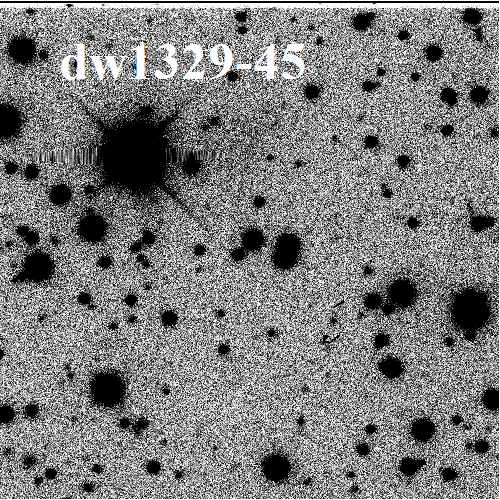}
\includegraphics[width=3.6cm]{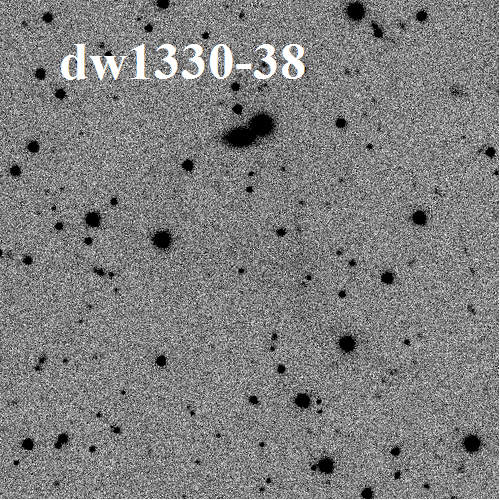}
\includegraphics[width=3.6cm]{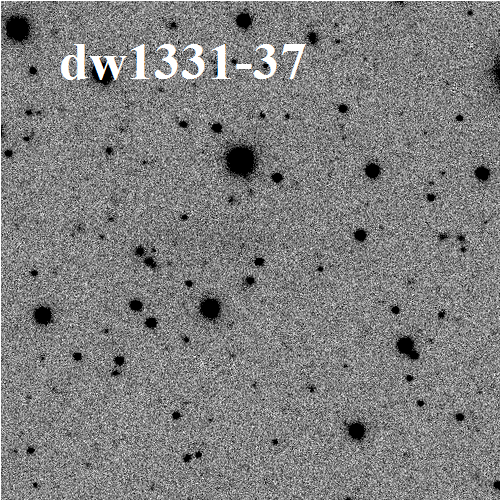}\\
\includegraphics[width=3.6cm]{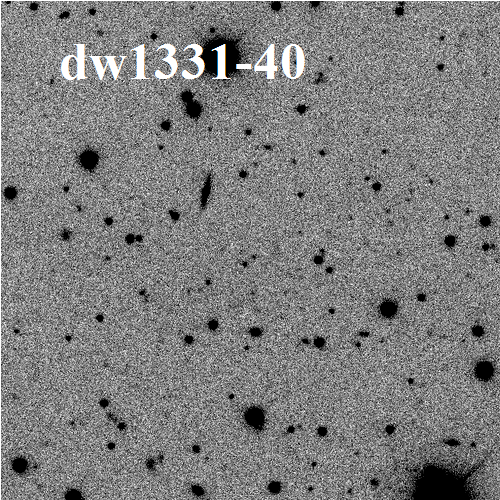}
\includegraphics[width=3.6cm]{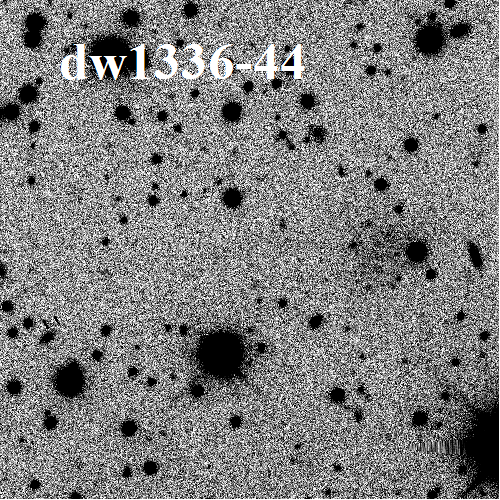}
\includegraphics[width=3.6cm]{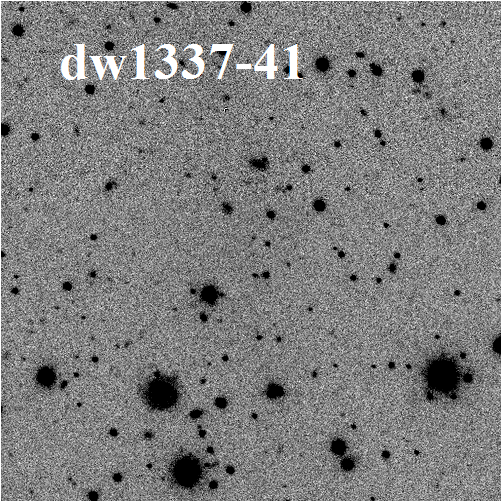}
\includegraphics[width=3.6cm]{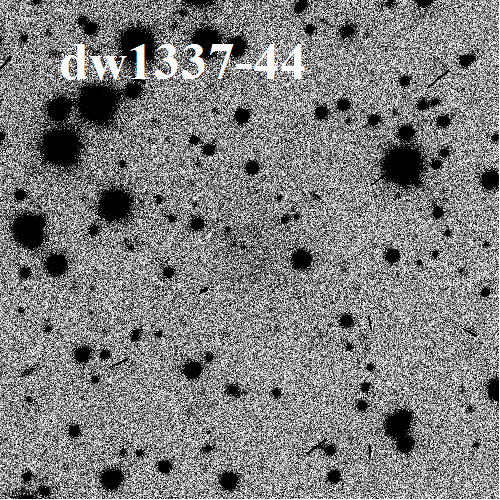}
\includegraphics[width=3.6cm]{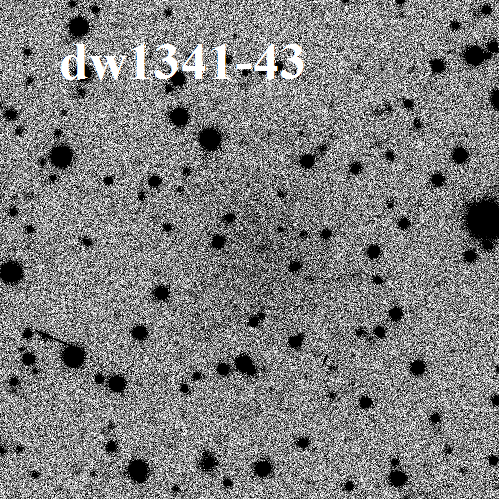}\\
\includegraphics[width=3.6cm]{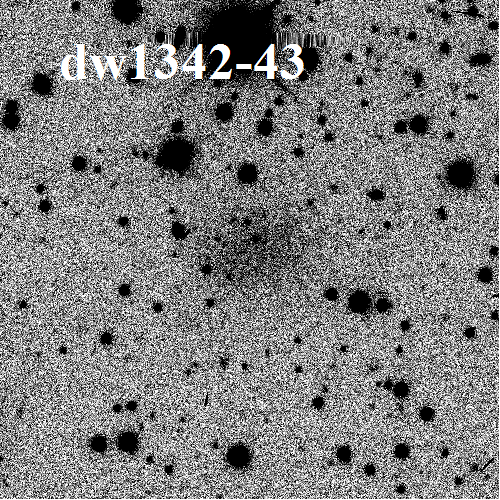}
\includegraphics[width=3.6cm]{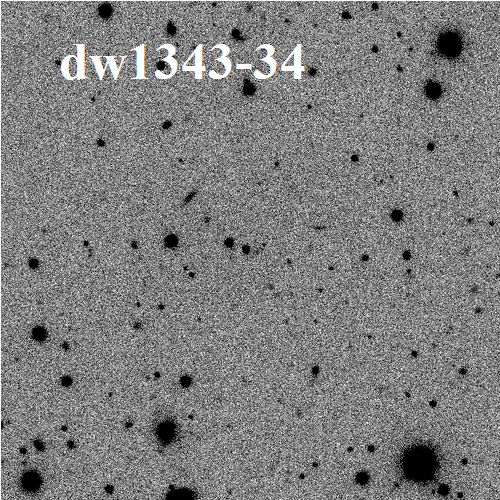}
\includegraphics[width=3.6cm]{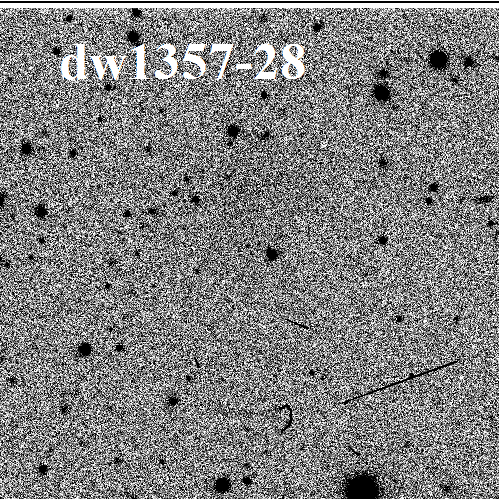}
\includegraphics[width=3.6cm]{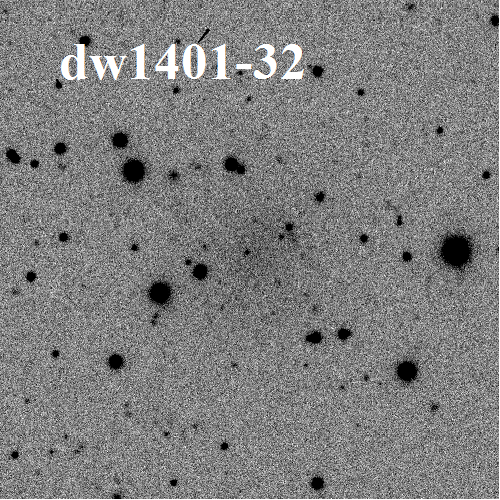}
\includegraphics[width=3.6cm]{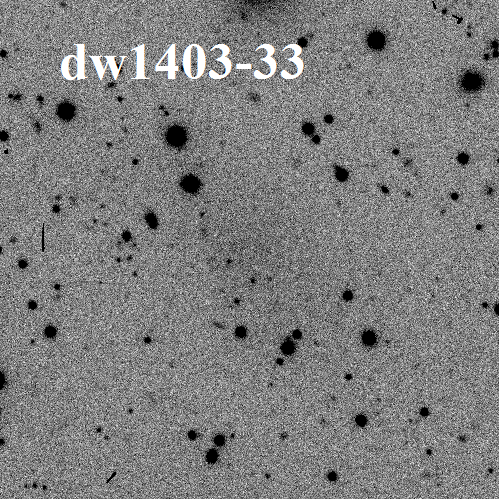}\\
\includegraphics[width=3.6cm]{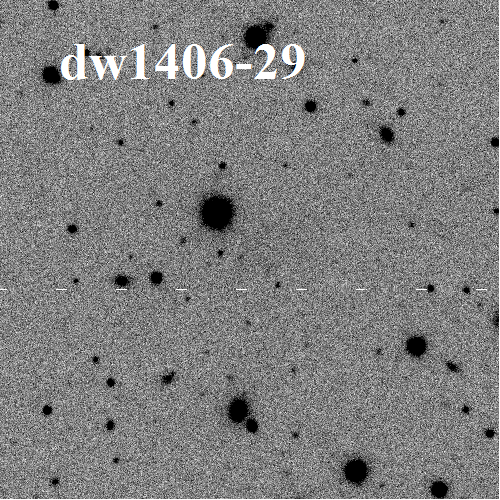}
\includegraphics[width=3.6cm]{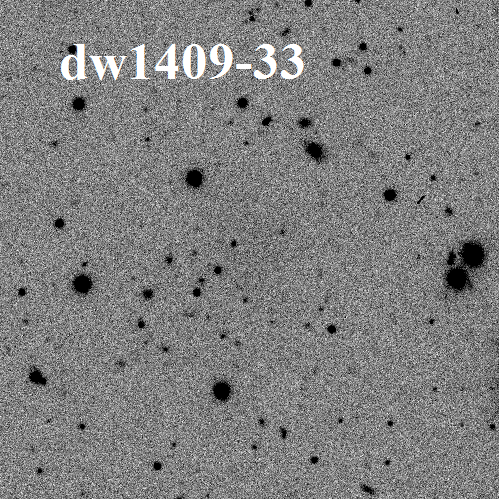}
\includegraphics[width=3.6cm]{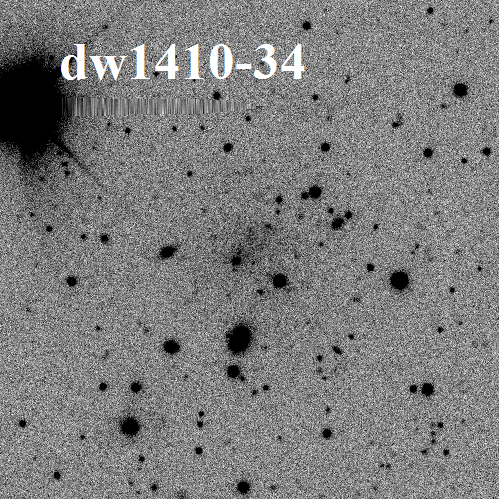}
\includegraphics[width=3.6cm]{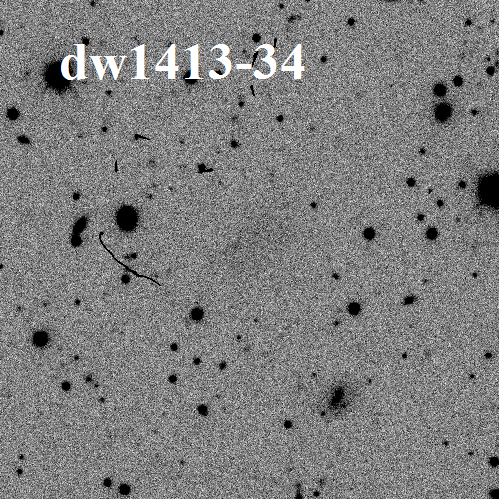}
\includegraphics[width=3.6cm]{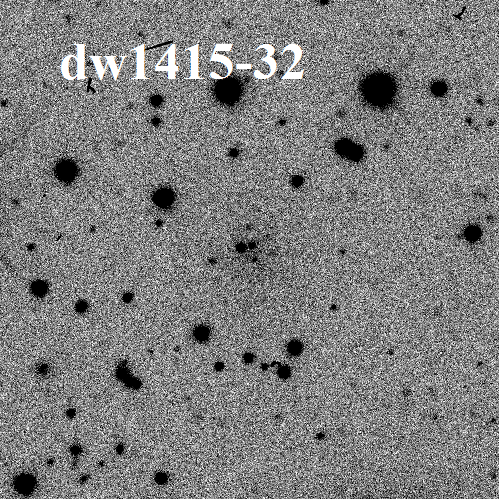}\\
\caption{Same as Fig.\,\ref{sample0}, continuing the image gallery.}
\label{sample1}
\end{figure*}

We found 41 new Centaurus group dwarf galaxy candidates. Their distribution among the known group members is shown in Fig.\,\ref{fieldDepth} (red dots). A gallery of the $r$-band images of the candidates is presented in Fig.\,\ref{sample0} and Fig.\,\ref{sample1}. The coordinates and morphological type
are compiled in Table 1. The morphological type is based on the assumption that the object is a member of the Centaurus group.
The photometric and structural parameters of the dwarf candidates are listed in Table 2.

 Three new candidates are in the MJB15 footprint. Just outside of the MJB15 footprint, dw1343-34 is visible when comparing Fig.\,1 and Fig.\,2; albeit fully visible in MJB15, dw1321-27 and dw1322-27 were then rejected and assumed to be satellites of NGC\,5101 in the background. With the background relation test carried out in this paper (see Fig.\,\ref{strucParameters}), we estimate that at least one of these candidates is too big in size relative to its surface brightness to be associated with the background galaxy, while the other  candidate can be argued as background or foreground dwarf. Still, we list both of them here as new candidates of the Centaurus group. Distance measurements will give a final answer to their membership.

We also checked for 21\,cm emission within 8\,arcmin of the direction of the candidates using the spectra from the HI Parkes All Sky Survey (HIPASS) 
survey \citep{2001MNRAS.322..486B}. None of the galaxies were detected in HI. Using the faint HI signal of the 
Centaurus group member HIPASSJ1348-37 as a reference ($S_{int}=2.5$\,Jy\,km\,s$^{-1}$), we derive
an upper limit for the HI content of the new dwarfs at $M_{HI}<8.5\times10^6M_\odot$.

%_________________________________________________________________

\section{Galaxy photometry}
%Photometry of candidates and all known galaxies in survey region

We measured $gr$ surface photometry for the new dwarf candidates and for the known Centaurus group dwarfs in 
the surveyed region, where possible. Pixels affected by foreground stars, background galaxies, and cosmic rays 
were replaced with patches of sky from the surrounding area to match the statistical properties of the local sky background. 
To find the center of the galaxy, we fitted a circle at the outer isophotes and took its center as the galaxy center.
For each photometric band, we computed the total  apparent magnitude, the mean effective surface 
brightness $\langle \mu\rangle_{eff} $, and the effective radius $r_{eff}$. 
To determine the sky brightness we varied the growth curve (cumulative intensity profile) of the galaxy 
until it became asymptotically flat at large radii. Radial surface brightness profiles were measured using a radial step 
size of $1\farcs35$ for galaxies visually larger than 13$\arcsec$ (radius) and $0\farcs54$ for smaller 
galaxies. A circular aperture was used for the photometry.
S\'ersic profiles  \citep{1968adga.book.....S} were fitted at the radial surface brightness profiles using the equation
$$\mu_{sersic}(r)= \mu_0+1.0857\cdot\left(\frac{r}{r_0}\right)^{n},$$
where $\mu_0$ is the S\'ersic central surface brightness, $r_0$ the S\'ersic scale length, and $n$ the S\'ersic curvature index. 
We note that some authors use $1/n$ instead of $n$. Although we clearly stated that we use $n$, even we were confused and used $1/n$ in Table 2 of 
MJB2015.  In Figure 7 of the same paper S\'ersic indices were plotted with $1/n$ 
instead of $n$ for our photometry. Our group membership argument does not change because most of the values are in the 
range between 0.8 and 1.2, still falling into the relation. We plot the correct values in Fig.~\ref{nToMr} here.

The combined uncertainty for the total magnitudes was estimated to be on the order of 0.3\,mag. Contributions to the 
error budget come from the star subtraction ($\approx$ 0.2\,mag), zero-point calibration 
(less than 0.04\,mag) and the estimated sky background  ($\approx$ 0.2\,mag); the star subtraction is estimated
by the average difference in magnitudes 
between the galaxy with star removal and without, assuming no bright star
is in the vicinity. An additional error for the absolute 
magnitudes (column 6 in Table 2) comes from the assumed distance ($\approx$ 0.25\,mag for an uncertainty of  $\pm0.5$ Mpc).
Uncertainties for the structural parameters  arise from the determination of the growth curve ($\Delta r_{eff}=1$\,arcsec, 
$\Delta \langle\mu\rangle_{eff}=0.3$\,mag ) and for the S\'ersic fit from numerics (see Table 2 for the corresponding errors).\\

One of our candidates (dw1326-35 from MJB15) is at the border of two different observation runs and is visible in both of 
them (see Fig.\,\ref{fieldDepth}). We performed photometry 
on the images of both runs to test the internal consistency of our photometry pipeline. We calculated the differences in apparent magnitude, which were $\Delta_r =0.21$\,mag and 
$\Delta_g =0.06$\,mag. These values are well within the estimated total uncertainty of 0.3\,mag. There exists an overlap of two known galaxies with the MJB15 region (KK200 and CenA-dE4). The photometric differences are 
$\Delta_{B}=0.095$\,mag and $\Delta_{B}=-0.144$\,mag, respectively, which is well within the estimated error.
\\

\begin{figure}
\includegraphics[width=9cm]{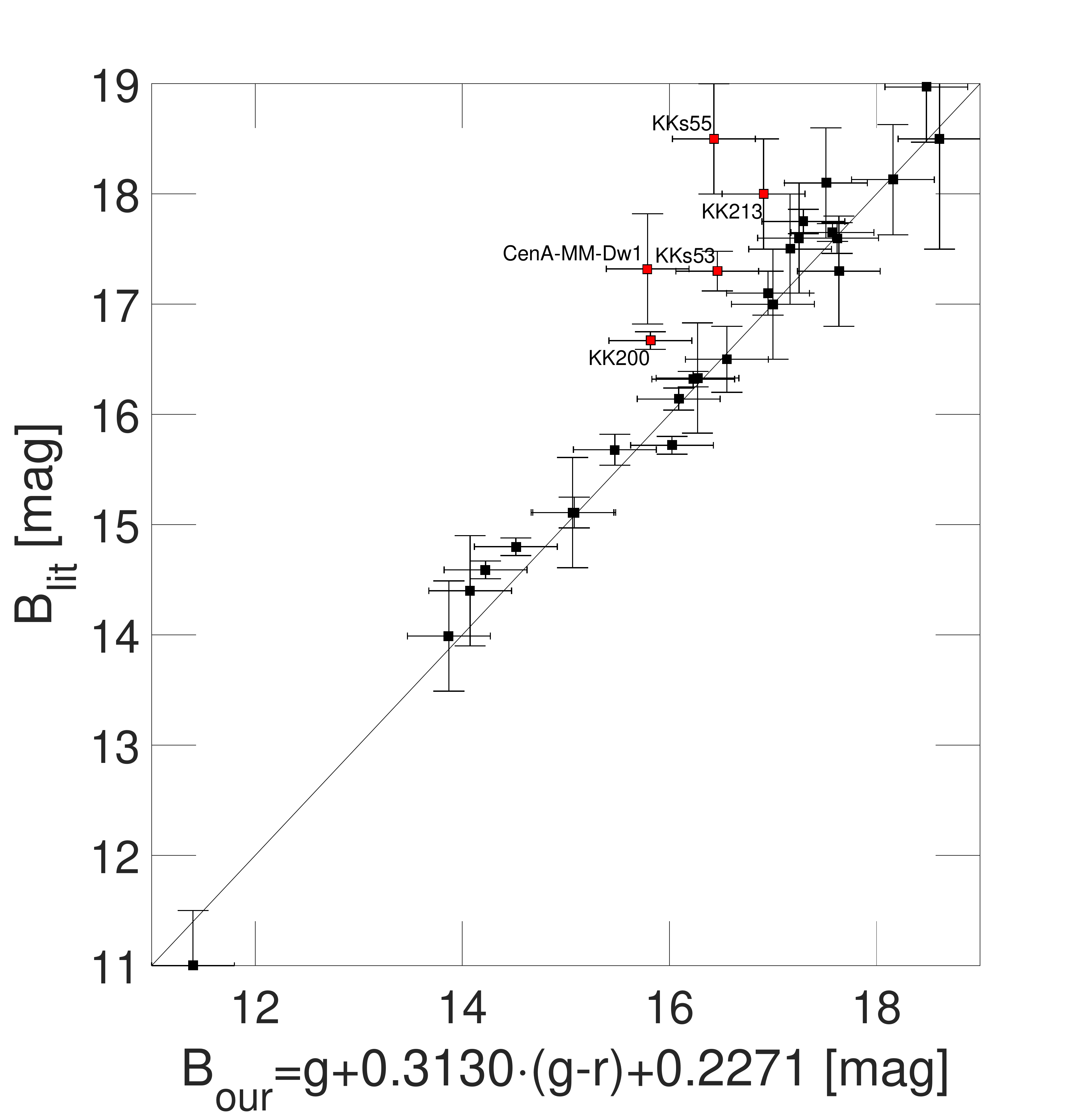}
\caption{Comparison of $B$-band photometry for known Centaurus group galaxies in our survey area.
Values from the literature (for references see Table 3) versus the $B$-band magnitude we derived from our $gr$
photometry. We adopt a conservative error of 0.4\,mag for our data, which includes the uncertainties for 
both filters. Photometric uncertainties for the literature values were taken from the publications. The unity line
is shown as a solid line. For five galaxies the magnitude difference is larger than the error tolerance (indicated in red). These discrepant cases are discussed in the text.}
  
  \label{diff}
\end{figure}

Another test of performance is to compare our photometry with literature values. For that purpose 
we transformed our $gr$ photometry into a $B$-band magnitude using the formula given in Section 2. We  
plot the total $B$ magnitude from the literature for 30 known dwarfs in our survey data versus the $B$\,magnitude 
from our photometry in Figure~\ref{diff}. The references for the literature values are given in Table\,\ref{table3}, and 
no adjustments were made for different methods to derive the photometry.
We note that 24 of the known dwarfs are missing in the lists provided in this paper and in MJB15 as they 
happen to lie either outside of our survey footprint (10), or are close to or even on the edge of a CCD field (7), are stretched over multiple CCD tiles (4), or in the case of the ultra-faint dwarfs are too faint to measure (3).  \\

 All but five galaxies (KK200, KKs53, KKs55, KK213, and CenA-MM-Dw1) 
agree within our estimated errors. If we exclude these five discrepant cases
the mean difference and standard deviation are $\langle \Delta\mu\rangle=0.10$\,mag  and $\sigma=0.37$\,mag, respectively. The discrepancy for the five objects can be explained as follows: (a) KK200 \citep{1994MNRAS.267..431M} 
was only integrated to the $\mu_b=26.75$ mag arcsec$^{-2}$ isophote, cutting the outskirts of the galaxy. The listed 
value is fainter than the real value (b) for KKs53, KKs55, and KK213 \citep{2001A&A...377..801H}, and the magnitudes were 
estimated by visual inspection only; no quantitative photometry was performed; and (c) for CenA-MM-Dw1, the case is described in detail in the following.  \\

Among the nine ultra-faint dwarfs found by \cite{2014ApJ...795L..35C,2015arXiv151205366C} photometry was 
possible for CenA-MM-Dw1, Dw4 and Dw9. We point out the good agreement for the photometric quantities of Dw4 and Dw9. For 
CenA-MM-Dw4, $\mu_{0,r}=25.1$ and $r_{eff,r}=20\farcs3$ versus $\mu_{0,r}=25.0$ and $r_{eff,r}=18\farcs6$ \citep{2015arXiv151205366C}; for
CenA-MM-Dw9, $\mu_{0,r}=25.9$ and $r_{eff,r}=24\farcs4$ versus $\mu_{0,r}=26.1$ and $r_{eff,r}=23\farcs4$ \citep{2015arXiv151205366C};
whereas for CenA-MM-Dw1, $\mu_{0,r}=25.1$ and $r_{eff,r}=65\farcs3$ versus $\mu_{0,r}=27.0$ and $r_{eff,r}=78\farcs6$ \citep{2014ApJ...795L..35C} clearly differs from our results. {Private communication with D.C. confirmed that our values are correct.}
CenA-MM-Dw2 is also visible on our DECam images but its small angular size and the presence of a number of bright foreground stars
(see Figure 2 of \cite{2015arXiv151205366C}) prevented us from conducting accurate photometry. Moreover, Dw3 is a tidal dwarf galaxy that is extended over 1.5 degrees making it impossible to perform aperture photometry, while Dw5 and Dw7, with central surface brightnesses
$\mu_{0,r}\approx 26.5$, were too faint. Dw6 is just visible when knowing the position, but too faint to be detected as dwarf galaxy. Dw8 is on the edge of an image.

\begin{figure}[Ht]
\centering
  \includegraphics[width=9cm]{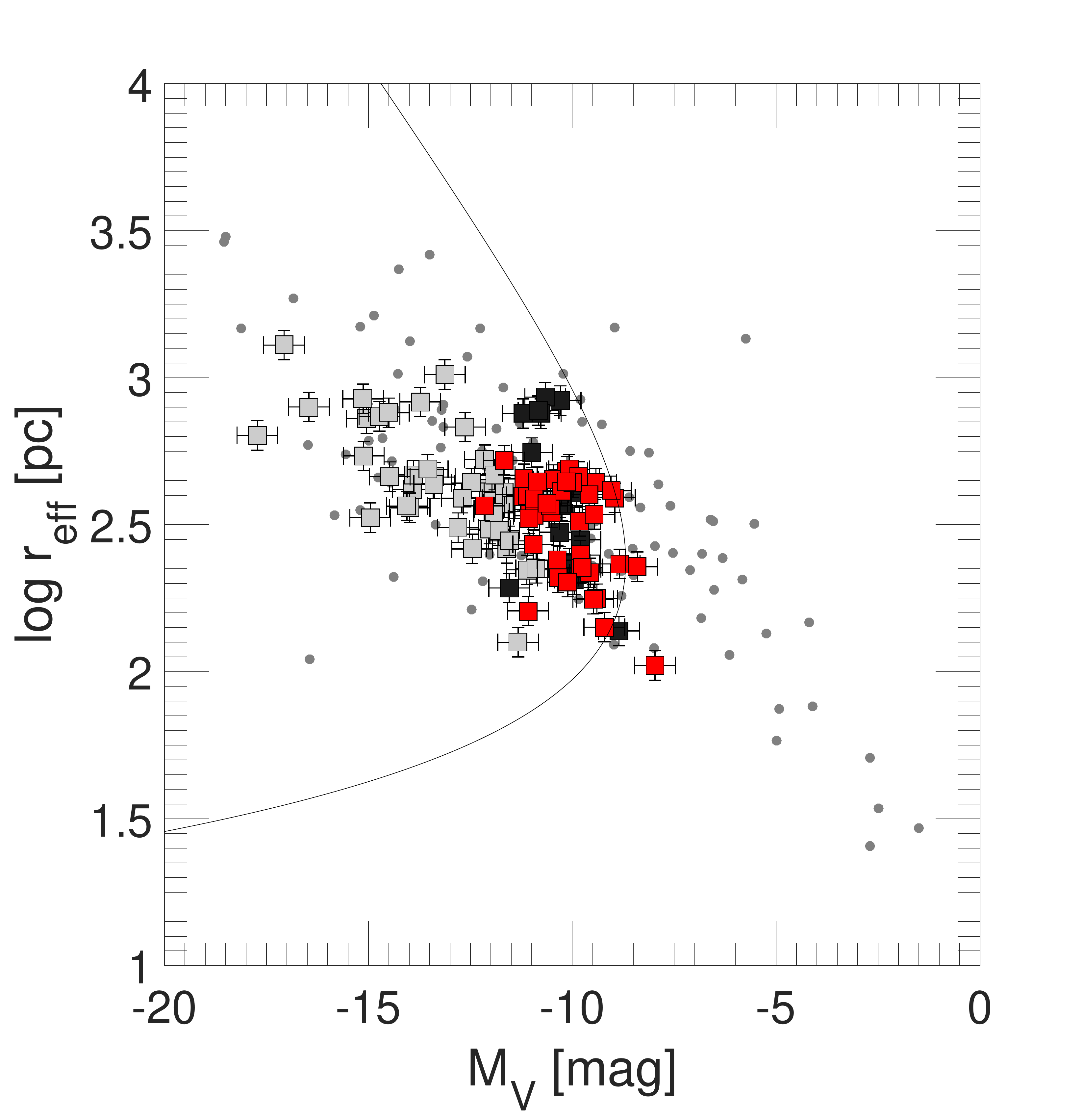}
  \caption{For our candidates, $\log(r_{eff})-M$ relation in the $V$ band (red squares), 
  all known dwarfs of the Centaurus group in the surveyed area for which photometry was 
  possible (gray filled squares), Local Group dwarf galaxies \citep[gray dots;\,][]{2012AJ....144....4M} and the 
  candidates from MJB15 (black squares).  Also plotted are conservative error bars of 0.5\,mag 
  for $M_V$ and 0.05 for $\log(r_{eff})$. Absolute magnitudes are based on a mean 
  distance of 4.5\,Mpc to all Centaurus galaxy candidates. The solid curve represents the completeness limit of the 
survey. It suggests that {most} low surface brightness {galaxies} larger than 26\arcsec in diameter at the isophotal 
magnitude of 28 mag arcsec$^{-2}$ {were} detected {(for more details see text)}.}
\label{completeness}
\end{figure}

In Table\,\ref{table2} we present the photometric data for the 41 newly detected dwarf galaxy candidates. 
Where possible we also performed photometry for known Centaurus group dwarfs in the survey area (Table\,\ref{table3}).
The quantities listed are as follows:
(1) name of candidate, or name, morphological type, and coordinates of the known galaxy;
(2+3) total apparent magnitude in the $g$ and $r$ bands;
(4+5) Galactic extinction values in $g$ and $r$ according to \citet{2011ApJ...737..103S};
(6) extinction-corrected absolute $r$-band magnitude. The assumed distance for the 
candidates is the mean distance of the Centaurus group (4.5\,Mpc). For the known 
galaxies we used the individual distances listed in the Updated Nearby Galaxy Catalog \citep{2013AJ....145..101K};
(7) extinction-corrected integrated $g-r$ color;
(8) S\'ersic central surface brightness in the $r$ band;
(9) S\'ersic scale length in the $r$ band;
(10) S\'ersic curvature index in the $r$ band;
(11) mean effective surface brightness in the $r$ band; and
(12) effective radius in the $r$ band.

\begin{table*}
\caption{Photometric and structural parameters of the new Centaurus group dwarf candidates in the surveyed region.\label{table2}}
\centering
\small
\setlength{\tabcolsep}{3pt}
\begin{tabular}{lccccrrcrccr}
\hline\hline 
Name & ${g_{tot}}$ & ${r_{tot}}$  & $A_{g}$ & $A_{r}$ & $M_{r}$ & $(g-r)_{0,tot}$ & $\mu_{0,r}$ & $r_{0,r}$ & $n_r$ & $\langle\mu\rangle_{eff,r}$ &  $r_{eff,r}$\\ 
 & mag & mag & mag & mag & mag & mag &  mag arcsec$^{-2}$ &arcsec & & mag arcsec$^{-2}$ &arcsec  \\ 
 (1)& (2) & (3) & (4) & (5) & (6) & (7) &  (8) & (9) &(10) & (11) & (12) \\ 
\hline \\
dw1240-42 & 17.94 & 17.21 & 0.366 & 0.253 & -11.29 & 0.613 & $24.61 \pm 0.06 $  & $15.35 \pm 0.70 $  & $1.45 \pm 0.10 $  & 25.24 & 16.0\\
dw1241-32 & 19.08 & 18.46 & 0.322 & 0.223 & -10.02 & 0.519 & $24.72 \pm 0.33 $  & $8.21 \pm 2.51 $  & $1.03 \pm 0.22 $  & 25.76 & 11.4\\
dw1243-42 & 18.40 & 17.76 & 0.329 & 0.227 & -10.72 & 0.542 & $24.88 \pm 0.14 $  & $13.04 \pm 1.77 $  & $1.08 \pm 0.16 $  & 25.78 & 15.8\\
dw1243-42b & 17.64 & 17.19 & 0.230 & 0.171 & -11.24 & 0.387 & $23.14 \pm 0.02 $  & $7.24 \pm 0.12 $  & $1.30 \pm 0.02 $  & 23.57 & 7.4\\
dw1251-40 & 19.54 & 19.21 & 0.358 & 0.248 & -9.30 & 0.220 & $24.33 \pm 0.20 $  & $4.56 \pm 0.84 $  & $1.09 \pm 0.17 $  & 25.32 & 6.5\\
dw1252-40 & 16.70 & 16.17 & 0.353 & 0.244 & -12.33 & 0.427 & $24.49 \pm 0.03 $  & $20.19 \pm 0.42 $  & $1.38 \pm 0.03 $  & 24.32 & 16.7\\
dw1252-43 & 19.39 & 18.94 & 0.321 & 0.222 & -9.54 & 0.353 & $24.88 \pm 0.06 $  & $8.13 \pm 0.44 $  & $1.37 \pm 0.10 $  & 25.53 & 8.11\\
dw1257-41 & 17.33 & 16.66 & 0.424 & 0.293 & -11.89 & 0.536 & $25.10 \pm 0.07 $  & $24.75 \pm 1.21 $  & $1.63 \pm 0.16 $  & 25.63 & 24.0\\
dw1258-37 & 18.63 & 18.04 & 0.182 & 0.126 & -10.34 & 0.532 & $26.48 \pm 0.09 $  & $26.50 \pm 1.44 $  & $3.30 \pm 1.28 $  & 26.78 & 22.0\\
dw1301-30 & 18.94 & 18.47 & 0.265 & 0.183 & -9.97 & 0.382 & $25.84 \pm 0.19 $  & $18.17 \pm 2.62 $  & $1.52 \pm 0.28 $  & 26.38 & 14.8\\
dw1302-40 & 18.58 & 17.78 & 0.372 & 0.258 & -10.74 & 0.684 & $25.79 \pm 0.13 $  & $21.87 \pm 2.00 $  & $1.73 \pm 0.35 $  & 26.40 & 20.6\\
dw1306-29 & 18.58 & 17.90 & 0.347 & 0.240 & -10.60 & 0.576 & $24.44 \pm 0.14 $  & $10.31 \pm 1.27 $  & $1.20 \pm 0.16 $  & 25.15 & 10.9\\
dw1314-28 & 17.58 & 17.04 & 0.260 & 0.180 & -11.39 & 0.453 & $24.74 \pm 0.08 $  & $18.34 \pm 1.09 $  & $1.49 \pm 0.13 $  & 25.35 & 18.1\\
dw1315-45 & 18.39 & 18.06 & 0.348 & 0.241 & -10.44 & 0.227 & $24.61 \pm 0.10 $  & $11.24 \pm 0.82 $  & $1.67 \pm 0.17 $  & 25.02 & 9.5\\
dw1318-21 & 18.06 & 17.26 & 0.353 & 0.245 & -11.24 & 0.691 & $23.93 \pm 0.49 $  & $8.14 \pm 3.68 $  & $0.93 \pm 0.25 $  & 24.72 & 12.4\\
dw1318-44 & 20.51 & 20.61 & 0.332 & 0.230 & -7.88 & -0.190 & $25.38 \pm 0.60 $  & $4.59 \pm 2.54 $  & $1.13 \pm 0.72 $  & 26.13 & 4.8\\
dw1321-27 & 18.67 & 18.13 & 0.210 & 0.145 & -10.27 & 0.473 & $26.47 \pm 0.14 $  & $32.94 \pm 2.72 $  & $1.87 \pm 0.71 $  & 26.89 & 22.3\\
dw1322-27 & 17.71 & 17.05 & 0.218 & 0.151 & -11.35 & 0.593 & $24.50 \pm 0.08 $  & $14.76 \pm 1.12 $  & $1.10 \pm 0.09 $  & 25.36 & 18.2\\
dw1322-39 & 17.60 & 17.12 & 0.293 & 0.203 & -11.34 & 0.387 & $25.03 \pm 0.21 $  & $18.70 \pm 3.42 $  & $1.15 \pm 0.19 $  & 25.74 & 20.7\\
dw1323-40 & 17.74 & 17.33 & 0.373 & 0.258 & -11.19 & 0.301 & $24.93 \pm 0.10 $  & $16.99 \pm 1.31 $  & $1.64 \pm 0.23  $  & 25.27 & 15.2\\
dw1323-40b & 18.16 & 17.84 & 0.401 & 0.277 & -10.69 & 0.193 & $25.44 \pm 0.13 $  & $18.09 \pm 2.03 $  & $1.35 \pm 0.19  $  & 26.06 & 17.1\\
dw1323-40c & 18.62 & 18.32 & 0.380 & 0.263 & -10.20 & 0.181 & $26.40 \pm 0.16 $  & $27.12 \pm 2.48 $  & $2.48 \pm 1.21  $  & 26.90 & 20.2\\
dw1326-37 & 18.90 & 18.47 & 0.226 & 0.156 & -9.95 & 0.358 & $25.49 \pm 0.11 $  & $14.28 \pm 1.17 $  & $1.72 \pm 0.27 $  & 25.57 & 10.2\\
dw1329-45 & 19.15 & 18.81 & 0.305 & 0.211 & -9.66 & 0.243 & $25.48 \pm 0.09 $  & $12.70 \pm 0.81 $  & $1.84 \pm 0.27 $  & 25.86 & 9.9\\
dw1330-38 & 19.44 & 18.63 & 0.154 & 0.107 & -9.74 & 0.758 & $25.91 \pm 0.42 $  & $11.81 \pm 5.42 $  & $0.99 \pm 0.47 $  & 27.14 & 20.1\\
dw1331-37 & 20.15 & 19.06 & 0.256 & 0.177 & -9.38 & 1.005 & $26.40 \pm 0.18 $  & $16.71 \pm 1.83 $  & $1.90 \pm 0.72 $  & 27.28 & 17.8\\
dw1331-40 & 20.53 & 19.80 & 0.297 & 0.206 & -8.67 & 0.637 & $26.38 \pm 0.35 $  & $13.11 \pm 3.69 $  & $1.41 \pm 0.60 $  & 26.89 & 10.4\\
dw1336-44 & 19.54 & 18.80 & 0.400 & 0.277 & -9.74 & 0.618 & $25.09 \pm 0.05 $  & $11.26 \pm 0.36 $  & $2.45 \pm 0.20 $  & 25.34 & 8.07\\
dw1337-41 & 18.98 & 18.88 & 0.301 & 0.208 & -9.59 & 0.006 & $26.81 \pm 0.14 $  & $28.19 \pm 2.43 $  & $2.04 \pm 0.51 $  & 27.29 & 18.3\\
dw1337-44 & 18.72 & 18.86 & 0.353 & 0.244 & -9.65 & -0.240 & $25.05 \pm 0.40 $  & $7.78 \pm 2.98 $  & $1.02 \pm 0.27 $  & 26.06 & 10.3\\
dw1341-43 & 17.92 & 17.47 & 0.309 & 0.214 & -11.00 & 0.348 & $25.58 \pm 0.10 $  & $21.37 \pm 1.10 $  & $2.14 \pm 0.49 $  & 26.06 & 20.2\\
dw1342-43 & 17.98 & 17.23 & 0.263 & 0.182 & -11.21 & 0.676 & $24.36 \pm 0.09 $  & $12.69 \pm 0.88 $  & $1.38 \pm 0.11 $  & 25.19 & 15.5\\
dw1343-34 & 19.81 & 19.07 & 0.205 & 0.142 & -9.34 & 0.681 & $26.82 \pm 0.20 $  & $25.42 \pm 3.52 $  & $1.82 \pm 0.52 $  & 27.45 & 18.9\\
dw1357-28 & 19.36 & 18.70 & 0.204 & 0.141 & -9.70 & 0.598 & $26.40 \pm 0.16 $  & $21.29 \pm 1.77 $  & $2.90 \pm 1.36 $  & 26.72 & 15.6\\
dw1401-32 & 18.37 & 17.59 & 0.217 & 0.150 & -10.82 & 0.715 & $25.16 \pm 0.08 $  & $17.15 \pm 1.24 $  & $1.37 \pm 0.15 $  & 25.69 & 16.8\\
dw1403-33 & 18.69 & 17.82 & 0.232 & 0.160 & -10.60 & 0.803 & $25.84 \pm 0.10 $  & $23.87 \pm 1.28 $  & $2.06 \pm 0.37 $  & 26.16 & 18.8\\
dw1406-29 & 18.62 & 18.56 & 0.197 & 0.137 & -9.83 & 0.000 & $26.12 \pm 0.22 $  & $16.42 \pm 2.51 $  & $1.69 \pm 0.63 $  & 27.27 & 21.1\\
dw1409-33 & 18.59 & 18.39 & 0.229 & 0.158 & -10.02 & 0.124 & $26.18 \pm 0.25 $  & $18.15 \pm 3.32 $  & $1.46 \pm 0.58 $  & 26.96 & 20.0\\
dw1410-34 & 17.68 & 17.41 & 0.239 & 0.165 & -11.01 & 0.204 & $23.42 \pm 0.58 $  & $2.90 \pm 2.37 $  & $0.55 \pm 0.14 $  & 25.70 & 17.7\\
dw1413-34 & 19.98 & 19.34 & 0.231 & 0.160 & -9.08 & 0.569 & $25.80 \pm 0.13 $  & $10.29 \pm 1.00 $  & $1.59 \pm 0.34 $  & 26.48 & 10.6\\
dw1415-32 & 18.69 & 18.09 & 0.227 & 0.157 & -10.32 & 0.525 & $23.89 \pm 0.18 $  & $4.91 \pm 1.01 $  & $0.85 \pm 0.11 $  & 24.94 & 9.2\\

\hline
\end{tabular}
\tablefoot{{Absolute {magnitudes} in column 6 assume a mean distance of 4.5\,Mpc. Total magnitudes have a mean uncertainty of 0.3 mag (see text). The last digit of the listed parameters (hundredth of magnitude) is therefore not significant. However, we leave this digit here so as not to introduce rounding errors, should these quantities be used in further arithmetic operations.}}
\end{table*}

%\begin{table*}
\begin{sidewaystable*}[TT]
\setlength{\tabcolsep}{3pt}
\vspace*{15cm}
\caption{Photometric parameters of known dwarf galaxies in the surveyed region. \label{table3}}
\centering
\small
\begin{tabular}{lccccccccrcrccc}
\hline\hline 
 & &$\alpha$ & $\delta$ & ${g_{tot}}$ & ${r_{tot}}$  & $A_{g}$ & $A_{r}$ & $M_{r}$ & $(g-r)_{0,tot}$ & $\mu_{0,r}$ & $r_{0,r}$ & $n_r$ & $\langle\mu\rangle_{eff,r}$ &  $r_{eff,r}$  \\ 
Names (alt. names) & type & (J2000.0) & (J2000.0) & mag & mag & mag & mag & mag & mag &  mag arcsec$^{-2}$ &arcsec & & mag arcsec$^{-2}$ &arcsec \\ 
 (1)& & &  &(2) & (3) & (4) & (5) & (6) & (7) & (8) &  (9) & (10) &(11) & (12) \\ 
\hline \\
KKs51 & dE & 12:44:21 & -42:56:23 & 17.22 & 16.61 & 0.288 & 0.199 & -11.36 & 0.513 & $22.98 \pm 0.07 $  & $7.37 \pm 0.47 $  & $0.98 \pm 0.04 $  & 24.16 & 12.7\\
ESO381-018 &dIrr& 12:44:42 & -35:57:59 & 15.47 & 14.42 & 0.208 & 0.144 & -14.39 & 0.978 & $21.29 \pm 0.02 $  & $11.60 \pm 0.19 $  & $1.27 \pm 0.02 $  & 22.06 & 13.6\\
ESO381-020 &dIrr &12:46:00 & -33:50:13& 13.81 & 13.69 & 0.217 & 0.150 & -15.14 & 0.051 & $22.77 \pm 0.02 $  & $31.36 \pm 0.38 $  & $1.38 \pm 0.02 $  & 23.29 & 31.9\\
ESO443-009&dIrr &12:54:54 & -28:20:27 & 16.70 & 16.50 & 0.212 & 0.147 & -12.52 & 0.142 & $24.23 \pm 0.03 $  & $18.93 \pm 0.33 $  & $2.12 \pm 0.08 $  & 24.45 & 15.1\\
Cen6 (KK182) &dIrr &13:05:02 & -40:04:58& 15.94 & 15.63 & 0.339 & 0.234 & -13.46 & 0.206 & $23.00 \pm 0.04 $  & $13.98 \pm 0.44 $  & $1.36 \pm 0.04 $  & 23.64 & 15.0\\
MCG-04-31-038 &dIrr& 13:09:36 & -27:08:26& 14.73 & 14.42 & 0.252 & 0.174 & -15.20 & 0.235 & $21.60 \pm 0.02 $  & $12.98 \pm 0.17 $  & $1.40 \pm 0.02 $  & 22.29 & 14.3\\
Cen7 (KKs53) & dE,N&13:11:14 & -38:54:22& 16.03 & 15.38 & 0.300 & 0.208 & -12.14 & 0.552 & $23.84 \pm 0.02 $  & $25.26 \pm 0.38 $  & $1.52 \pm 0.03 $  & 24.33 & 24.5\\
CenA-dE1 (KK189) & dE&13:12:45 & -41:49:55& 16.93 & 16.49 & 0.366 & 0.253 & -11.93 & 0.319 & $23.54 \pm 0.03 $  & $12.35 \pm 0.28 $  & $1.23 \pm 0.03 $  & 24.32 & 14.4\\
ESO269-066 &dE,N &13:13:08 & -44:53:21& 13.78 & 13.12 & 0.314 & 0.217 & -14.96 & 0.570 & $22.46 \pm 0.01 $  & $34.37 \pm 0.25 $  & $1.20 \pm 0.01 $  & 23.17 & 40.6\\
NGC5011C &dE,N &13:13:11 & -43:15:55& 14.08 & 13.43 & 0.398 & 0.276 & -14.69 & 0.528 & $20.92 \pm 0.05 $  & $9.51 \pm 0.41 $  & $0.79 \pm 0.01 $  & 22.48 & 25.4\\
KK195&dIrr? &13:21:08 & -31:31:45 & 16.77 & 16.91 & 0.203 & 0.140 & -11.81 & -0.200 & $24.19 \pm 0.09 $  & $11.26 \pm 1.13 $  & $0.91 \pm 0.06 $  & 25.34 & 18.4\\
CenA-dE2 (KKs54) &dSph &13:21:32 & -31:53:11 & 17.76 & 17.24 & 0.215 & 0.148 & -11.35 & 0.453 & $24.98 \pm 0.08 $  & $20.29 \pm 1.06 $  & $1.74 \pm 0.17 $  & 25.38 & 16.6\\
KK196 & dIrr?&13:21:47 & -45:03:47& 15.70 & 15.20 & 0.276 & 0.191 & -12.97 & 0.412 & $22.69 \pm 0.03 $  & $13.03 \pm 0.37 $  & $1.11 \pm 0.03 $  & 23.27 & 16.0\\
KK197 (SGC1319.1-4216)& dE,N&13:22:01 & -42:32:08 & 15.02 & 14.31 & 0.511 & 0.353 & -13.95 & 0.548 & $23.87 \pm 0.02 $  & $39.46 \pm 0.63 $  & $1.21 \pm 0.02 $  & 24.56 & 44.4\\
KKs55&dSph &13:22:12 & -42:43:44& 16.02 & 15.44 & 0.472 & 0.327 & -12.80 & 0.428 & $24.92 \pm 0.02 $  & $45.55 \pm 0.74 $  & $1.90 \pm 0.07 $  & 25.28 & 36.4\\
Cen8 (KK198) &dIrr &13:22:56 & -33:34:22& 17.20 & 16.74 & 0.043 & 0.030 & -11.73 & 0.445 & $23.64 \pm 0.04 $  & $12.48 \pm 0.36 $  & $1.52 \pm 0.06 $  & 24.12 & 11.7\\
CenA-MM-Dw4 & dSph&13:23:02 & -41:47:10& 18.16 & 17.48 & 0.415 & 0.287 & -10.75 & 0.551 & $25.10 \pm 0.37 $  & $14.40 \pm 5.32 $  & $0.99 \pm 0.29 $  & 26.03 & 20.3\\
AM1320-230 &dE &13:23:29 & -23:23:35& 16.83 & 16.51 & 0.267 & 0.184 & -12.11 & 0.237 & $23.02 \pm 0.05 $  & $8.28 \pm 0.41 $  & $1.04 \pm 0.04 $  & 24.10 & 12.8\\
AM1321-304 (KK200) &dE/dIrr &13:24:36 & -30:58:18& 15.39 & 14.76 & 0.228 & 0.158 & -13.77 & 0.554 & $22.23 \pm 0.02 $  & $11.57 \pm 0.28 $  & $0.92 \pm 0.01 $  & 23.40 & 21.1\\
CenA-MM-Dw1 &dSph &13:30:14 & -41:53:25& 15.37 & 14.76 & 0.419 & 0.290 & -13.33 & 0.605 & $25.11  \pm  0.02 $  & $65.29 \pm  1.03 $  & $1.65 \pm  0.04 $  & 25.61 & 58.3\\
CenA-MM-Dw9 &dSph &13:33:01 & -42:31:48& 18.12 & 17.72 & 0.382 & 0.265 & -10.44 & 0.288 & $25.89 \pm 0.14 $  & $23.41 \pm 2.41 $  & $1.72 \pm 0.34 $  & 26.71 & 24.4\\
HIPASSJ1337-39  &dIrr &13:37:25 & -37:53:48& 16.31 & 16.26 & 0.247 & 0.171 & -12.43 & -0.020 & $23.01 \pm 0.05 $  & $11.04 \pm 0.35 $  & $1.48 \pm 0.05 $  & 23.48 & 10.6\\
KKs57& dSph&13:41:38 & -42:34:55 & 17.28 & 17.28 & 0.297 & 0.206 & -10.84 & -0.090 & $24.18 \pm 0.10 $  & $12.23 \pm 0.91 $  & $1.46 \pm 0.13 $  & 24.80 & 12.0\\
KK211 (AM1339-445) &dE,N &13:42:05 & -45:12:20& 15.80 & 15.15 & 0.368 & 0.254 & -12.92 & 0.530 & $23.42 \pm 0.02 $  & $24.68 \pm 0.33 $  & $1.63 \pm 0.04 $  & 23.86 & 21.8\\
KK213 &dE &13:43:35 & -43:46:09& 16.60 & 16.37 & 0.322 & 0.223 & -11.72 & 0.131 & $23.89 \pm 0.07 $  & $12.63 \pm 0.91 $  & $1.06 \pm 0.07 $  & 25.01 & 20.5\\
ESO325-011 & S/Irr&13:45:00 & -41:51:37& 13.55 & 13.28 & 0.291 & 0.201 & -14.57 & 0.181 & $23.22 \pm 0.01 $  & $50.28 \pm 0.46 $  & $1.56 \pm 0.02 $  & 23.67 & 46.1\\
CenA-dE4 (KK218)&dE &13:46:40 & -29:58:41& 17.24 & 16.77 & 0.198 & 0.137 & -11.83 & 0.409 & $24.16 \pm 0.10 $  & $13.95 \pm 1.28 $  & $1.13 \pm 0.10 $  & 25.01 & 17.0\\
HIPASSJ1348-37 &dIrr &13:48:47 & -37:58:29& 16.95 & 16.73 & 0.257 & 0.178 & -12.20 & 0.140 & $24.25 \pm 0.13 $  & $13.54 \pm 1.81 $  & $0.97 \pm 0.09 $  & 25.23 & 19.2\\
ESO383-087 &S/Irr &13:49:17 & -36:03:48& 10.98 & 10.40 & 0.237 & 0.164 & -17.28 & 0.513 & $21.03 \pm 0.01 $  & $48.84 \pm 0.59 $  & $1.06 \pm 0.01 $  & 22.04 & 83.4\\
ESO384-016 &dE/dIrr& 13:45:04 & -35:05:21& 14.69 & 14.19 & 0.245 & 0.169 & -14.23 & 0.427 & $21.01 \pm 0.04 $  & $7.72 \pm 0.29 $  & $0.85 \pm 0.02 $  & 22.37 & 16.7\\
\hline
\end{tabular}
\tablefoot{
References for the $B$-band photometry: ESO443-009 and ESO383-087 \citep{1989spce.book.....L}; AM1321-304/KK200 \citep{1994MNRAS.267..431M}; HIPASSJ1337-39 and HIPASSJ1348-37 \citep{1999ApJ...524..612B}; CenA-dE1/KK189, KK197/SGC1319.1-4216, Cen8/KK198, KK211/AM1339-445, CenA-dE2/KKs54, ESO381-018, CenA-dE4/KK218, ESO384-016, ESO269-066 and KK196 \citep{2000AJ....119..593J}; Cen7/KKs53 and KKs55 \citep{2001A&A...377..801H}; AM1320-230, ESO325-011, ESO381-020, Cen6/KK182, KK195, KK213, KKs51 and KKs57 \citep{2002A&A...385...21K,2004AJ....127.2031K,2013AJ....145..101K}; NGC5101C \citep{2007AJ....133.1756S}; MCG-04-31-038 \citep{2009MNRAS.397.1672M}; CenA-MM-Dw1, CenA-MM-Dw4 and CenA-MM-Dw9 \citep{2014ApJ...795L..35C,2015arXiv151205366C}. No adaption for different photometry techniques were made.}
%\end{table*}
\end{sidewaystable*}

\begin{figure*}
\centering
\includegraphics[width=3.6cm]{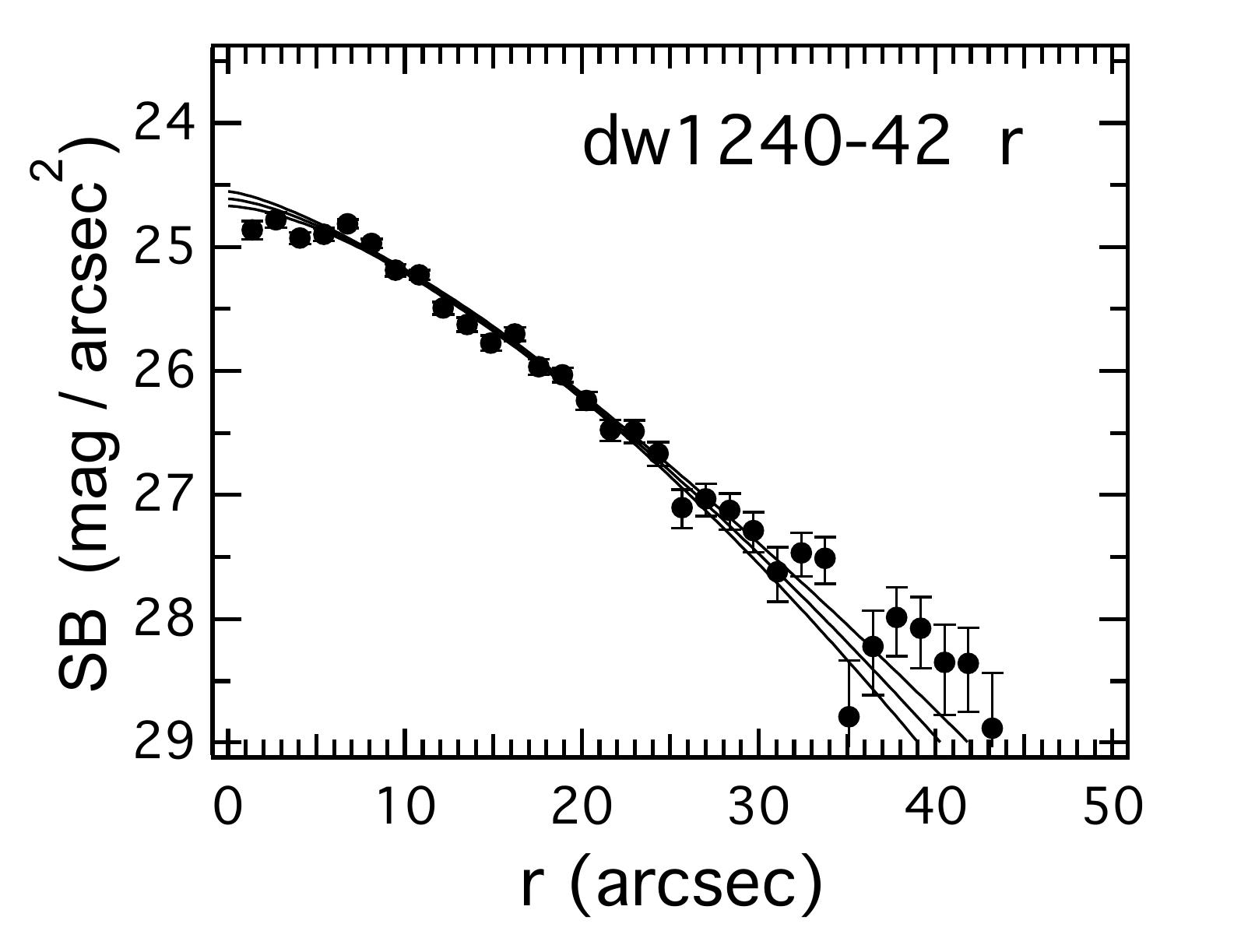}
\includegraphics[width=3.6cm]{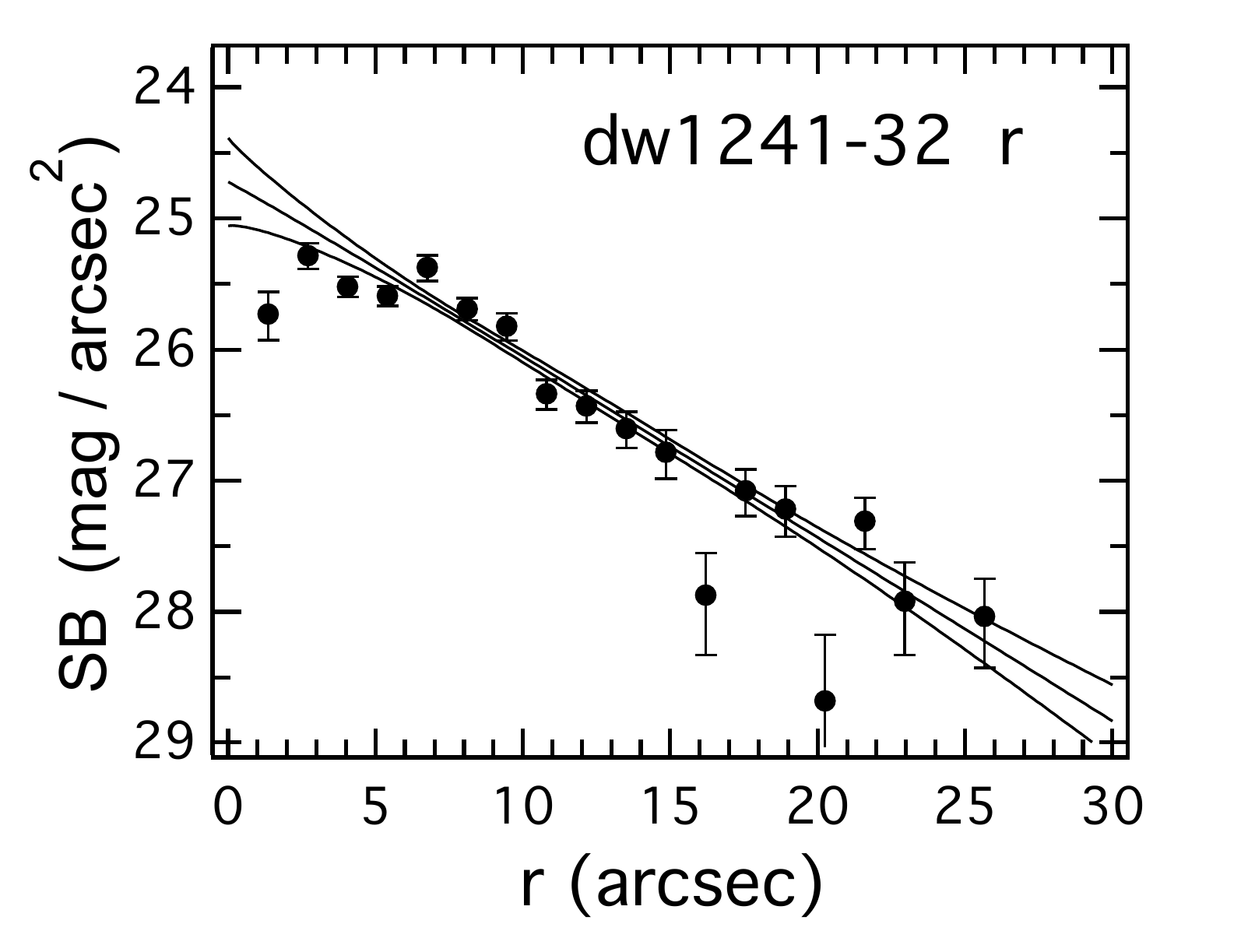}
\includegraphics[width=3.6cm]{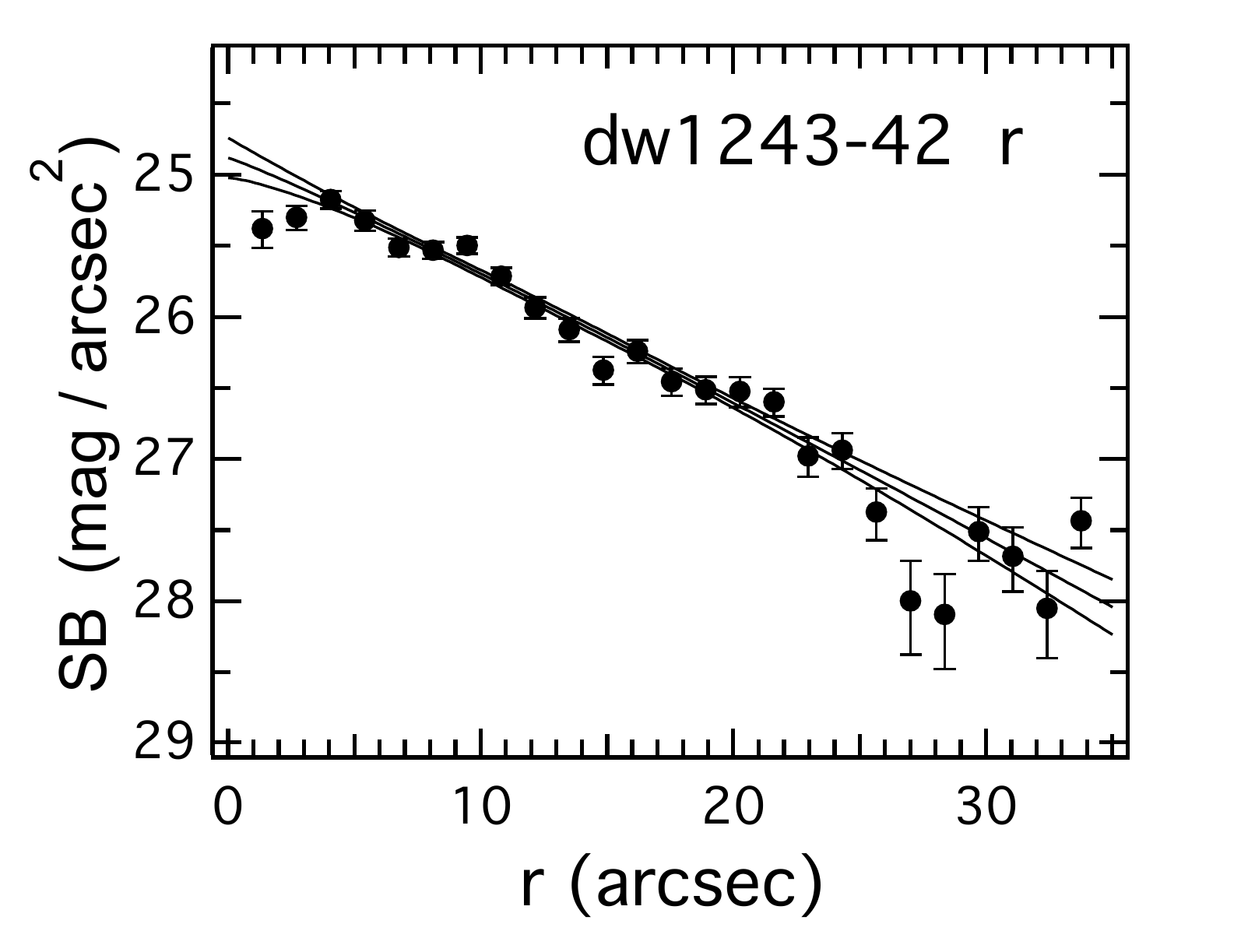}
\includegraphics[width=3.6cm]{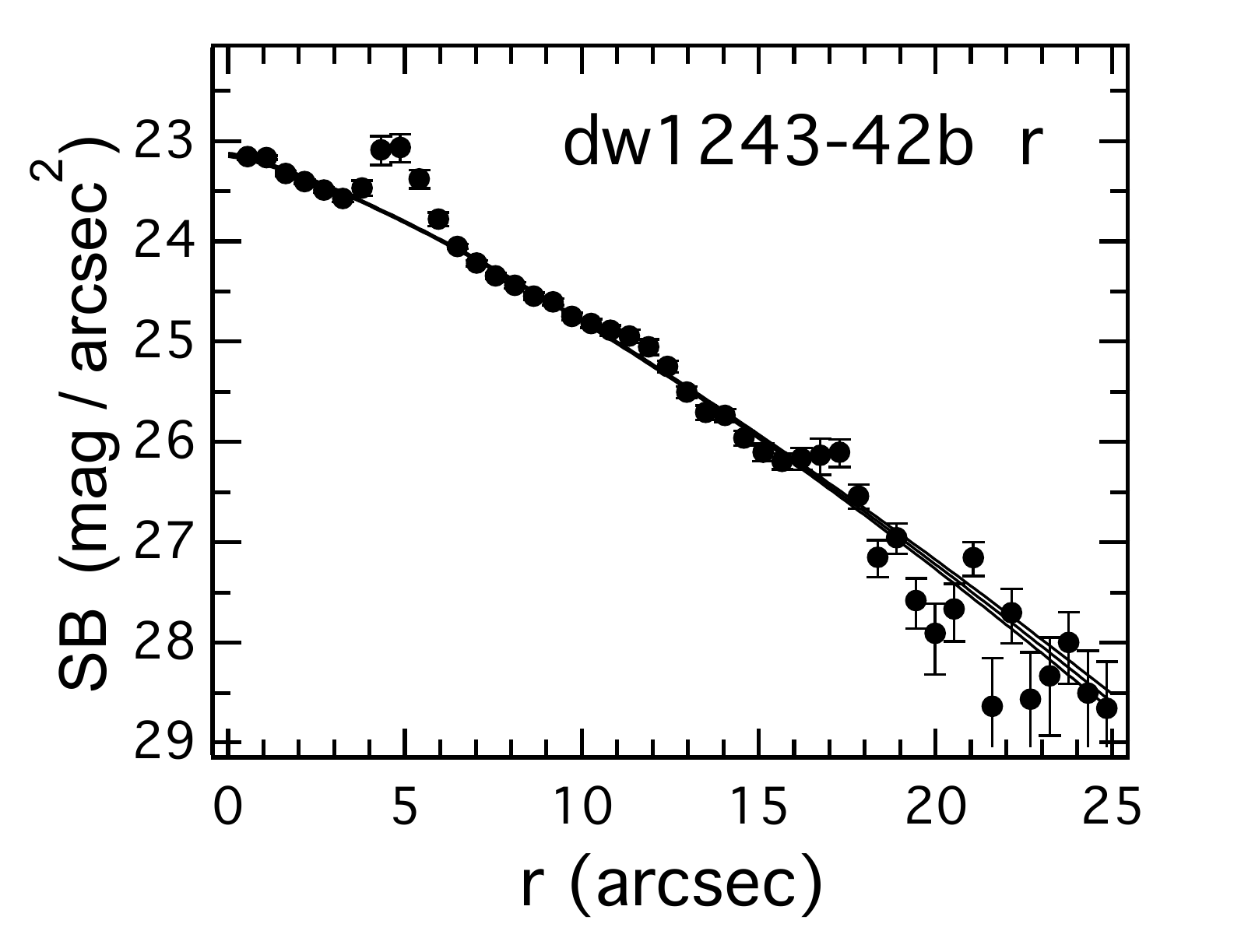}
\includegraphics[width=3.6cm]{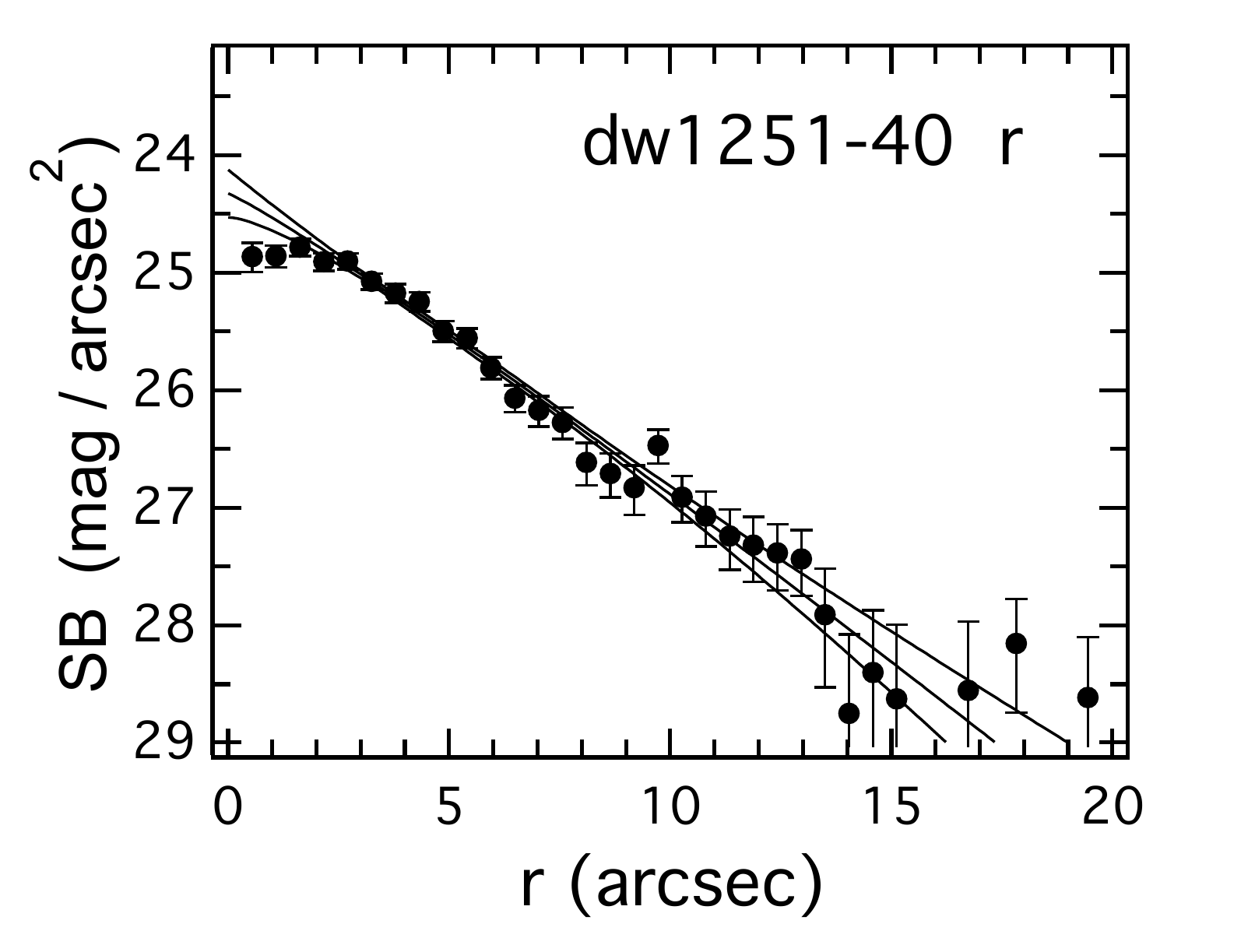}\\
\includegraphics[width=3.6cm]{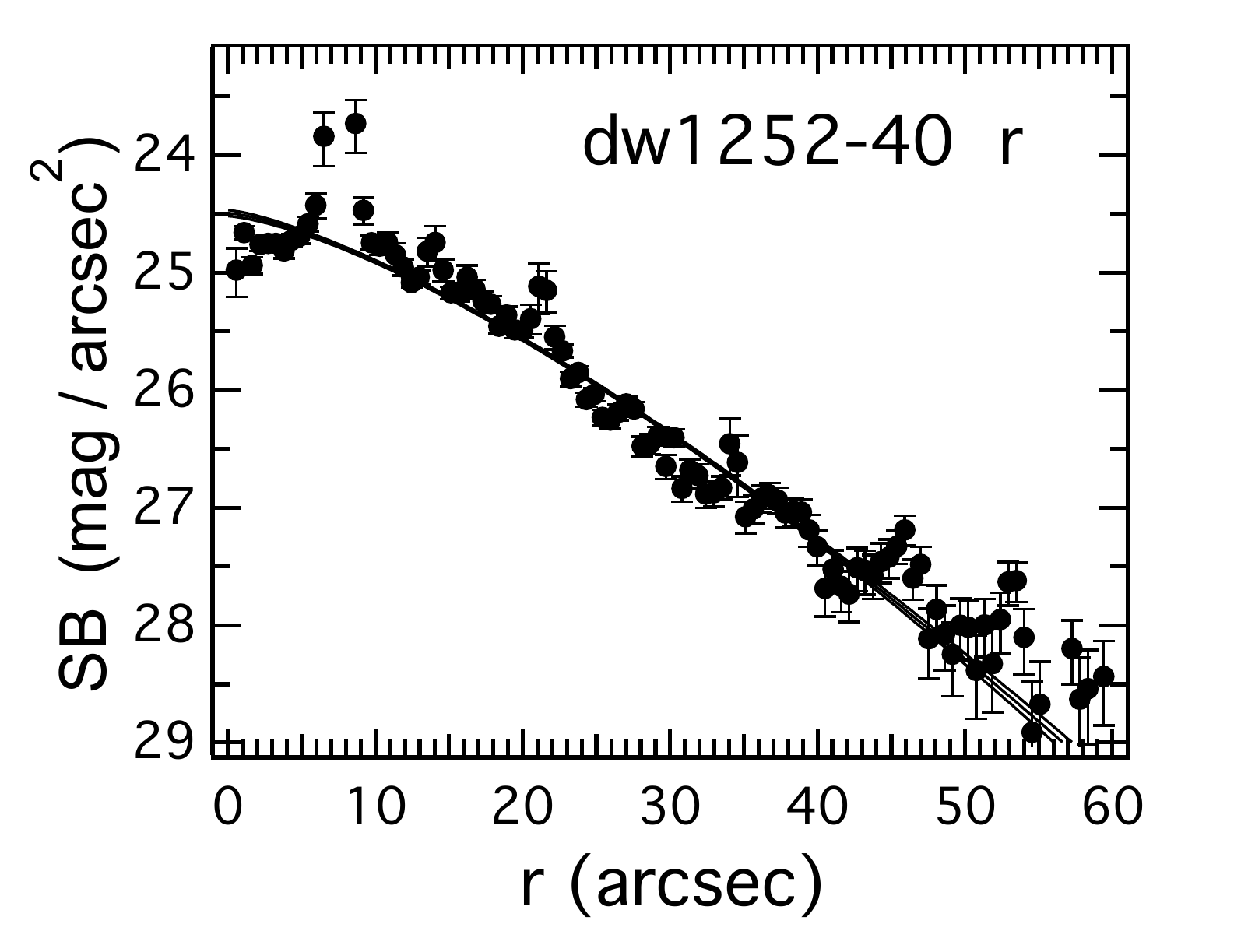}
\includegraphics[width=3.6cm]{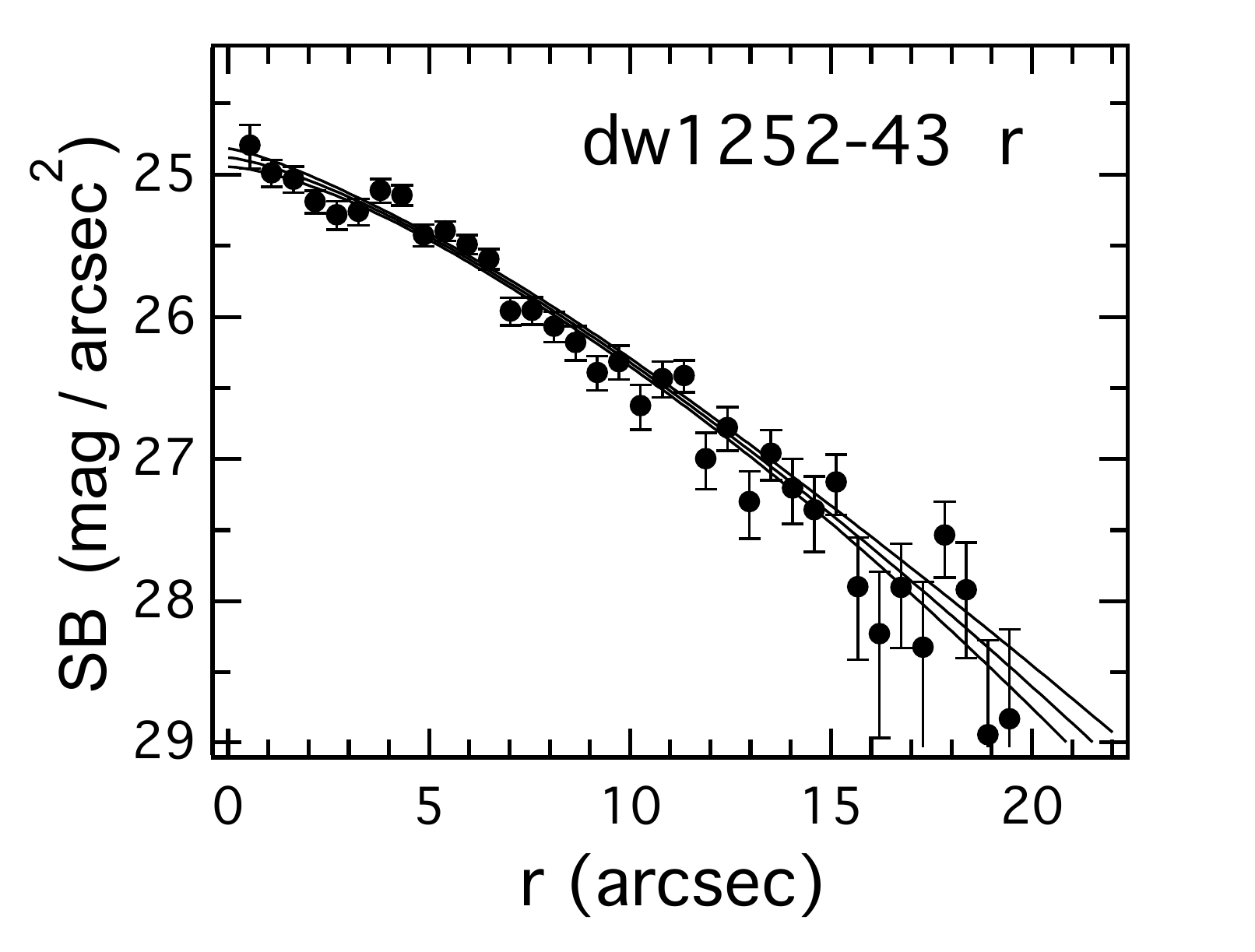}
\includegraphics[width=3.6cm]{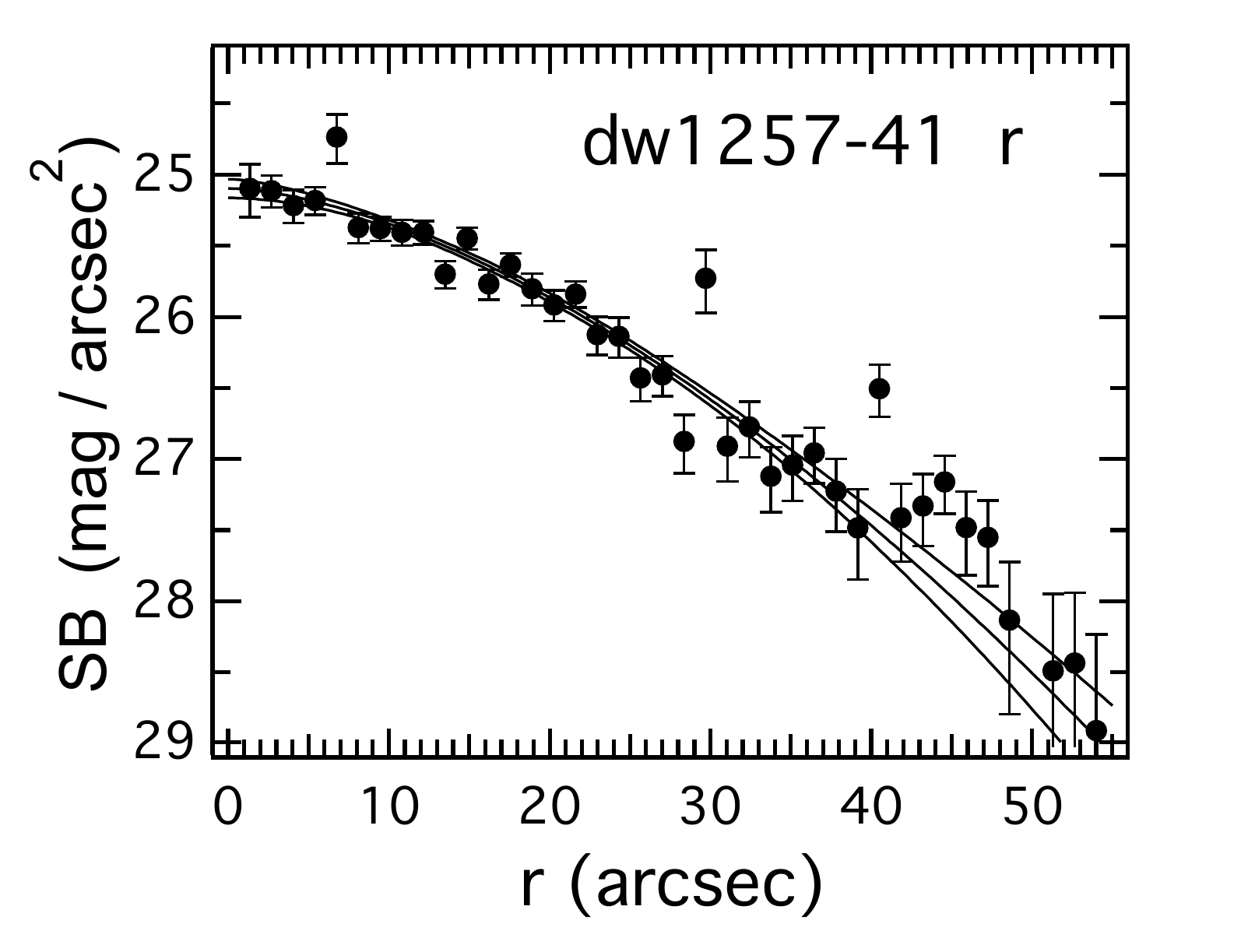}
\includegraphics[width=3.6cm]{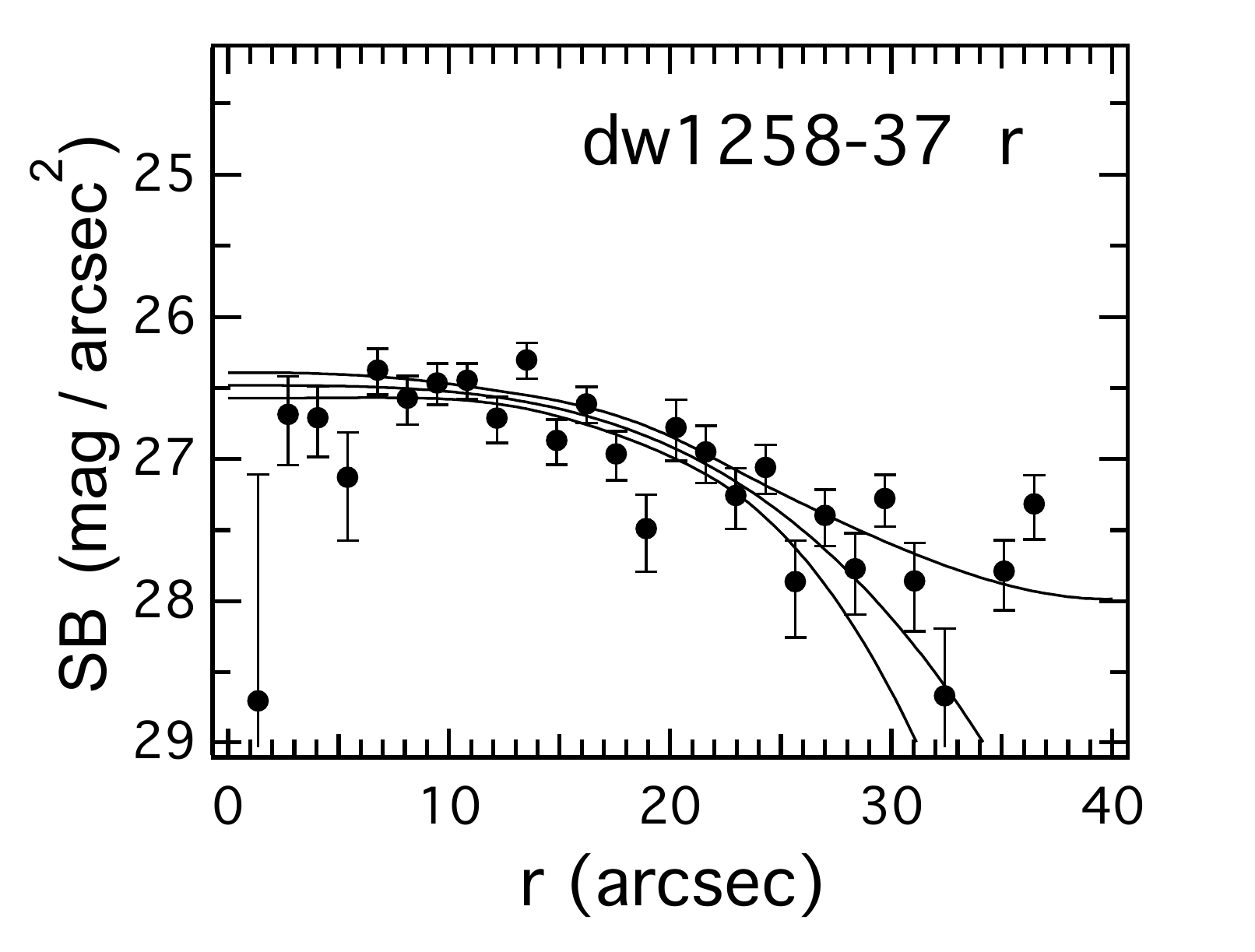}  
\includegraphics[width=3.6cm]{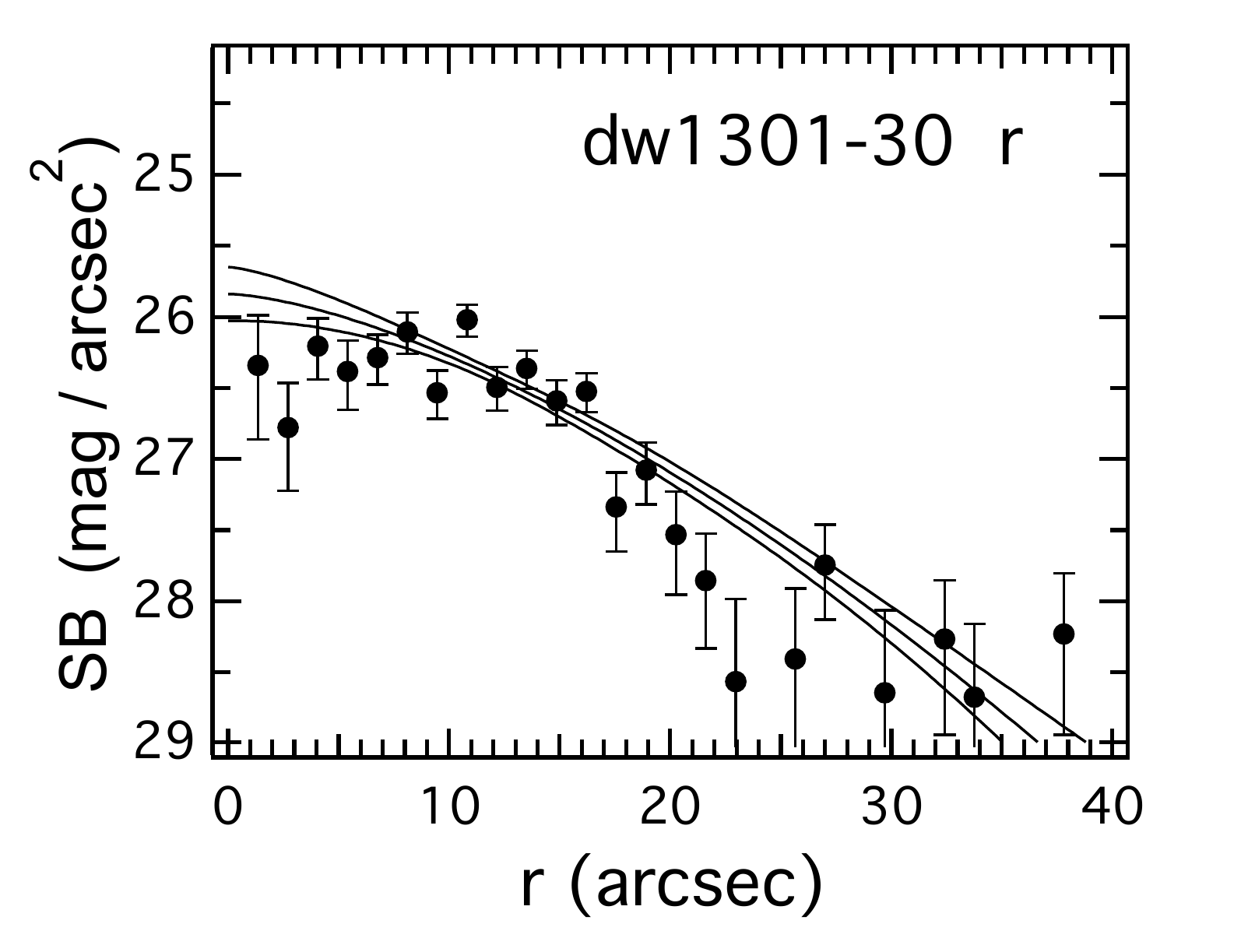}\\
\includegraphics[width=3.6cm]{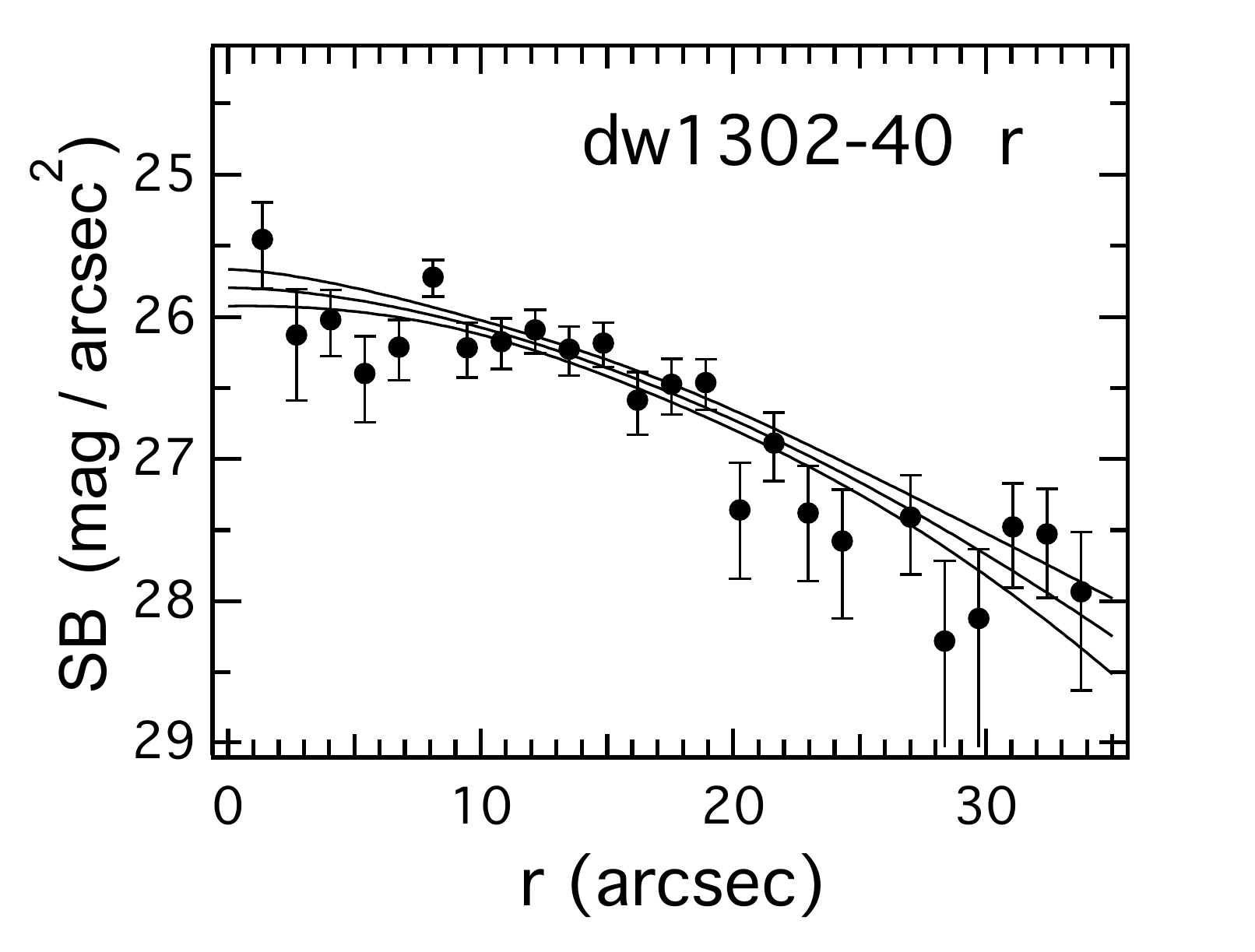}
\includegraphics[width=3.6cm]{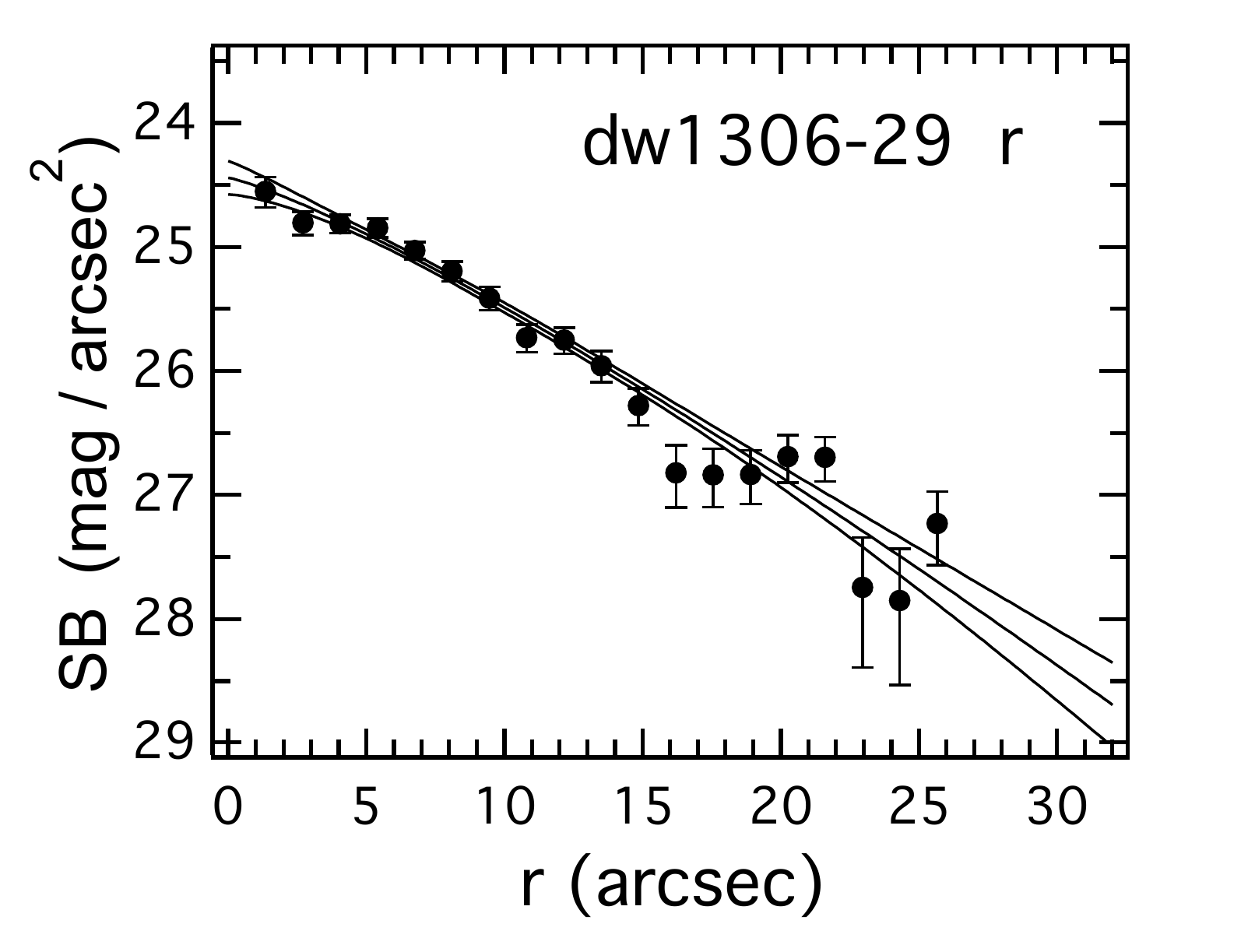}
\includegraphics[width=3.6cm]{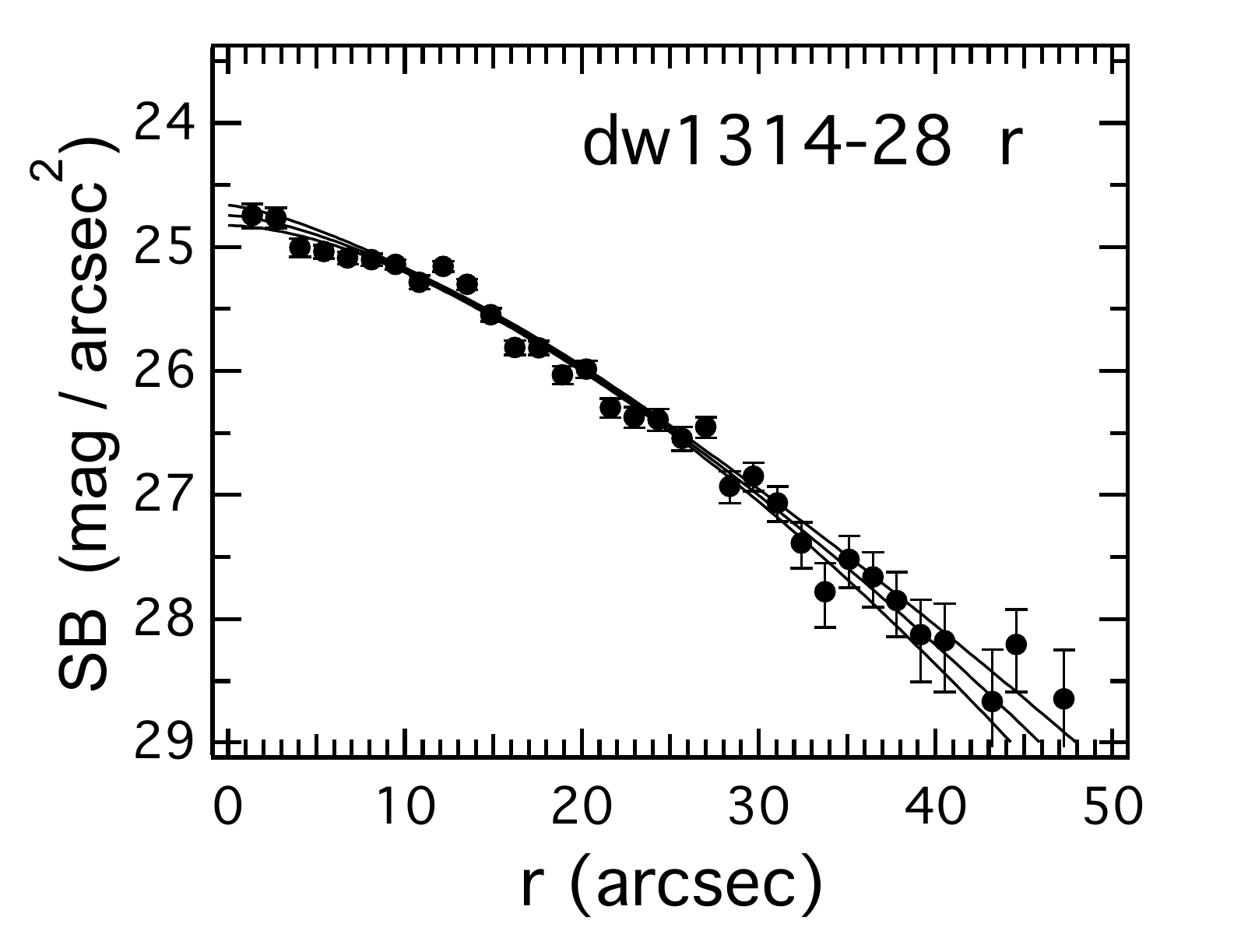} 
\includegraphics[width=3.6cm]{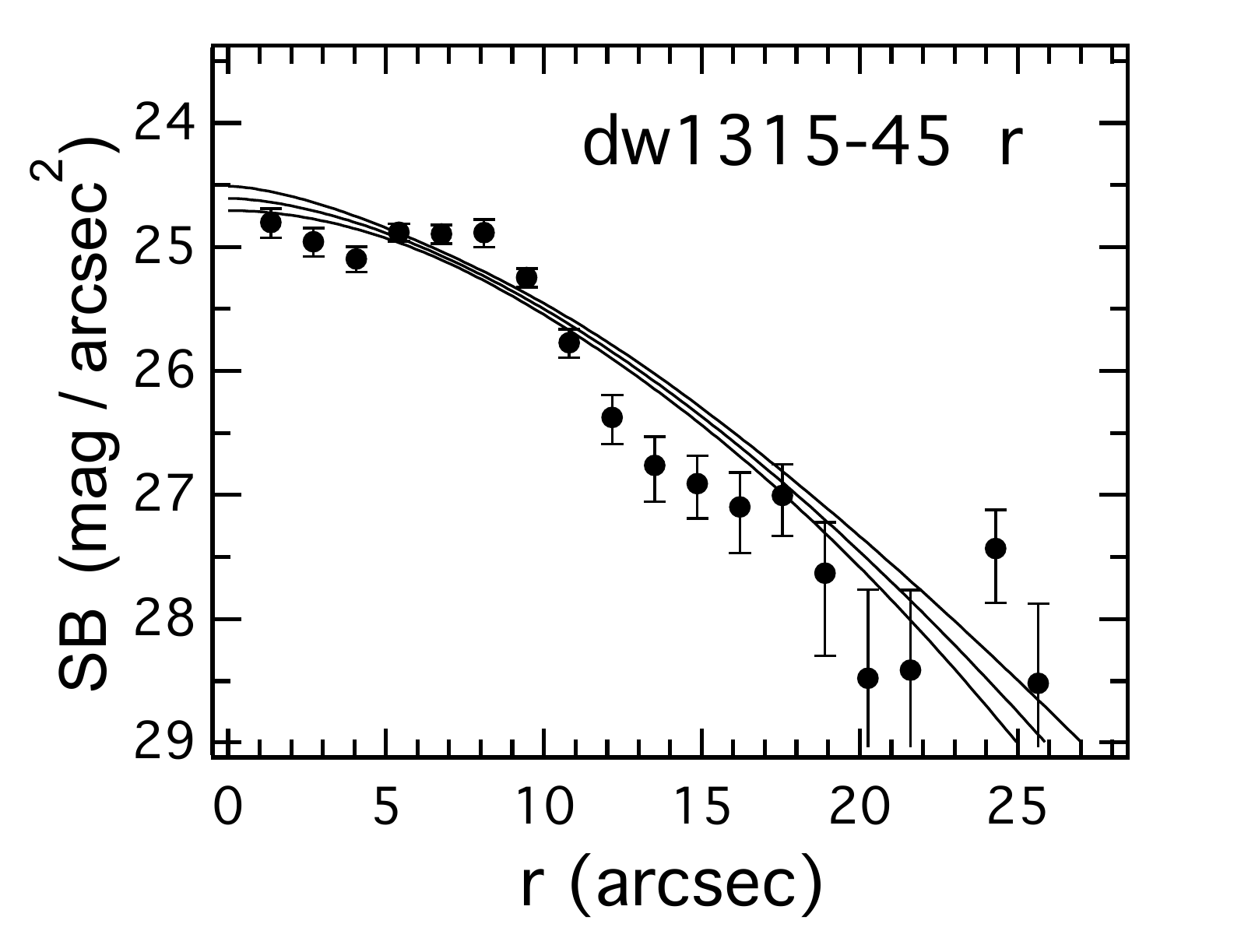}
\includegraphics[width=3.6cm]{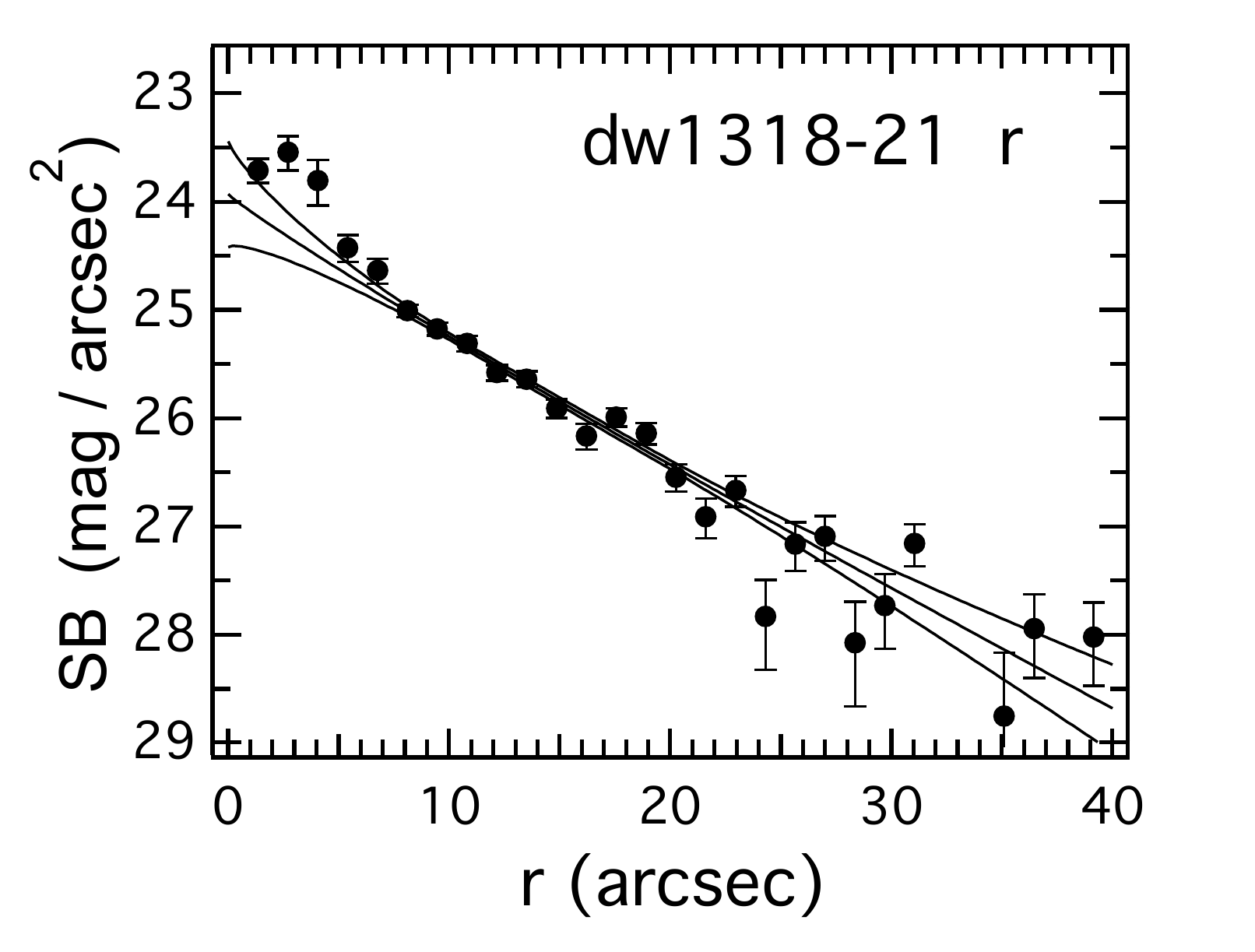}\\ 
\includegraphics[width=3.6cm]{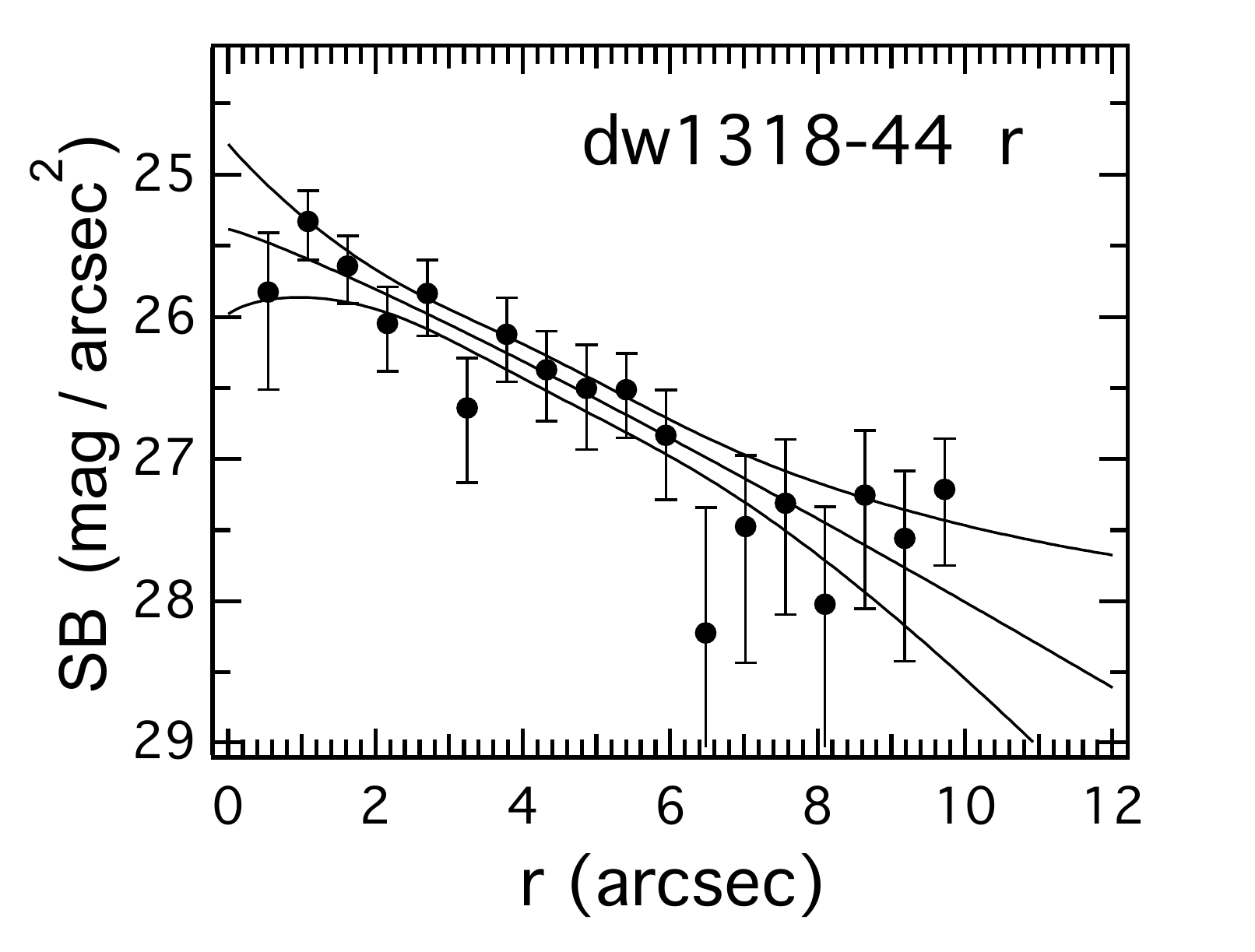}
\includegraphics[width=3.6cm]{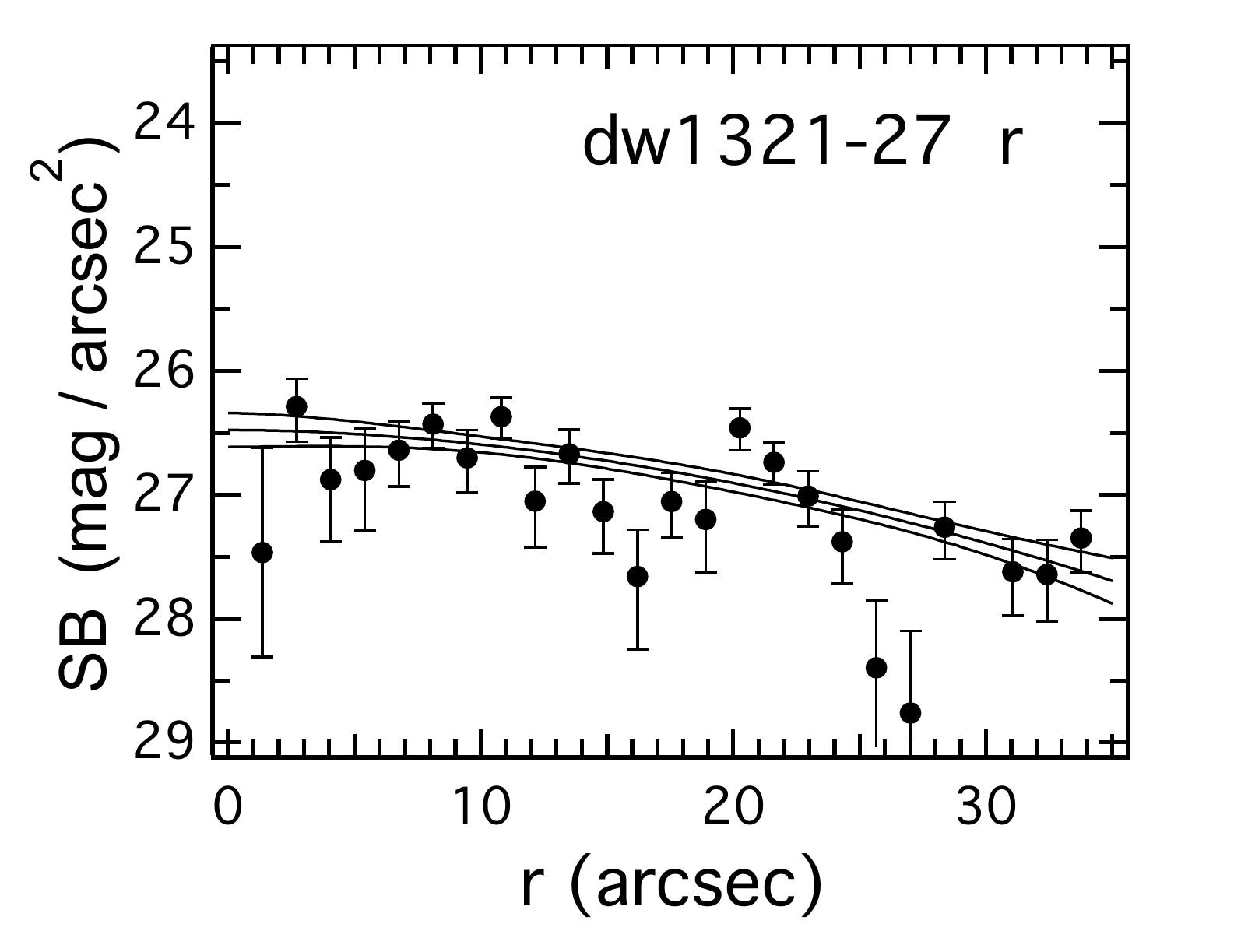}
\includegraphics[width=3.6cm]{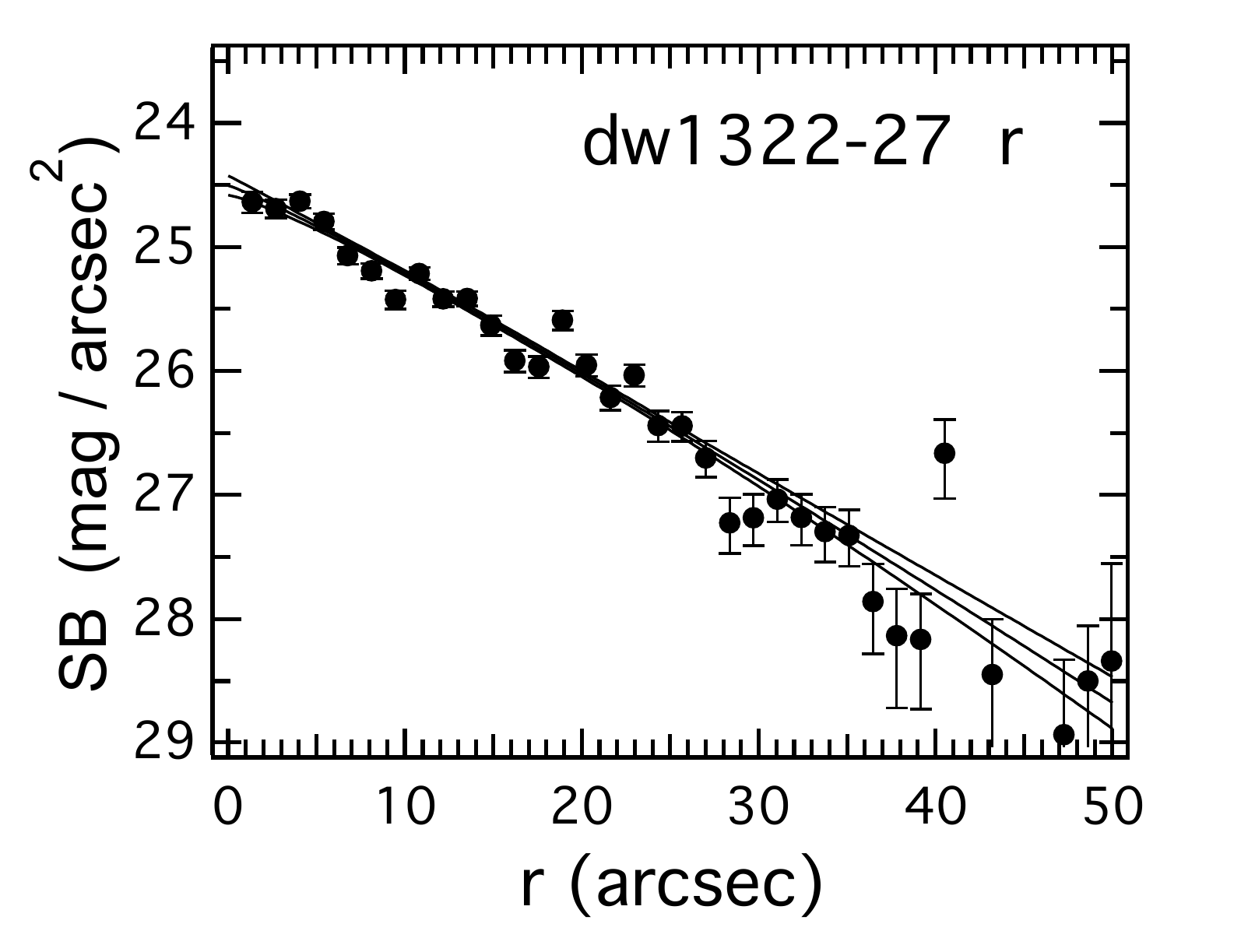}
\includegraphics[width=3.6cm]{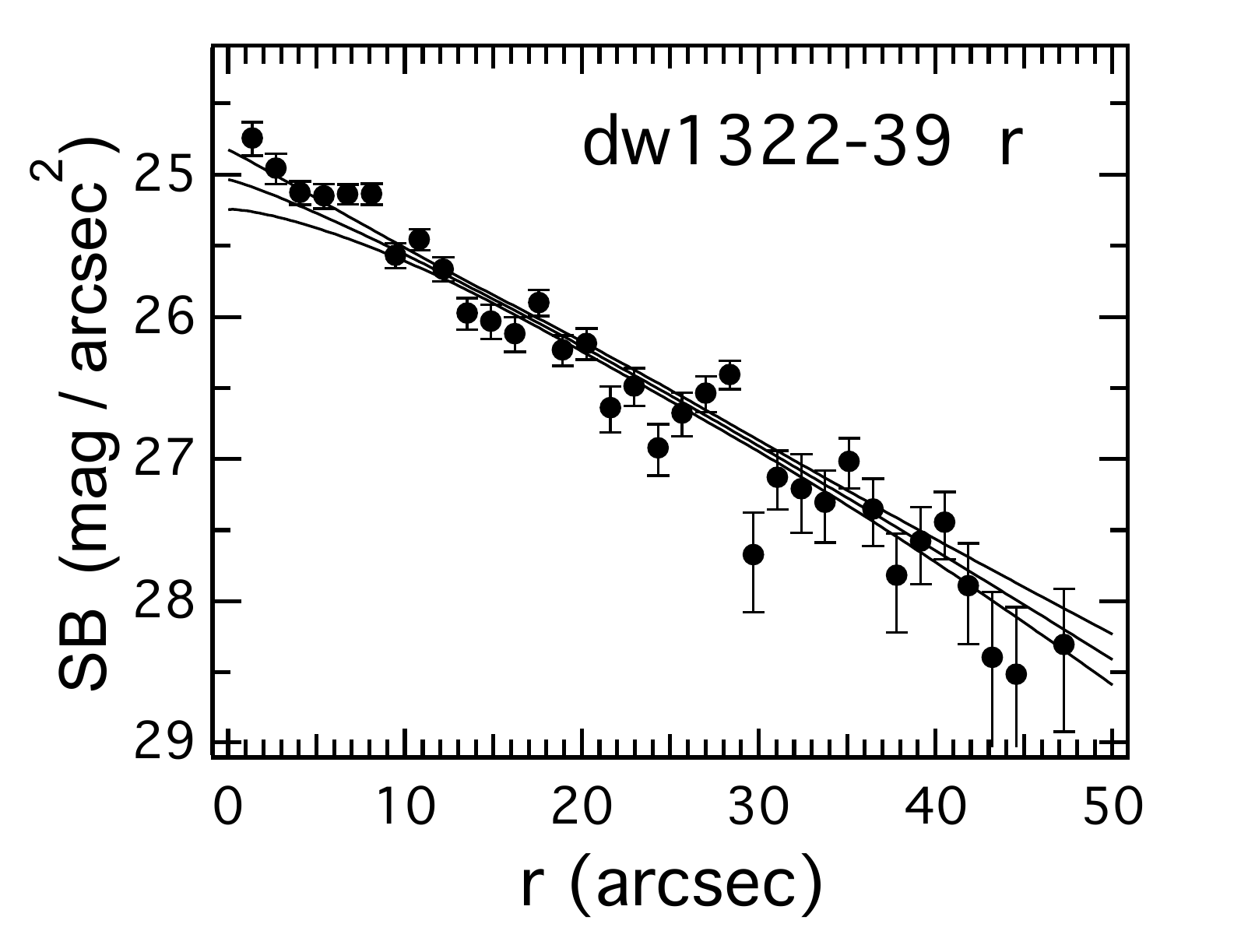} 
\includegraphics[width=3.6cm]{dw1322-39_r.pdf} \\
\includegraphics[width=3.6cm]{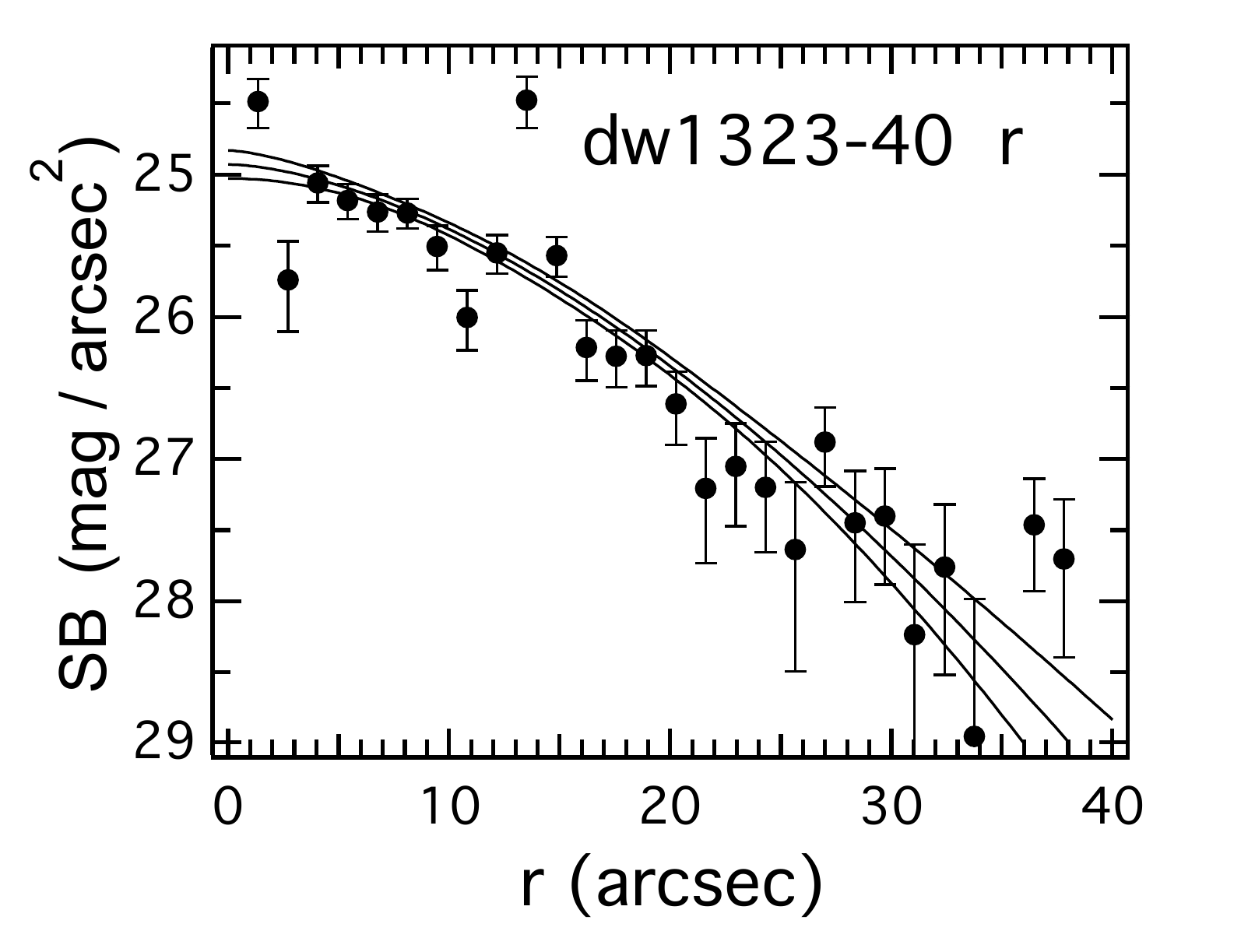} 
\includegraphics[width=3.6cm]{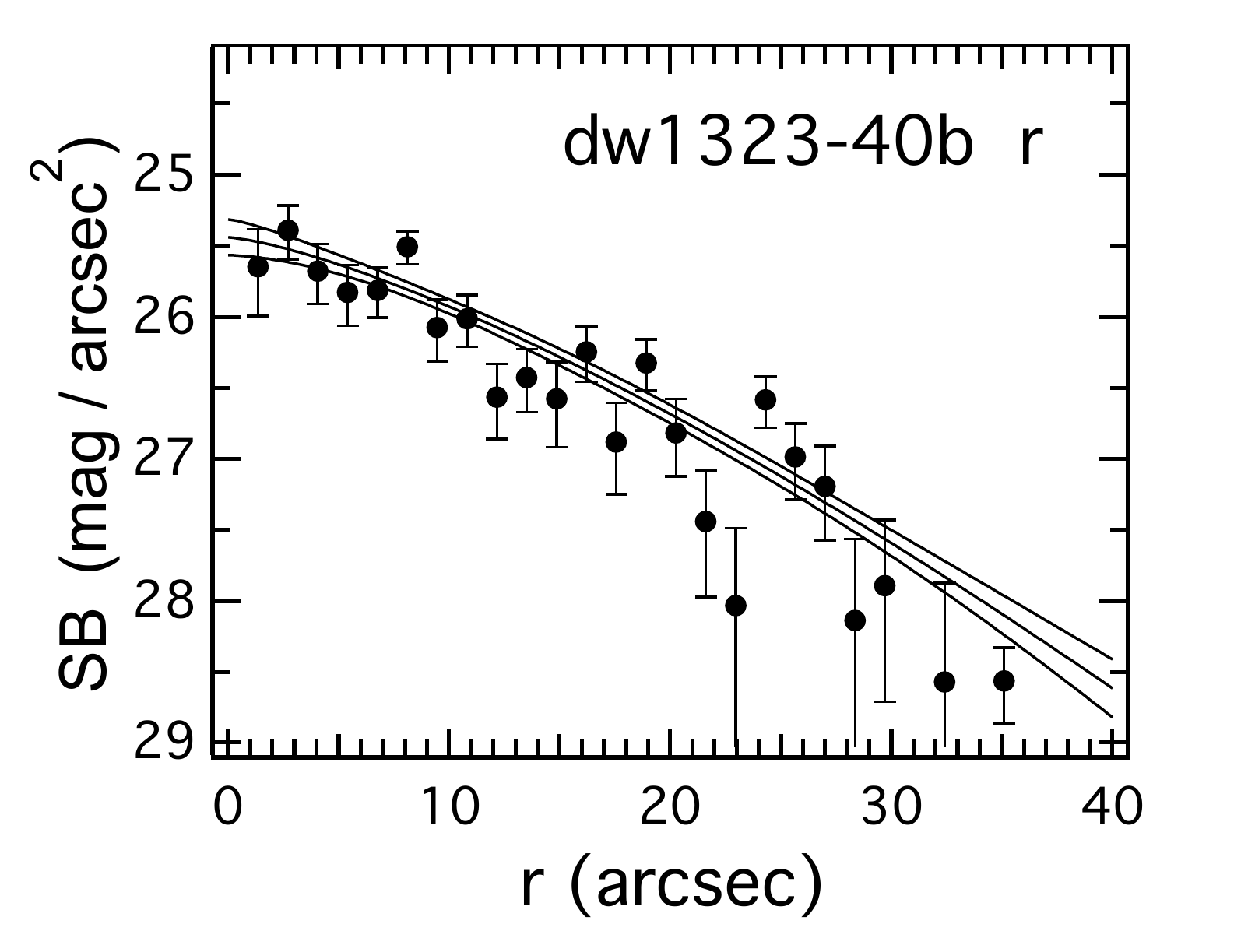} 
\includegraphics[width=3.6cm]{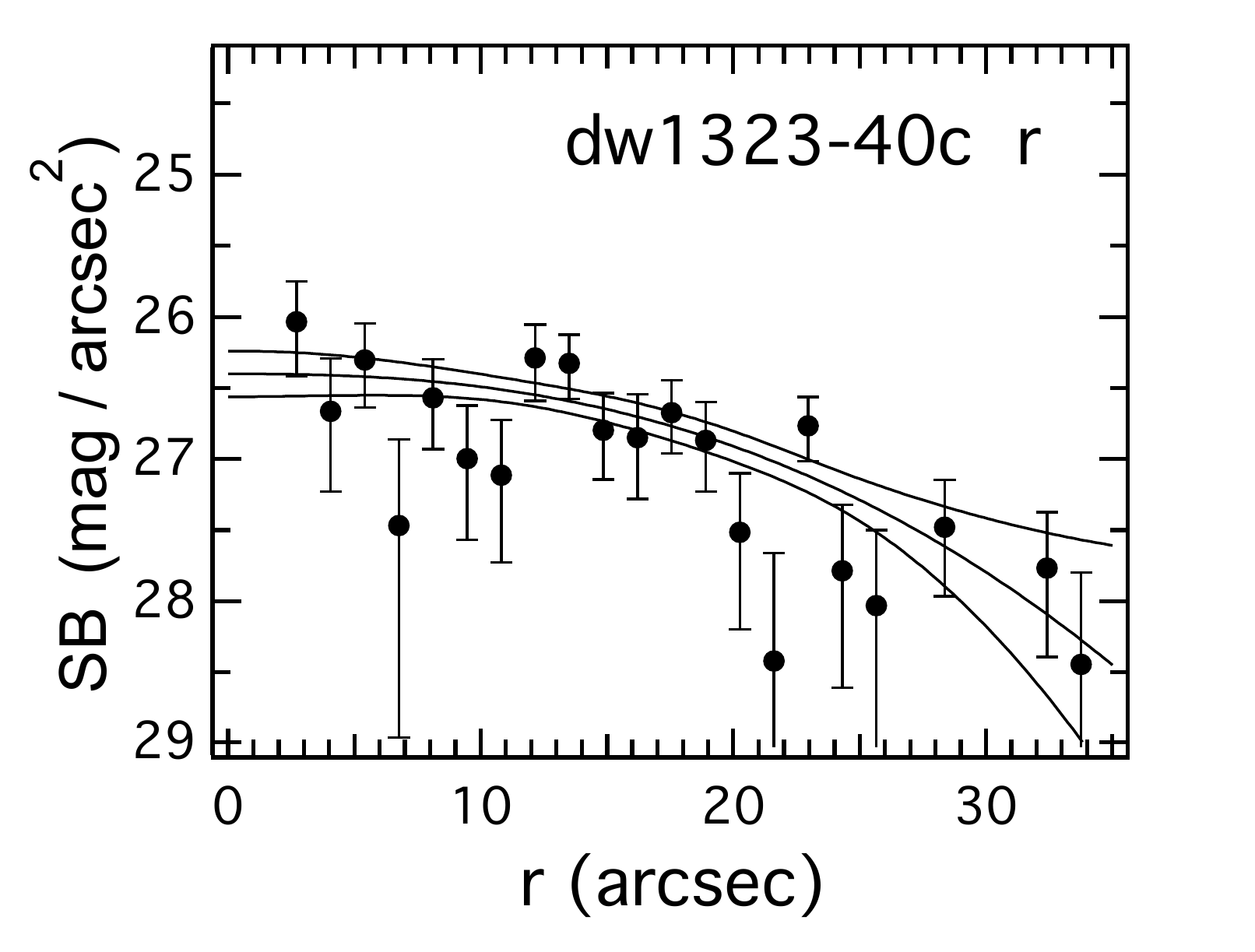}
\includegraphics[width=3.6cm]{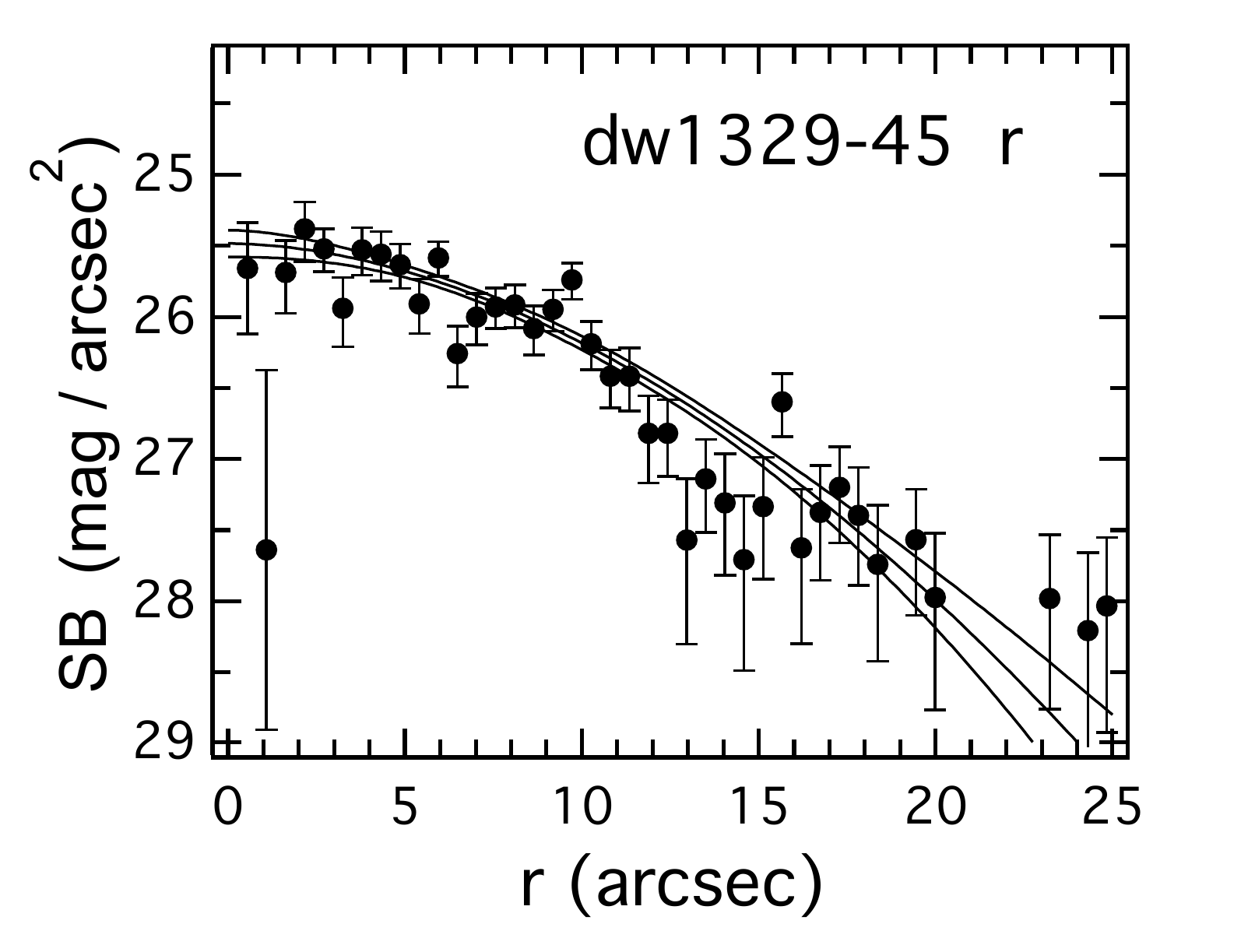}
\includegraphics[width=3.6cm]{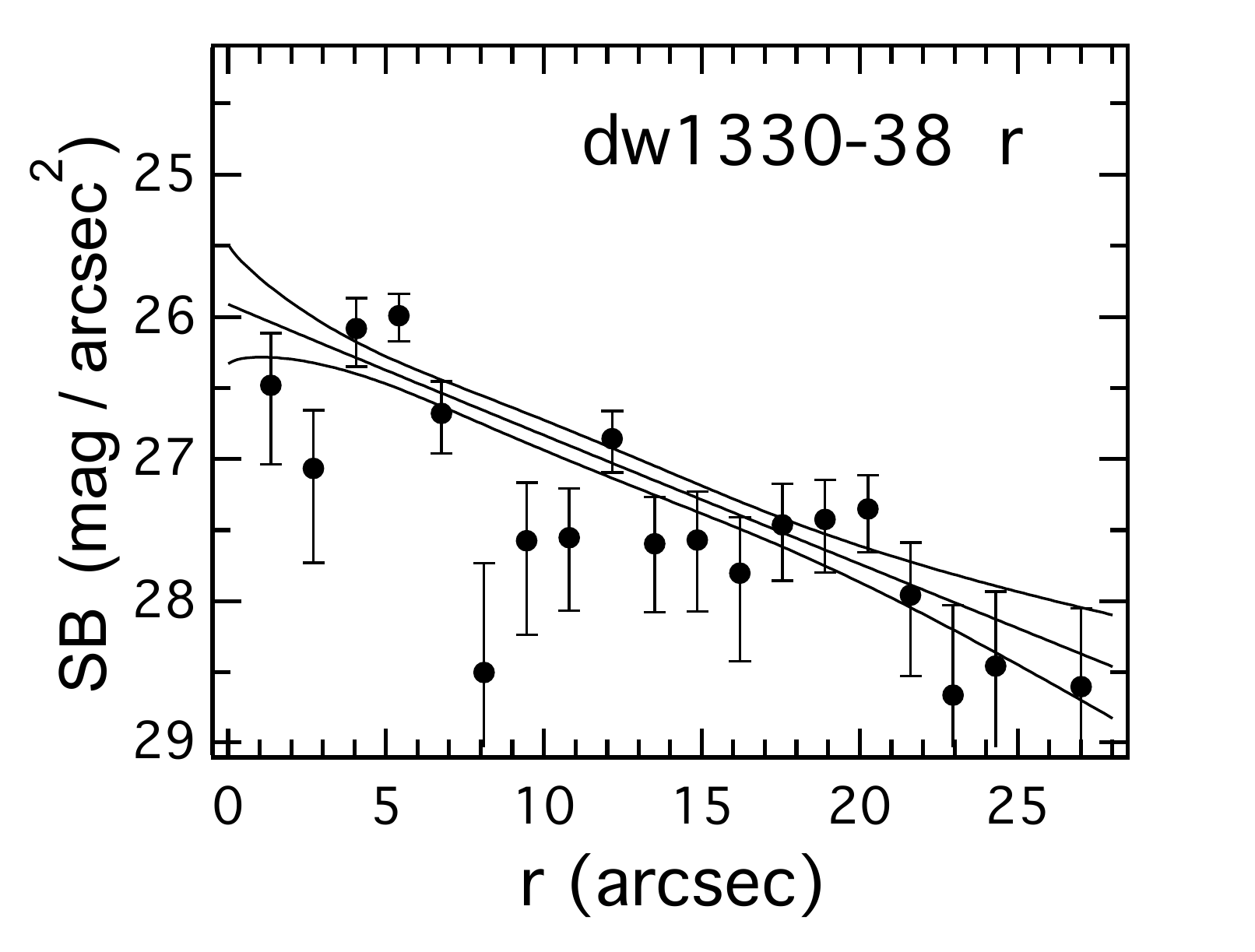}\\
\includegraphics[width=3.6cm]{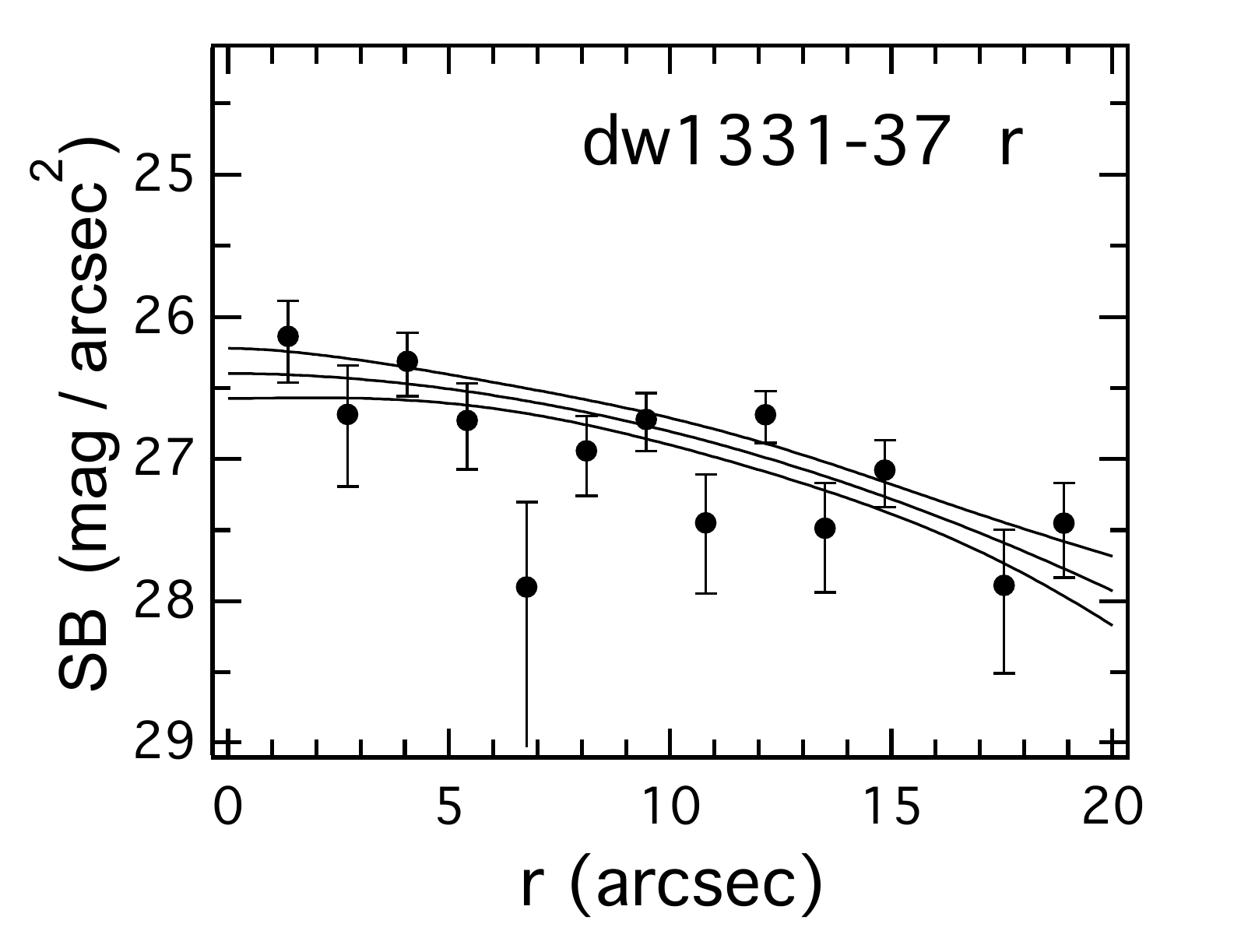}
\includegraphics[width=3.6cm]{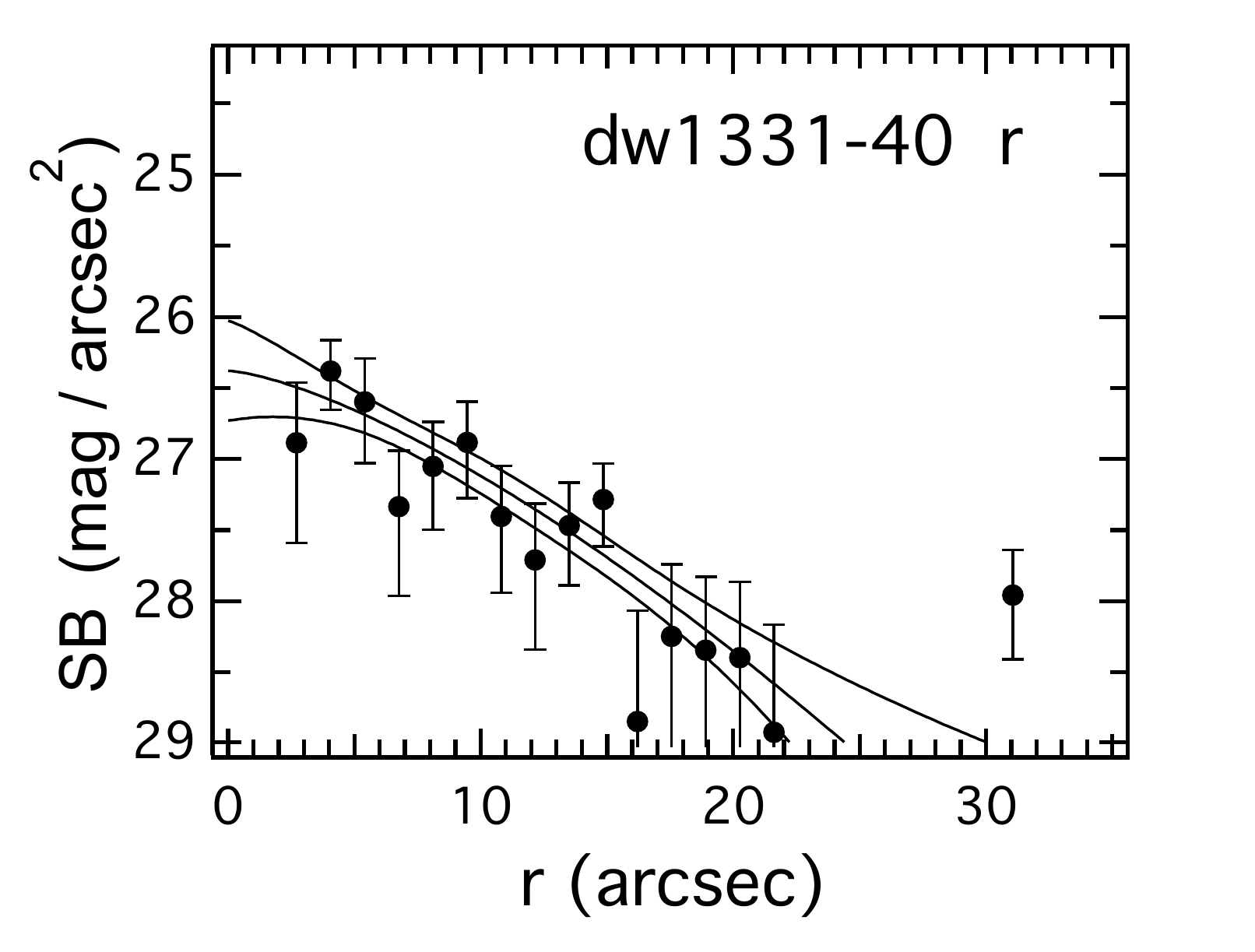}
\includegraphics[width=3.6cm]{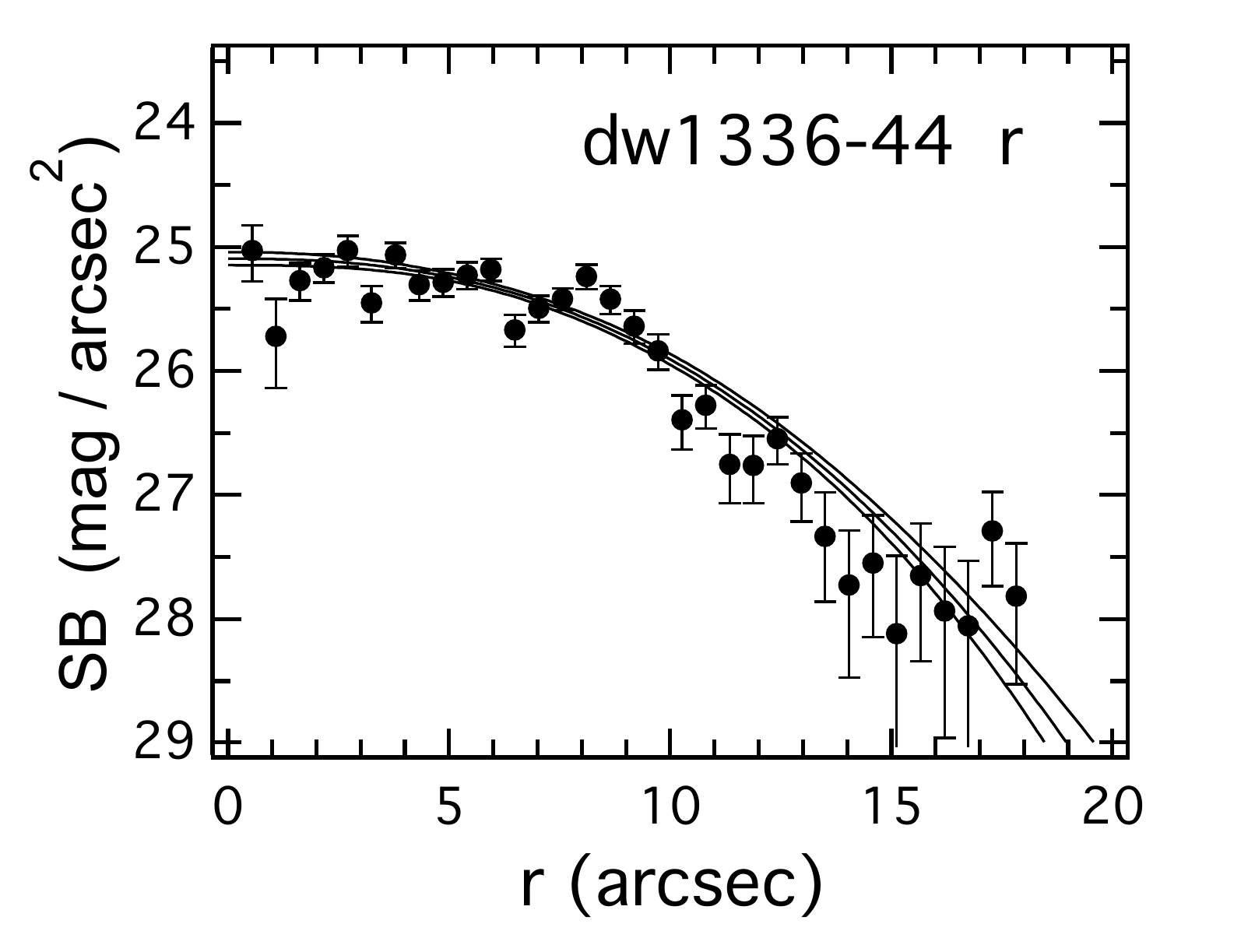} 
\includegraphics[width=3.6cm]{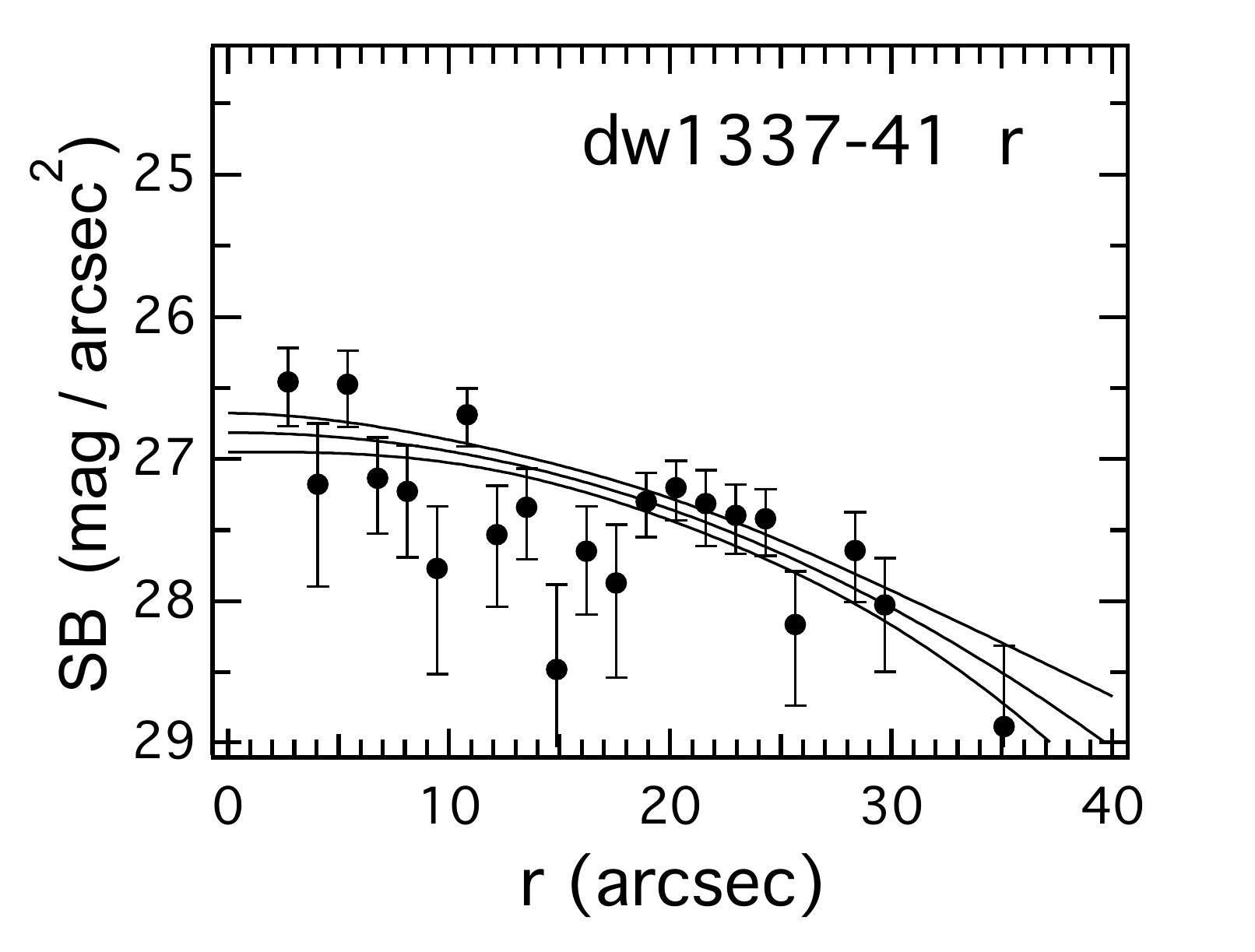}
\includegraphics[width=3.6cm]{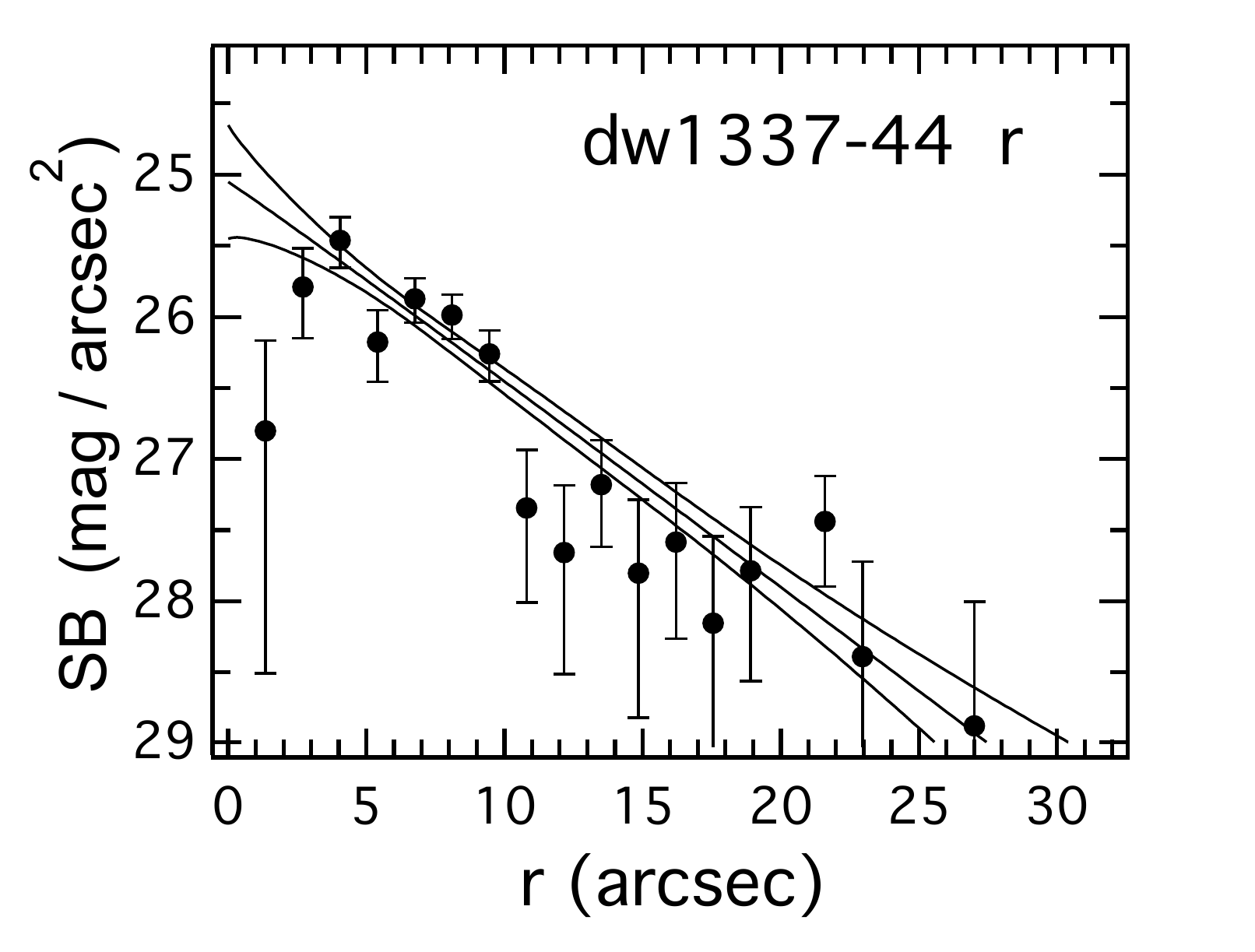}\\
\includegraphics[width=3.6cm]{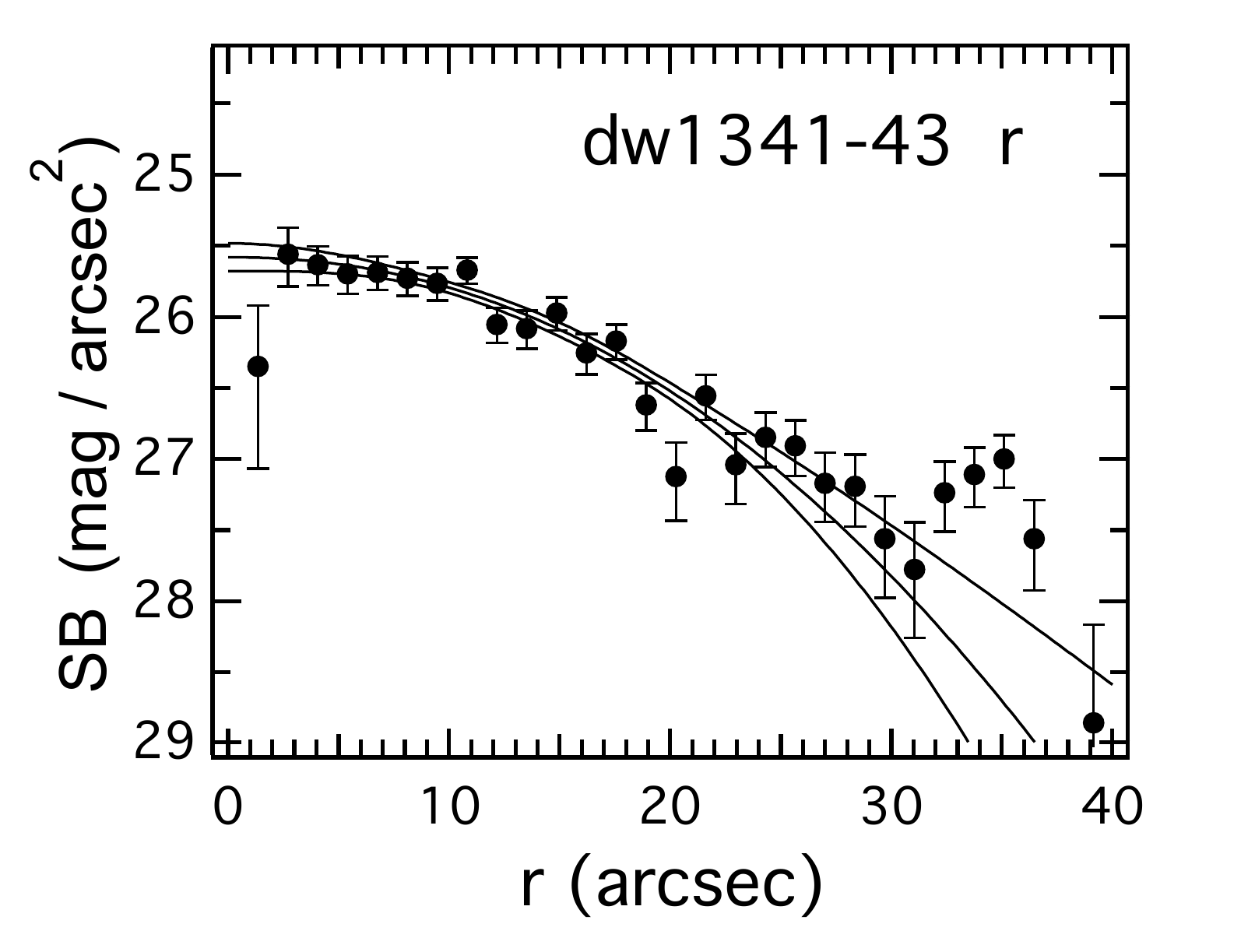}
\includegraphics[width=3.6cm]{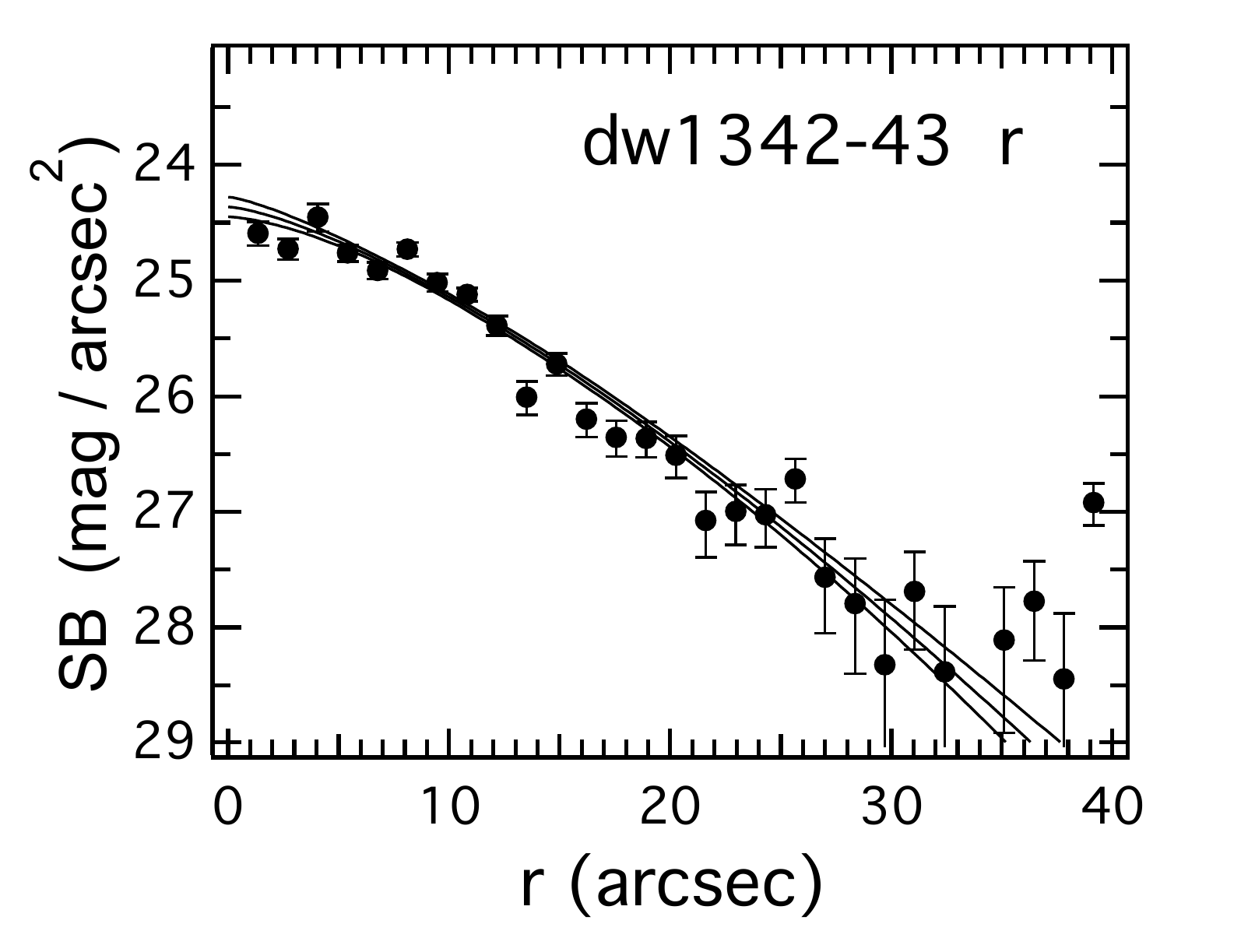}
\includegraphics[width=3.6cm]{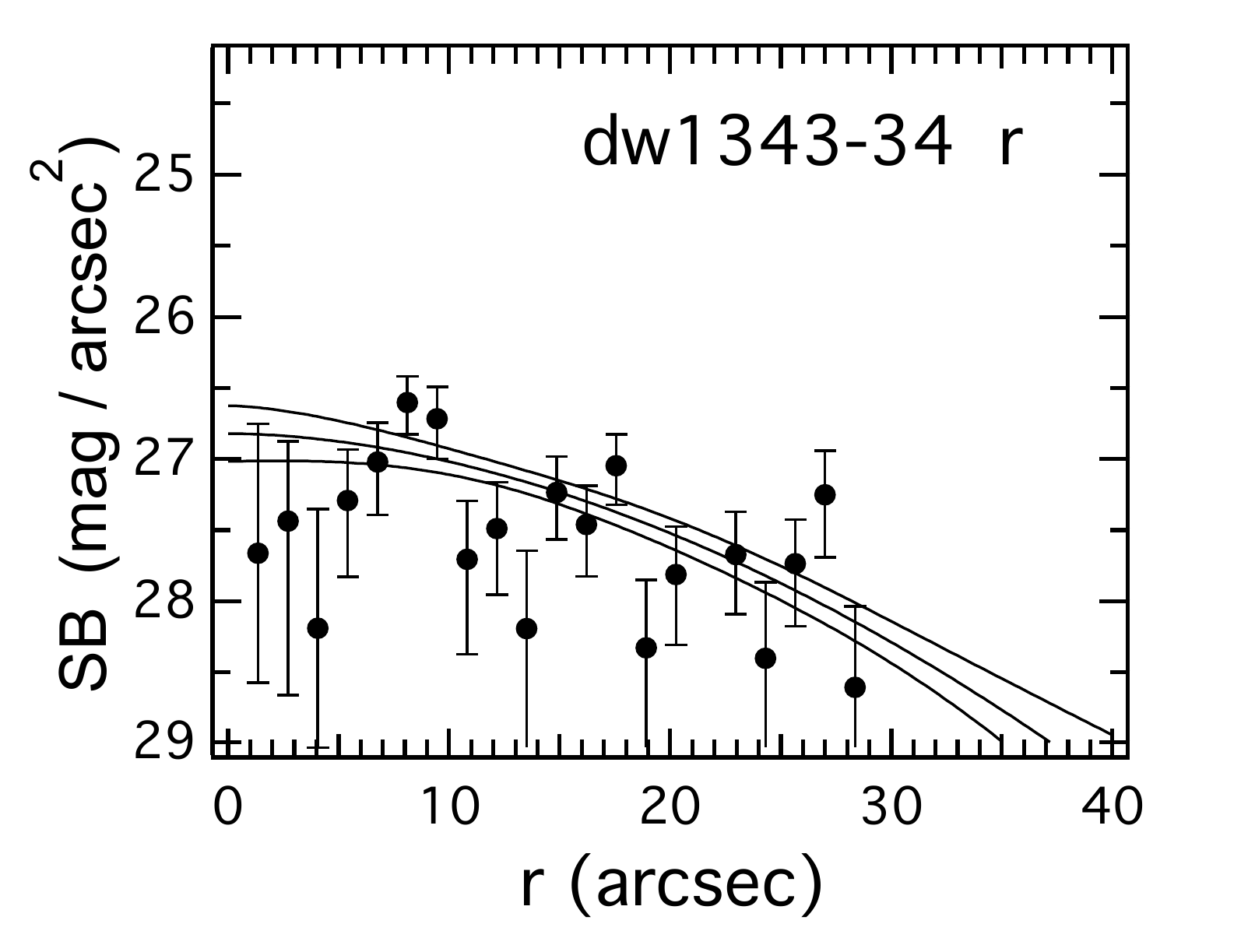}
\includegraphics[width=3.6cm]{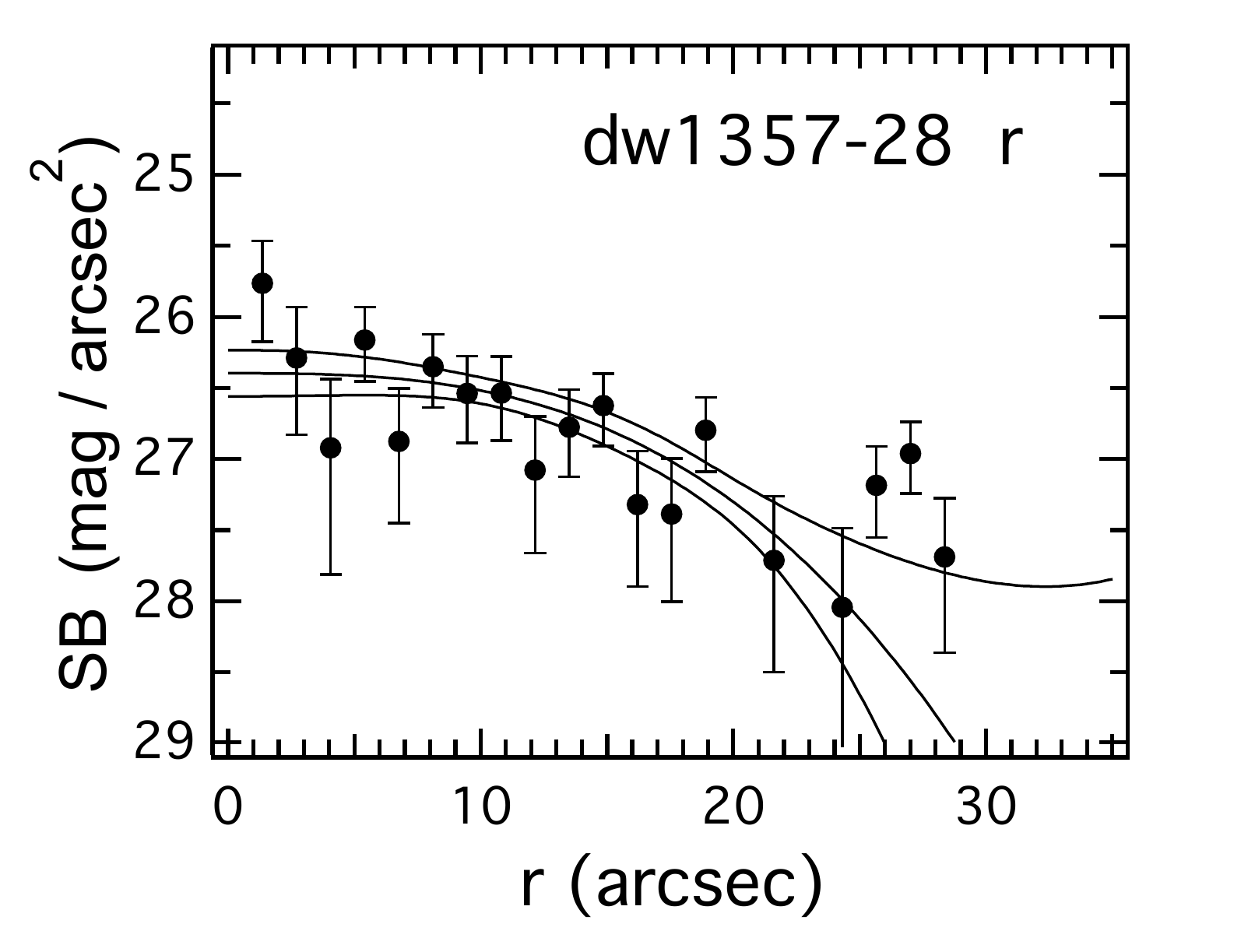}
\includegraphics[width=3.6cm]{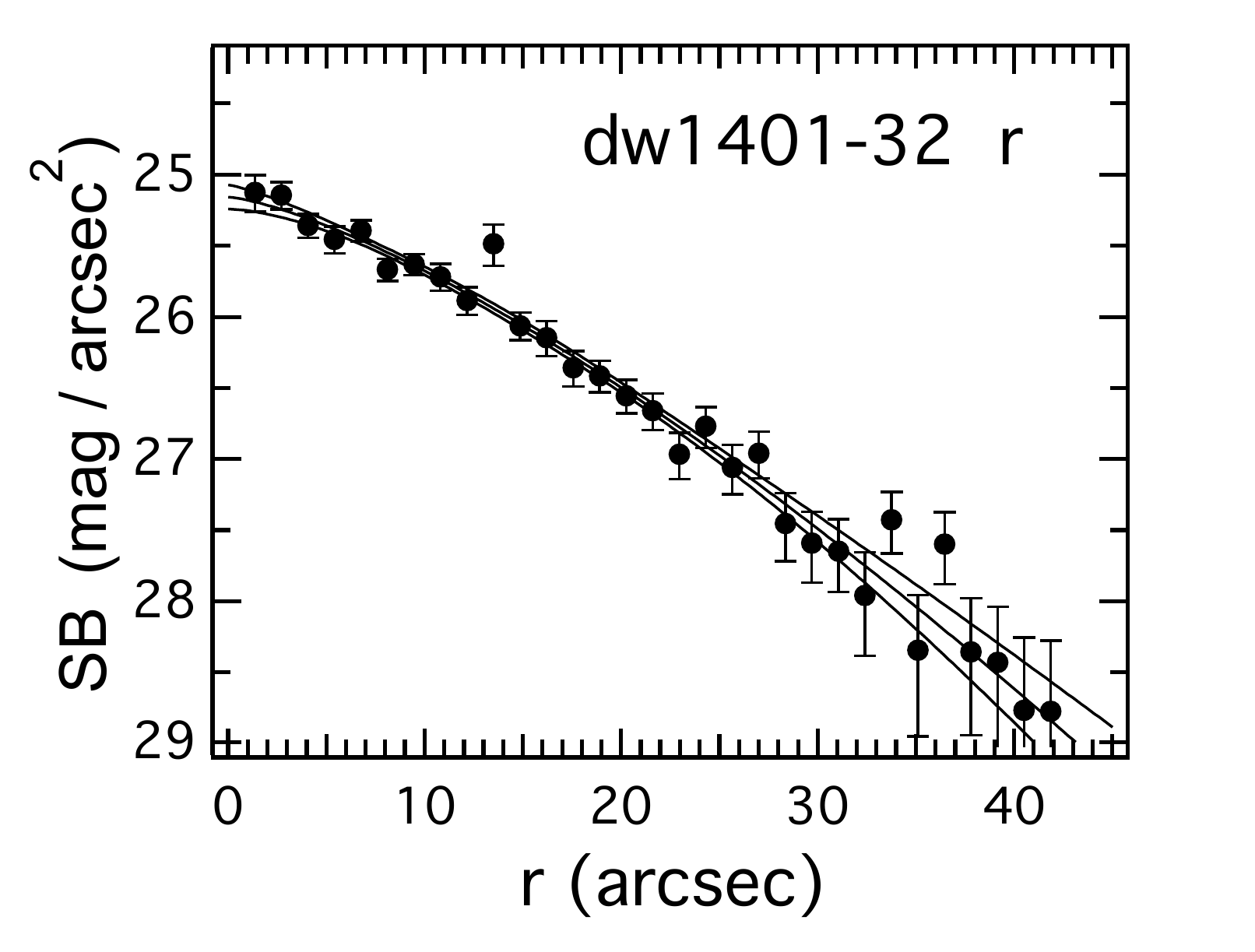}\\
\includegraphics[width=3.6cm]{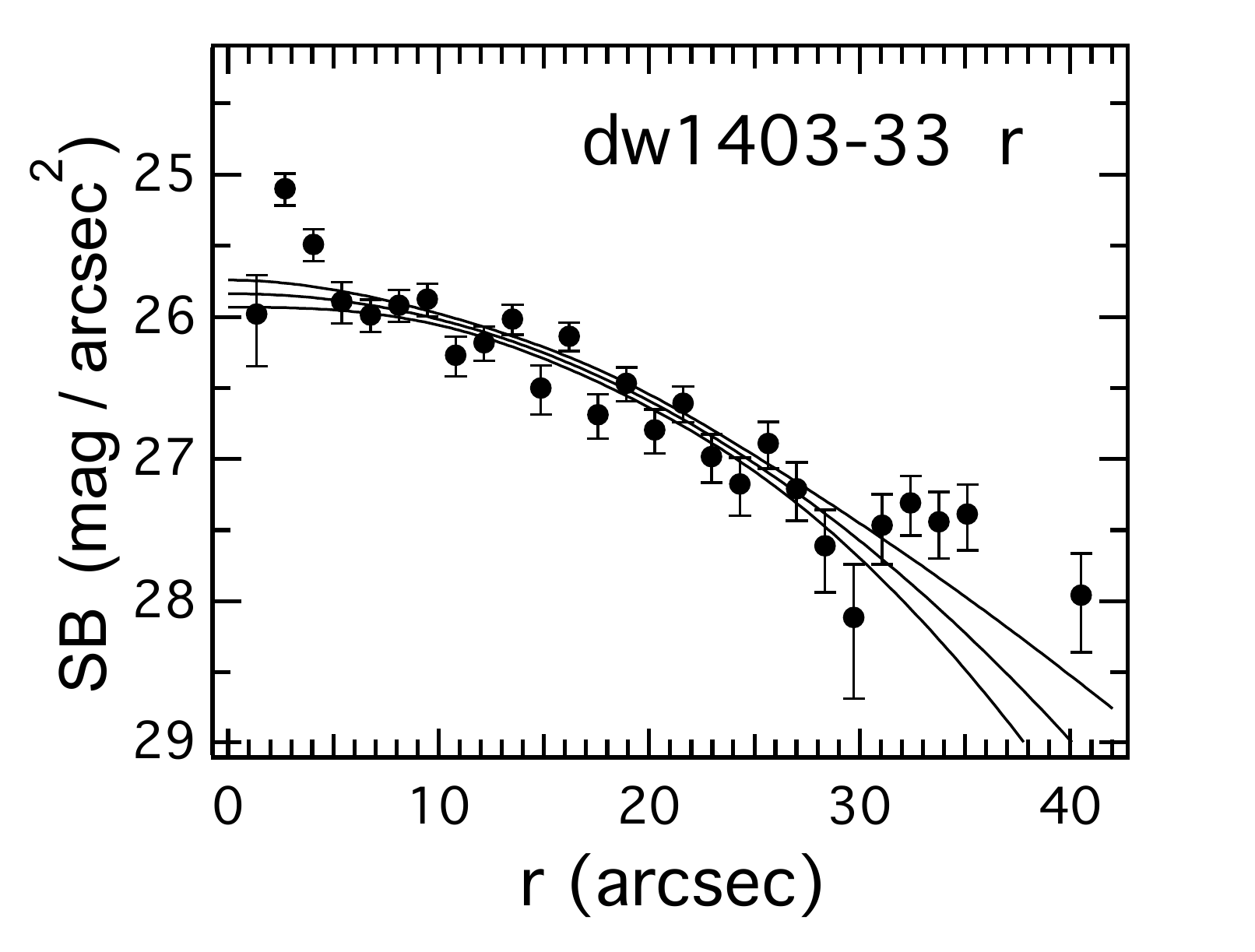} 
\includegraphics[width=3.6cm]{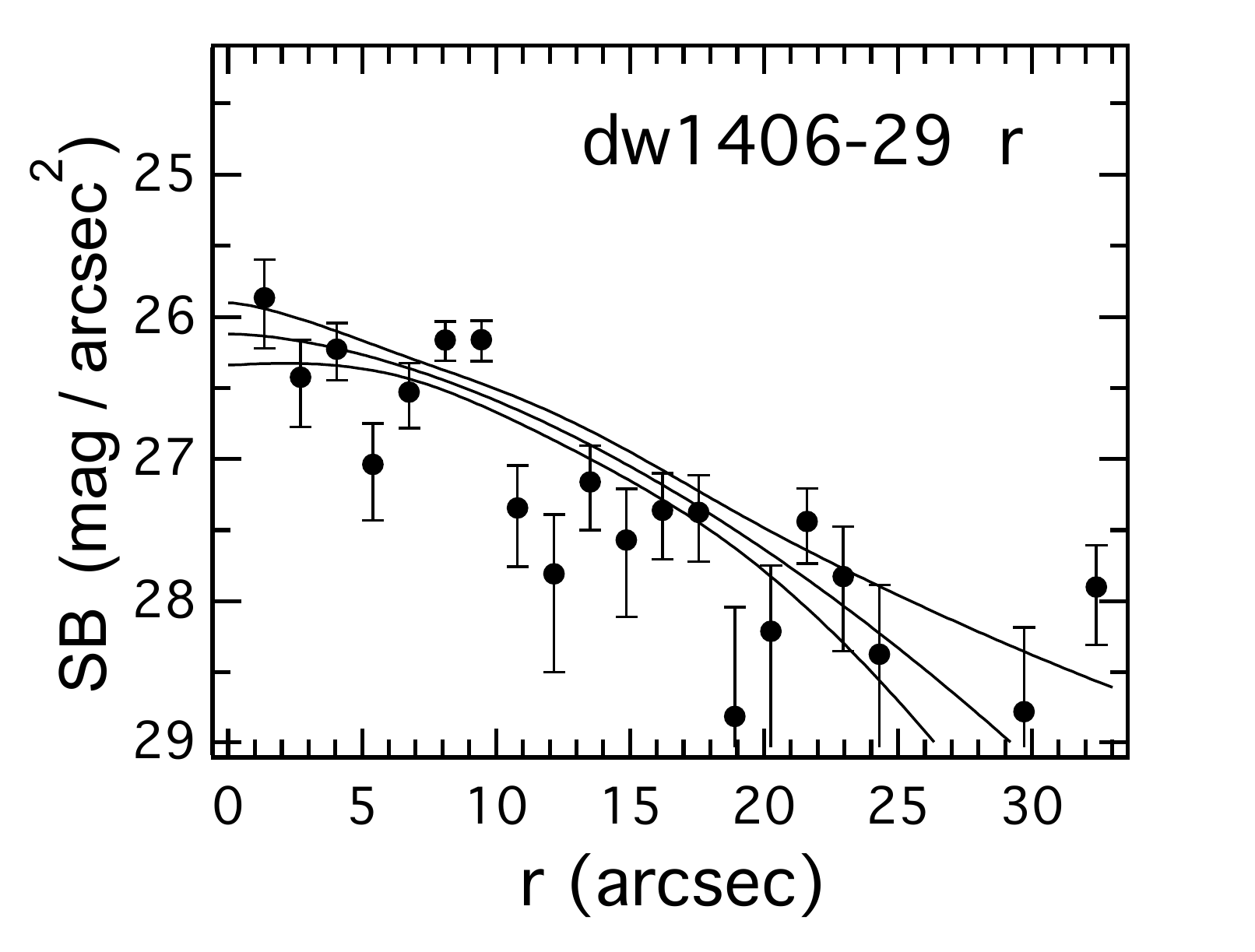}
\includegraphics[width=3.6cm]{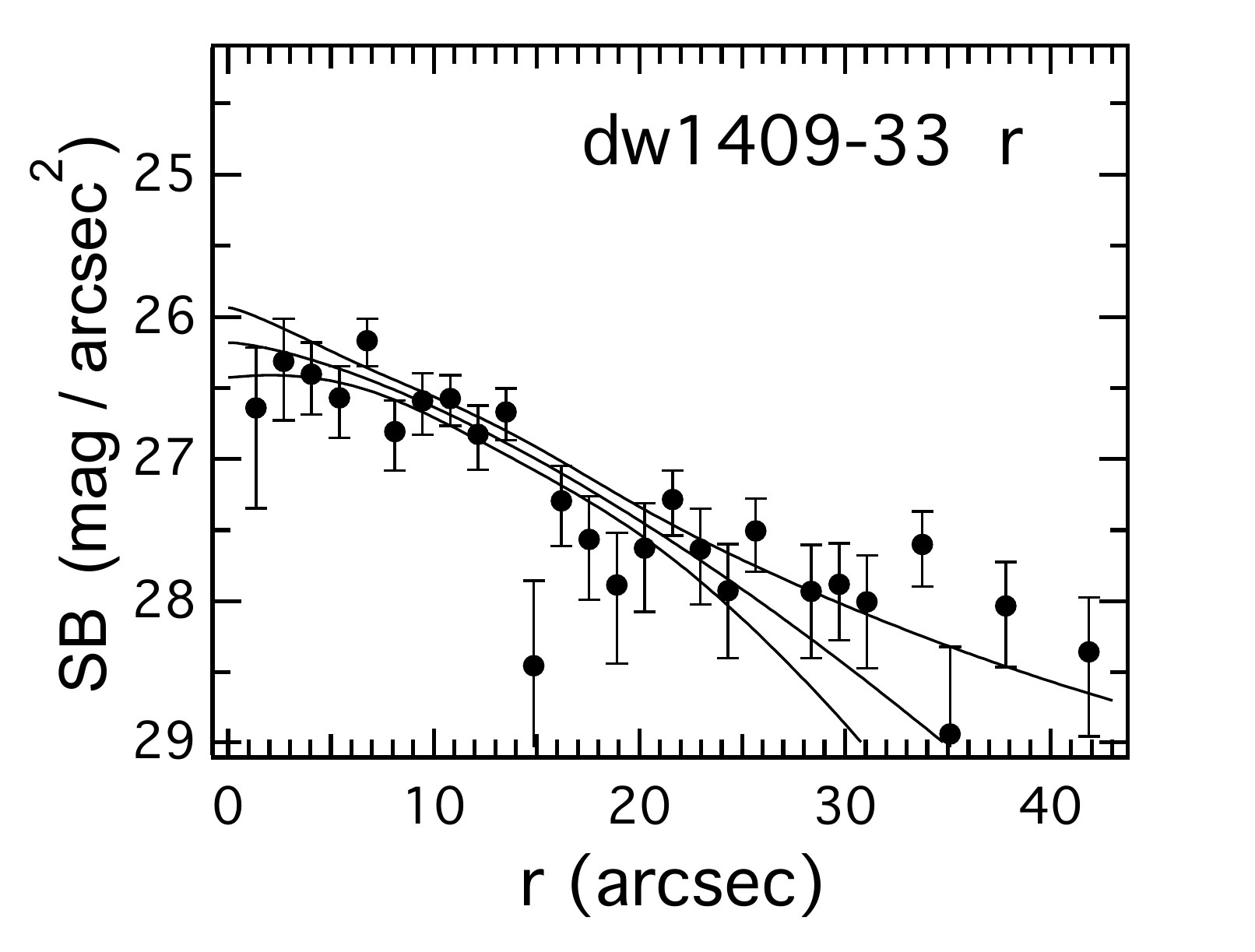}
\includegraphics[width=3.6cm]{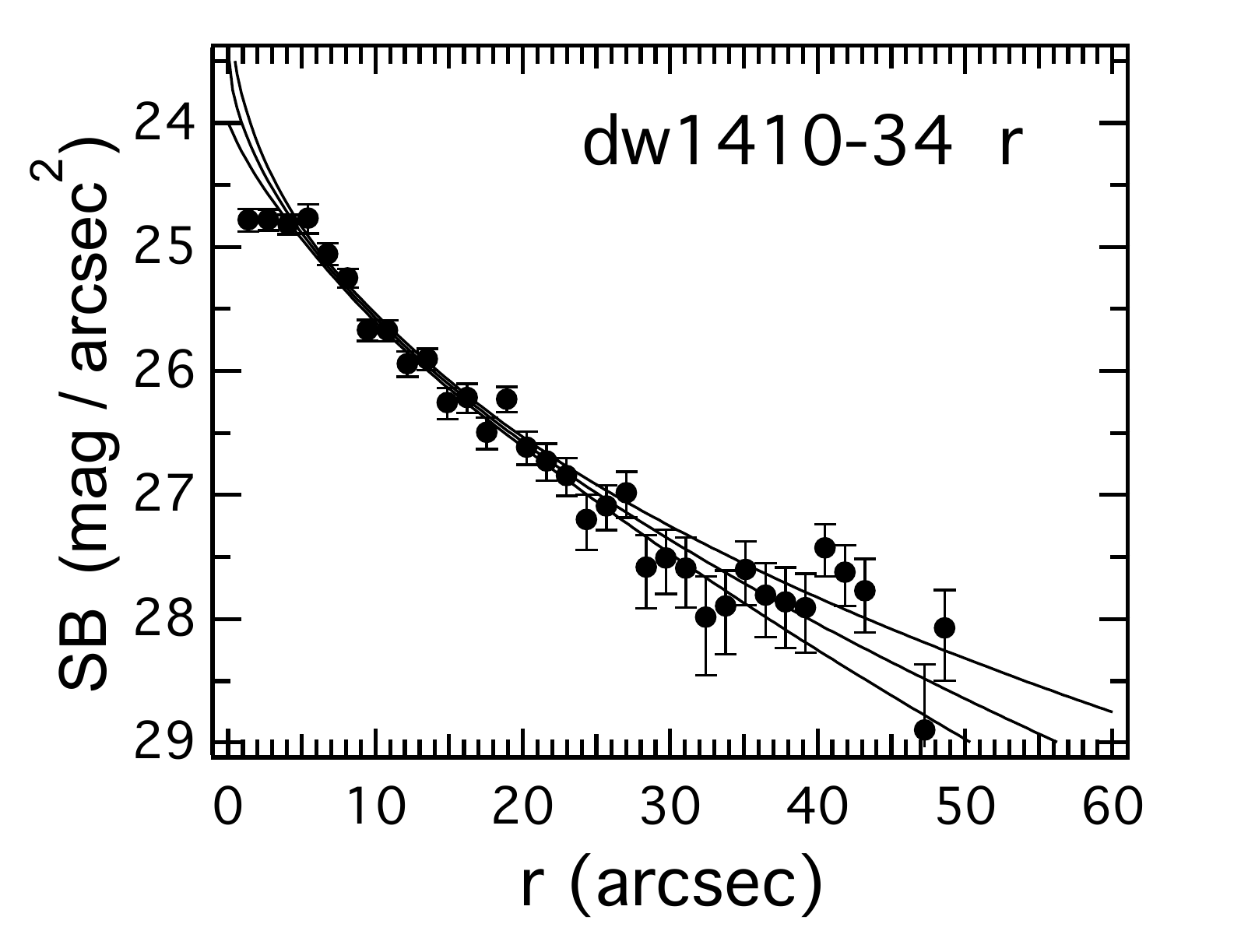}
\includegraphics[width=3.6cm]{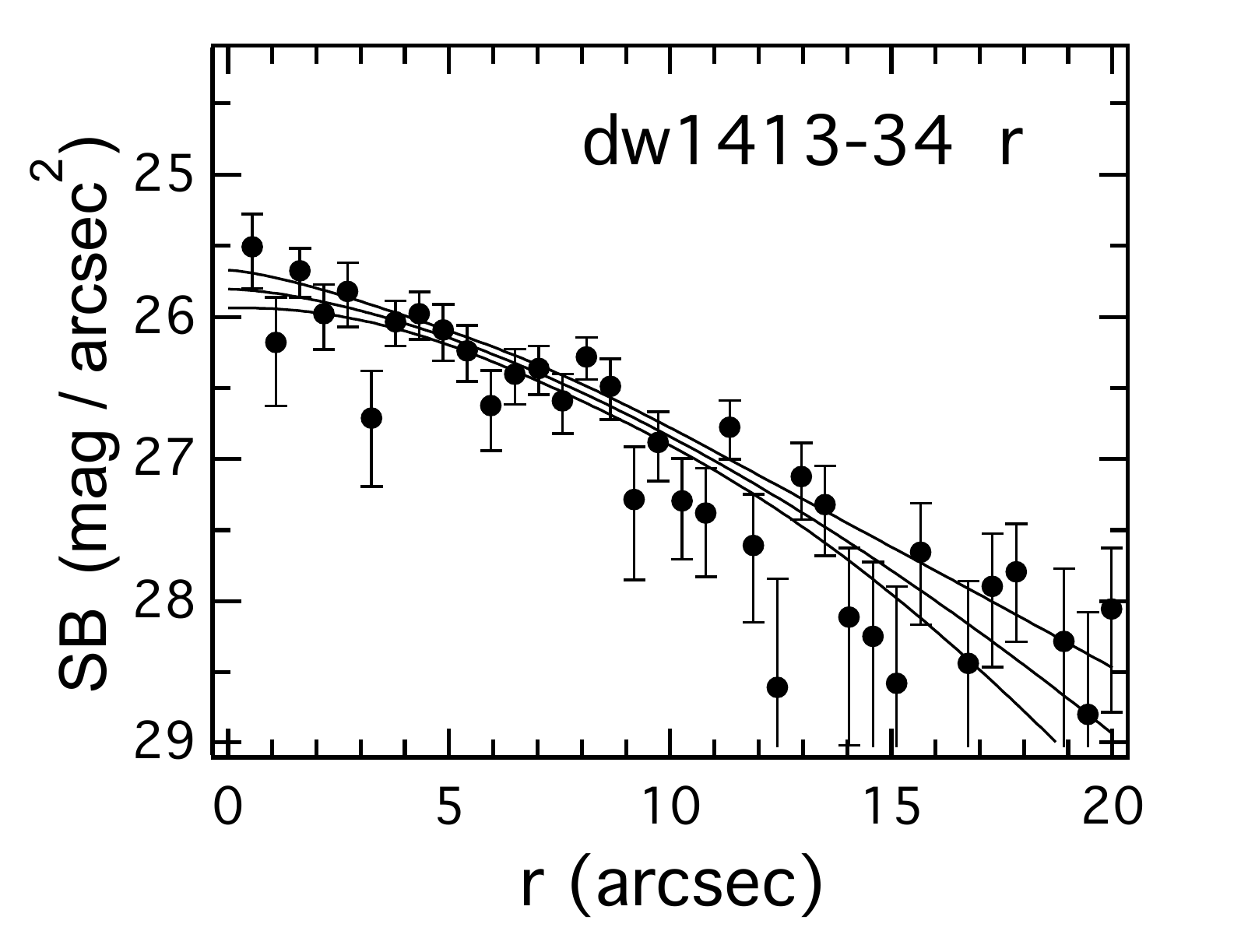}\\  
\includegraphics[width=3.6cm]{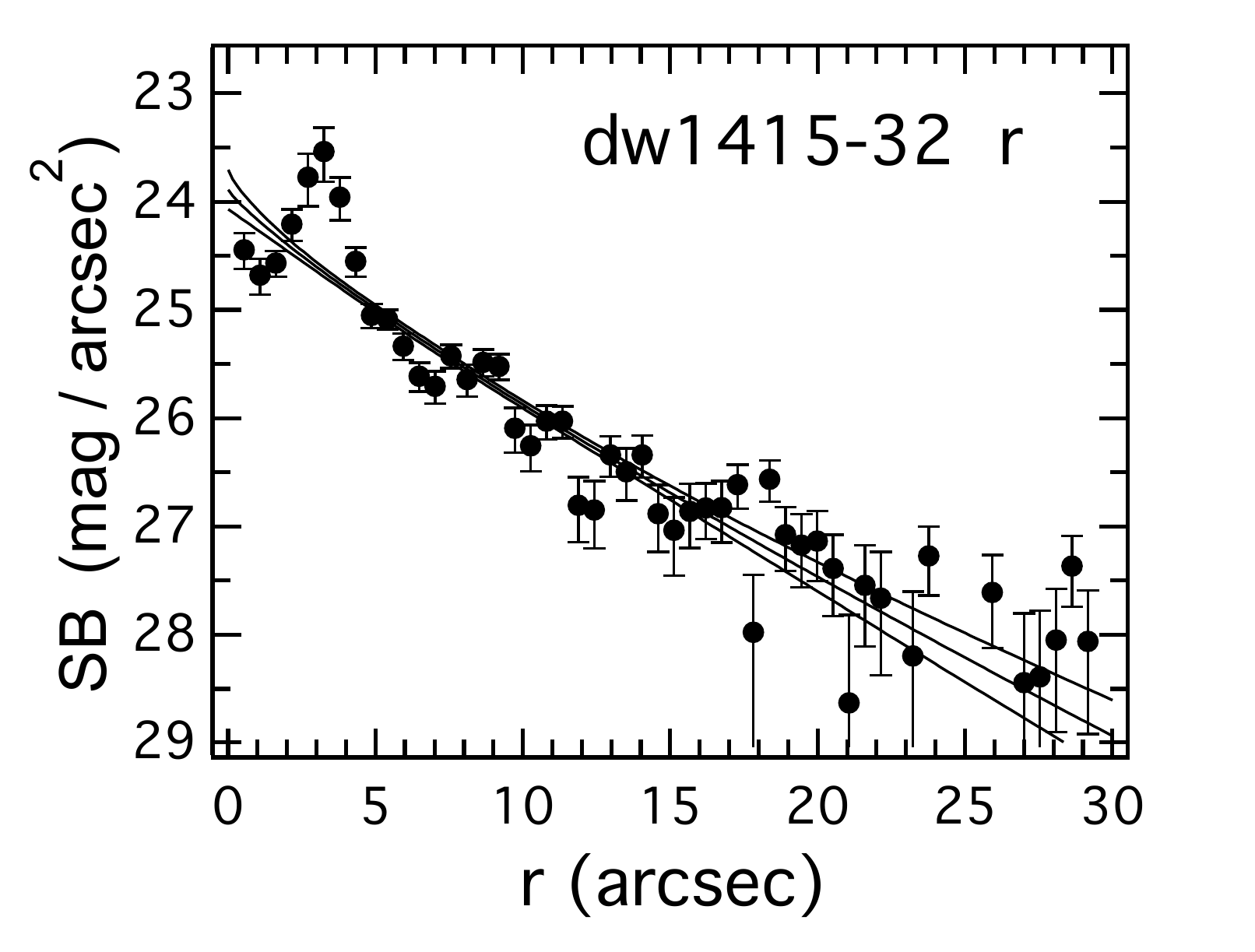}
\caption{Radial surface brightness profiles and best-fitting S\'ersic profiles with $1\sigma$ confidence intervals for all dwarf candidates in the $r$ band. }
\label{sbp}
\end{figure*}

\section{Discussion}
Prior to our study there were about 60 group members known in the whole Centaurus group \citep{1998A&AS..127..409K,2002A&A...385...21K,2013AJ....145..101K,2000AJ....119..593J,2000AJ....119..166J, 2014ApJ...795L..35C,2015arXiv151205366C}, half of these have accurate distances while the others got preliminary membership 
based on morphology or velocity measurements. 
This galaxy population has the potential to almost double in size if the majority of the 41 + 16 (MJB15) new dwarf galaxies are confirmed as group members.
In this context a number of interesting questions arise: What are our detection limits? How plausible are the new candidates? 
Are their photometric and structural properties comparable to the Local Group and known Centaurus group dwarfs, or are they perhaps associated  with background galaxies? Due to the lack of distance information these questions shall be addressed with the help of the available photometric results.

\subsection{Detection limits}
{As mentioned at the end of Sect.\,2, the {photometric} depth for the various DECam observing campaigns, including the one our previous study (MJB15) was based on, is uniform within a range of 0.3 mag. In MJB15 we conducted extensive artificial galaxy tests to determine the detection limits and efficiency of our search for low surface brightness objects. The results of this testing, also valid for the present study, are shown in Fig.\,4 of MJB15. The figure shows the fraction of detected artificial galaxies as a function 
of total magnitude and central surface brightness. The detection efficiency is generally above 80\% for galaxies brighter 
than m=19 r mag and with a central surface brightness $\mu<26.5$\,V\,mag. An alternative way to represent detection efficiency is the completeness boundary curve of \cite{1990PhDT.........1F} and \cite{1988AJ.....96.1520F}, assuming exponential surface brightness profiles
(S\'ersic Index $n=1$) for the objects. The corresponding equation for this completeness curve is
$$m_{tot}=\mu_{lim} - \frac{r_{lim}}{0.5487 r_{eff}} - 2.5\log[2\pi\cdot (0.5958\cdot r_{eff})^2],$$
meaning that (nearly) all, or most objects with a diameter larger than 2$r_{lim}$ at the surface brightness level of $\mu_{lim}$ should have been detected. Our best estimates for the two free limiting parameters in MJB15 (see Figs.\,4 and 6 there) was $r_{lim}\approx 20$ arcsec and $\mu_{lim} \approx 28$ $V$ mag arcsec$^{-2}$. For the present study we found a slightly smaller radius of $r_{lim} \approx 13$ arcsec gives a boundary curve that better fits the data.} 

{To allow for a comparison of our results with the Local Group dwarfs \citep[data from][]{2012AJ....144....4M}}
we used Eq.\,(1) to transform our $gr$ photometry to 
the $V$ band. Having all the galaxies on the same photometric system, in Fig.\ref{completeness} we plot the effective radius versus total $V$-band luminosity relation
for our candidates, all known Centaurus dwarfs in the survey area for which we have photometry
in Table\,\ref{table3}, Local Group dwarf galaxies, and the candidates from MJB15. Absolute magnitudes for the Centaurus galaxy candidates are based on a mean distance of 4.5\,Mpc. The solid curve represents {the completeness boundary curve given above 
with best estimates} $r_{lim}\approx13$ arcsec and $\mu_{lim} \approx 28$ $V$ mag arcsec$^{-2}$, suggesting
that we detected {most} dwarf galaxy candidates in our survey footprint with diameters larger than 
26\,arcsec ($\approx 600$\,pc) at a surface brightness of $28\,V\,$mag\,arcsec$^{-2}$. These quantities 
translate roughly into a luminosity limit of $M_V\approx -10$
or $M_r\approx -9.5$. {The completeness boundary curve, properly transformed to the $\mu-M$ plane, is also shown in Fig.\,9.}

\subsection{Centaurus group membership}
\begin{figure}[Ht]
\centering
  \includegraphics[width=9cm]{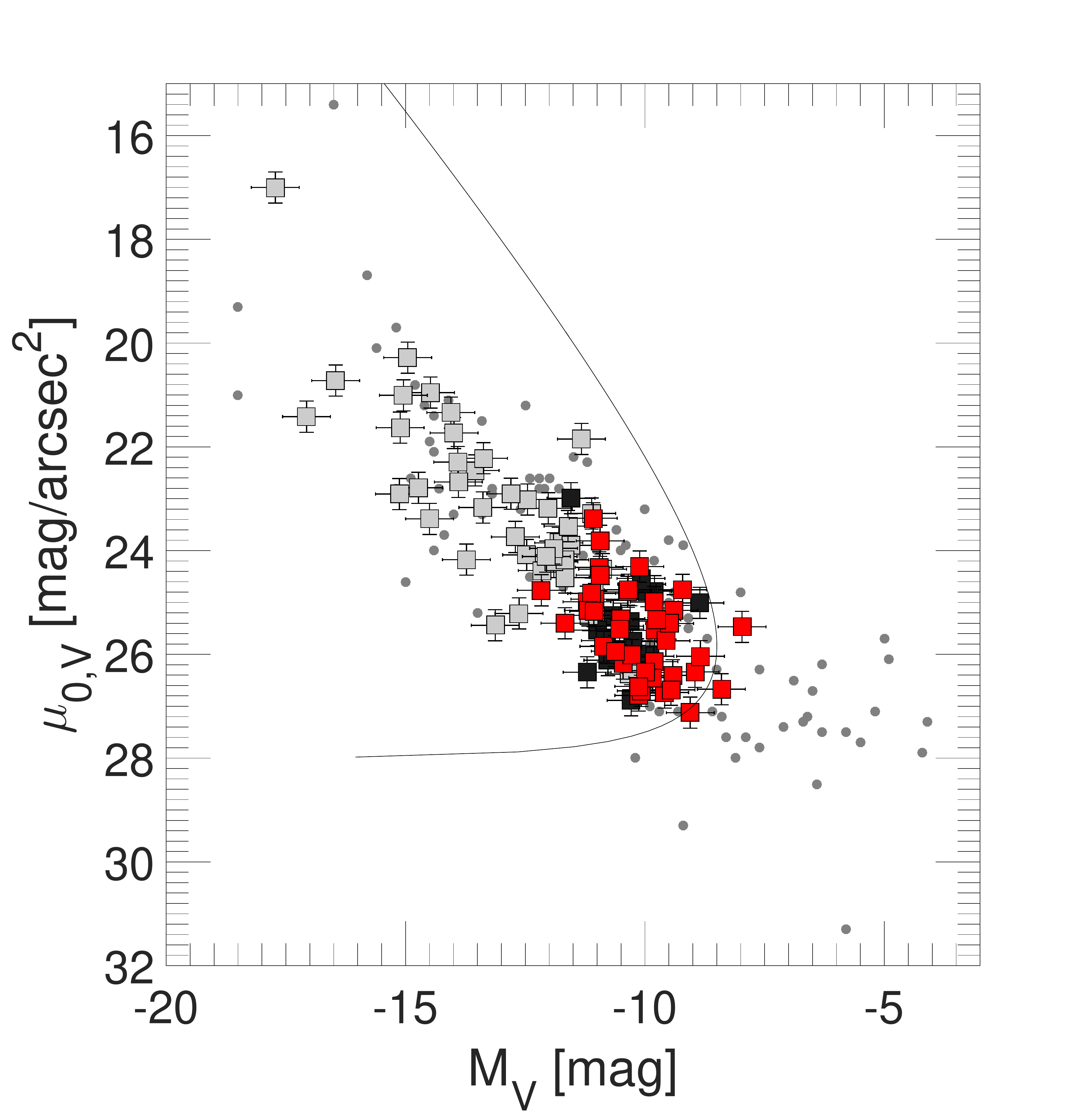}
  \caption{For our dwarf candidates, $\mu_0-M$ relation in the $V$ band (red squares), all known dwarfs 
  of the Centaurus group in the survey area (gray filled squares), Local Group dwarf galaxies \citep[dots;\,][]{2012AJ....144....4M} 
  and the candidates from MJB15 (black squares). The newly discovered Centaurus dwarf candidates have similar properties 
  to those of known Centaurus dwarfs and Local Group dwarfs.}
  \label{strucParameters}
\end{figure}

\begin{figure}
  \resizebox{\hsize}{!}{\includegraphics{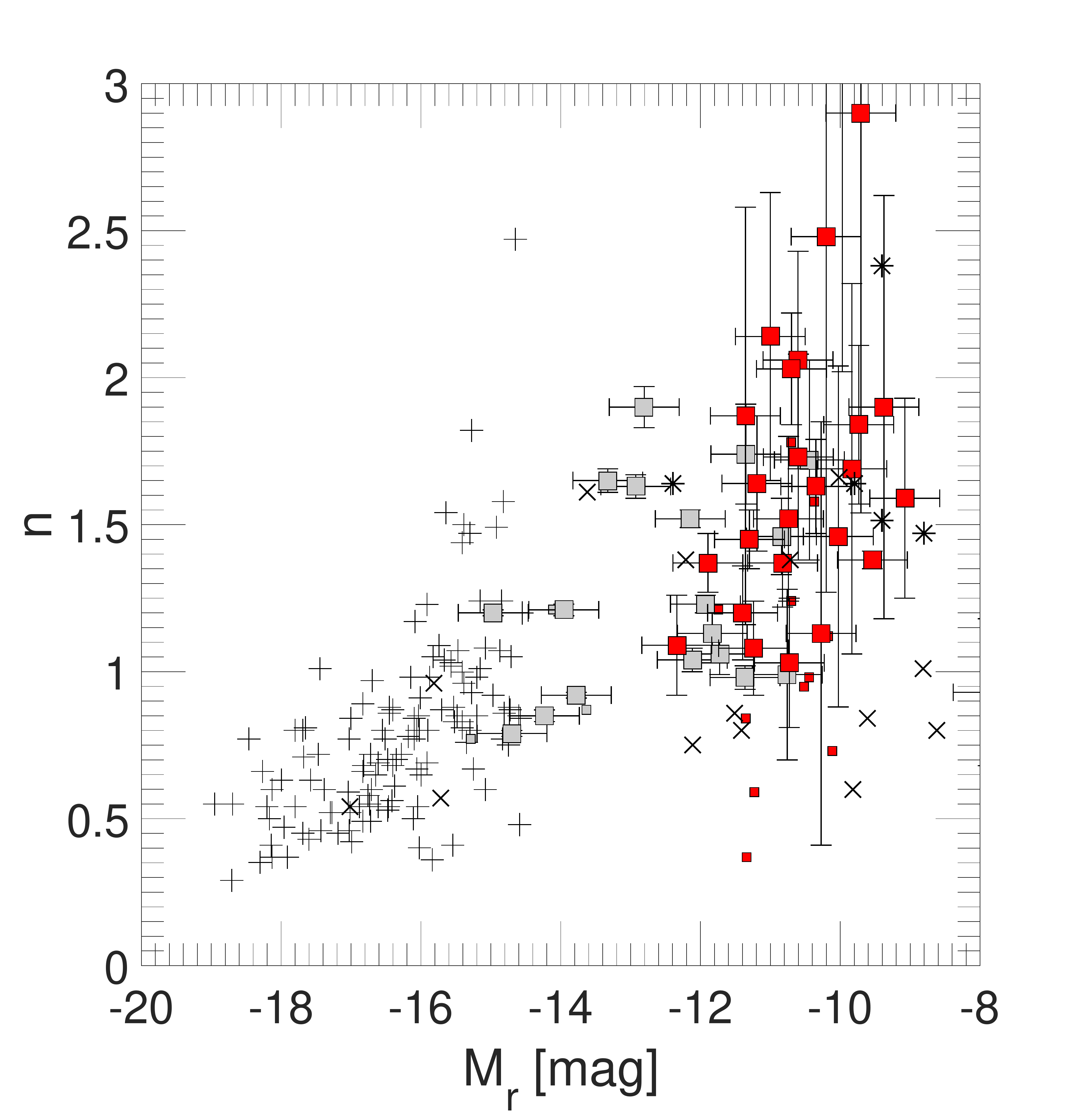}}
  \caption{Shape parameter - luminosity (S\'ersic $n-M_r$) relation for early-type dwarf galaxies. The known 
  Centaurus dSph dwarfs in the survey region (gray squares), Virgo dEs  \citep[plus signs; ][]{1998A&A...333...17B}, M\,81 dSph 
  dwarfs \citep[stars; ][]{2009AJ....137.3009C}, Local Group dSph \citep[crosses; ][]{2000AJ....119..593J}, MJB15 known and candidates dSph (small gray and red squares) and the new dSph known dwarfs and candidates of this study (gray and red squares), respectively. 
  Error bars for our candidates and previously known dwarfs in the survey area come from numerics and are listed in Table 2. The error bars for the 
  absolute magnitudes is globally chosen to be $\pm$0.5 mag. We compare $n_B$ (Virgo) with $n_r$ (M\,81, Local Group, MJB15, and this work). 
  }
  \label{nToMr}
\end{figure}

As we discussed in section 3, the photometric depth of our survey data reached the brightest stars in some of the dwarf candidates, but
photometry of red giant branch stars at least 1\,mag fainter would be necessary to establish TRGB distances for the dwarf candidates. 
Another means to measure distances of galaxies is the surface brightness fluctuation method \citep[SBF;][]{1988AJ.....96..807T}. This method was 
successfully tested for early-type dwarfs by \cite{2000AJ....119..166J} and used to measure distances to five dE galaxies in the Centaurus group and  
many more in the Local Volume \citep{2001A&A...380...90J,2005A&A...437..823R}. The minimum exposure time required for the SBF method to work can be 
calculate using equation (1)
 in \cite{2006AJ....132.1384D}. Using $\mu_{gal}=25$ mag arcsec$^{-2}$ for the mean surface brightness of a typical dwarf candidate in this 
 survey, the sky surface brightness $\mu_{sky}=21$ mag arcsec$^{-2}$, 
 a distance modulus of 28.0 for the Cen\,A subgroup,  the fluctuation luminosity of the underlying stellar population $\overline{M_r}=-1.3$\,mag, 
 and the photometric zero point $m_{1}=24$ mag gives an integration time of 2400\,sec (S/N=5),
 which is six times longer than the exposure times of our DECam images. 

As the new dwarf candidates are not resolved into stars and the SBF method requires longer integration times, the only 
way to test (or rather suggest) group membership at the moment is to compare the photometric and structural properties of the galaxy candidates with the known dwarfs 
in the Centaurus group and Local Group. This can be achieved with the surface brightness -- luminosity relation. 
To calculate the luminosities of the candidates we placed them at the mean distance of the Centaurus group (4.5\,Mpc). 
Because the surface brightness is a distance independent quantity, the only parameter that decides how well a 
candidate fits into the $\mu - M$ relation is the luminosity and thus the assumed distance. We plot the central surface 
brightness $\mu_0$ for all galaxies versus their estimated absolute magnitude $M_V$ for Local Group dwarfs, the known 
Centaurus dwarfs,  candidates from MJB15, and candidates from this work in Fig.\,\ref{strucParameters}.  
The Local Group dwarf $\mu_0$ values come from King or exponential profiles, while our photometric parameter comes from S\'ersic fits.
Our candidates are in good accord with the photometric values of known dwarfs. They all fit into the relation 
outlined by the Local Group dwarfs and naturally bridge the gap to the more luminous dwarfs in the Centaurus group.
This agreement provides qualitative evidence that the majority of the new dwarf candidates are indeed Centaurus group members. 

Complementary to this we can compare the shape parameter $n$ from the best-fitting S\'ersic profiles of our dwarfs with 
the S\'ersic indices of Local Group, Virgo, M\,81, and the known Cen\,A dwarfs (Fig.\,\ref{nToMr}). The faint end of the 
shape parameter - luminosity relation is notably widespread. Still, the S\'ersic indices of the candidates are in good 
agreement with the known dwarfs and fit into the relation. \\

\begin{figure}
  \resizebox{\hsize}{!}{\includegraphics{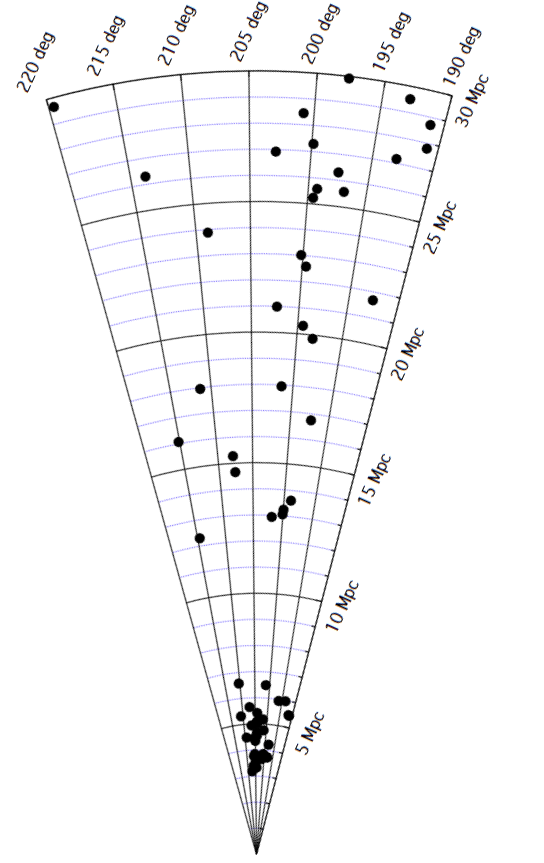}}
  \caption{Wedge diagram in right ascension for all galaxies with measured distances 
  in the direction of the survey region ($190^\circ<\alpha<220^\circ$ and -45\degree $<\delta<$ -20\degree). 
  Data taken from the Cosmicflows-2 catalog \citep{2013AJ....146...86T}.
  The Centaurus group is the prominent overdensity covering the distance interval $3<D<6.5$\,Mpc. The region behind 
  the Centaurus group is the Local Void followed by a low density environment made up of a small number
  of field galaxies with distances $D>12$\,Mpc. }
  \label{velocity}
\end{figure}

We can also look into the membership question by studying the 3D distribution of galaxies in the direction of the 
survey region. No massive galaxies are known in the immediate vicinity behind the Centaurus group. This is illustrated 
in Fig.\,\ref{velocity}, where we plot the wedge diagram in right ascension for the galaxies with measured distances. 
Data were taken from the Cosmicflows-2 catalog \citep{2013AJ....146...86T}. The Centaurus group is 
the prominent overdensity covering the distance range $3.0<D<6.5$\,Mpc. Behind the group is the 
Local Void \citep{2008ApJ...676..184T, 2015ApJ...802L..25T} followed by a low density environment made up of a 
population of field galaxies and small groups. There is no larger concentration of galaxies within 30\,Mpc.
The conclusion is that galaxies found in our survey area either belong to the Centaurus group or must be background galaxies 
at least $2-3$ times further away. 

\begin{figure}
  \resizebox{\hsize}{!}{\includegraphics{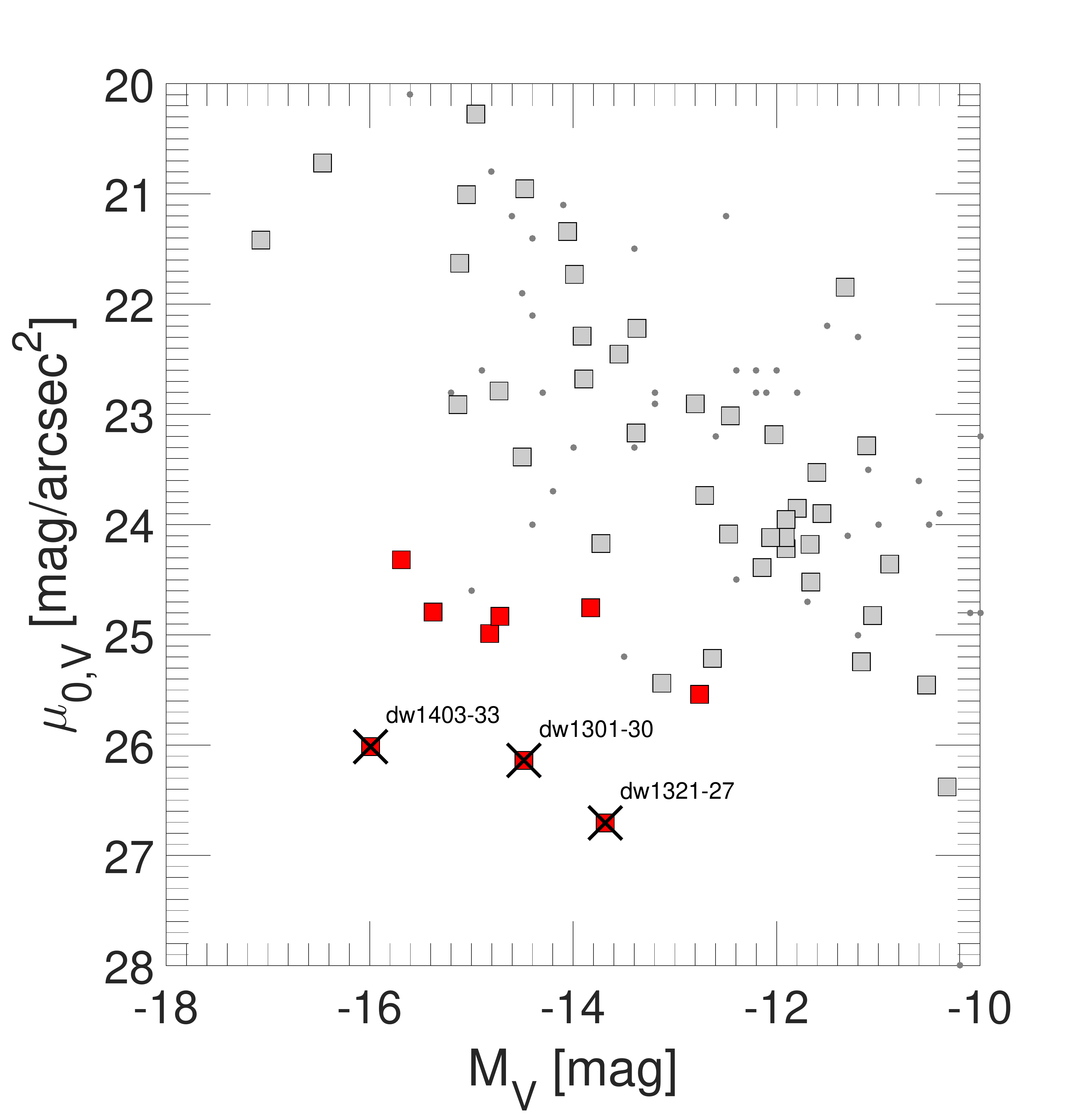}}
 \caption{ $\mu_0-M$ relation for the Local Group and Centaurus group galaxies. We added nine dwarf candidates (red squares) 
 adopting the distances of background galaxies that were close to the dwarf in the sky. If a candidate is a satellite of the background 
 galaxy, it should follow the same relation as defined by the Centaurus and Local Group dwarfs. We crossed out  the candidates that we dismissed 
 as possible background galaxies, meaning that they are more probably Centaurus group members. The candidates that are closer
 than 1\,mag (radial distance) from any known dwarf in the relation could be background dwarf galaxies and are indicated as
 \textit{bg?} in Table\,\ref{table:1}. All such \textit{bg?} candidates are still outside of the relation, making them unlikely to be 
 background dwarfs; this is why we still present them as possible Centaurus group members.}
  \label{backrelation}
\end{figure}

We further tested the hypothesis that some of our candidates are satellites of luminous background galaxies.
For example, dw1321-27 and dw1322-27 are approximately 20\,arcmin away from the barred spiral galaxy NGC\,5101, 
which has a velocity of 1868\,km\,s$^{-1}$ \citep{2004AJ....128...16K} and a luminosity-line width distance of 27.4\,Mpc \citep{1988ang..book.....T}. At that distance the linear separation between these galaxies would 
be around 164\,kpc. This is comparable with the distance between Fornax and the Milky Way \citep{2012AJ....144....4M}. Nine of our candidates have a background galaxy within a radius of 60\,arcmin. We 
plotted these candidates again in the $\mu-M$ diagram (Fig.\,\ref{backrelation}) this time with an absolute magnitude 
that corresponds to the velocity distance of the background galaxy. The three candidates, dw1301-30, dw1321-27, and dw1403-33,
now fall outside of the relation defined by the known dwarfs, making their association to a background galaxy unlikely. 
The situation for the other six candidates remains ambiguous in this test, and thus they got a \textit{bg?} label in Table\,\ref{table:1}. 
However, given that these candidates are located at the edge of the general trend makes them more likely to be Centaurus group 
members than background galaxies.

\cite{2014ApJ...795L..35C} found the first close pair of faint dwarf galaxies outside the Local Group 
in the halo of Cen\,A, namely the pair CenA-MM-Dw1 and CenA-MM-Dw2, with a projected distance 
of 3\,kpc. Our galaxy sample contains two other potential pairs of dwarf galaxies in the same group, 
(dw1243-42, dw1243-42b) and (dw1251-40, dw1252-40). They are separated by 75\,arcsec 
(1.6\,kpc at 4.5\,Mpc) and 135\,arcsec (2.9\,kpc at 4.5\,Mpc), respectively.

\subsection{Galaxy distribution}
\cite{2015ApJ...802L..25T} reported that almost all members of the Cen\,A subgroup with known distances are 
distributed in two thin parallel planes. The authors further noted that one of these planes points in the direction 
of M\,83. Interestingly, all but one of the nine dwarfs detected in the PISCeS survey also belong to the two 
planes \citep{2015arXiv151205366C}. Looking at the 2D distribution of our new dwarf candidates, we find that 
a significant number of those candidates in the vicinity of Cen\,A are aligned in the Cen\,A-M\,83 direction (see Fig.\,\ref{fieldDepth}). 
This confirms the result of the PISCeS survey that most of their galaxies were found in the northward direction.
The opposite situation is observed in the M\,83 subgroup where the MJB15 candidates are preferentially found southward 
of M\,83 in the direction of Cen\,A. Intriguingly, even when looking on the galactic scale there is evidence of asymmetry, such 
as the lopsided distribution of star-forming regions in the outer disk of M83 with a large number detected on the southern side and only a few
on the northern side as evident in deep GALEX images \citep{2005ApJ...619L..79T}. Is this a hint of some kind of dwarf galaxy substructure 
between the two main galaxies, possibly a filament of dwarf galaxy infall, or an extension of the Cen\,A plane? 
Another interesting feature revealed by the new candidates is an elongated, filamentary structure 
that runs diagonally through the group, {from ($\alpha$/$\delta$) $\approx$ (14:20,$-$33) to $\approx$ (12:40,$-$43)}, seemingly separating the two subgroups (see Fig.\,\ref{fieldDepth}). 
At the distance of 4.5\,Mpc this structure extends over 1.8\,Mpc. Accurate distances to the new dwarf 
galaxy candidates will be needed for a more quantitative assessment of the substructural properties of the Centaurus group.

\section{Conclusions}
We have conducted the first CCD-based, large-scale survey of the nearby Centaurus group covering 
an area of over 500 square degrees or 3.3\,Mpc$^{2}$. We found a total of 41 new dwarf galaxy candidates (in addition to 16 new candidates reported on previously in MJB15)
in the magnitude range 17  $< r <$ 20.5 mag and surface brightness range of $24< \langle \mu \rangle_{eff,r} <27$\,mag asec$^{-2}$ 
pushing the absolute magnitude limit of the galaxy population down to $M_r\approx -9.5$. Although no distance information is currently 
available, except for dw1335-29 \citep{2016AAS...22713625C}, the comparison of the photometric and structural parameters of the candidates 
with the known dwarf galaxies strongly suggests that the majority of the galaxies belong to the Centaurus group. Follow-up measurements 
of the distances are crucial to confirm their membership. There are a number of research areas that will greatly benefit from further
analysis of the new galaxies. The mere abundance and spatial distribution of the Centaurus galaxies will be a new empirical benchmark to test 
structure formation processes and the cosmological models behind them. Is the Local Group a statistical outlier or does the conflict with
$\Lambda$CDM also apply to the Centaurus group? In this context, understanding the two galaxy planes will play a central role. 
How were they formed and why are they almost parallel? Is there a dwarf galaxy bridge from Cen\,A to the M\,83 subgroup? 
The expected small distance uncertainties in the range of 0.2-0.5\,Mpc from the TRGB method will be instrumental to trace the 
3D galaxy distribution along the 2\,Mpc line-of-sight depth of the Centaurus group. 
It will be intriguing to see how the new dwarf galaxies are distributed in the double planar structure.
One possibility is that they will increase the statistical significance of the bimodality, proving the 
double structure to be real beyond any doubt. The exact significance level will depend 
on the intrinsic thickness of the planes, the plane orientations relative to the line-of-sight, and the 
number of galaxies in each component.  Another possible outcome is that the new galaxies 
fill the $\approx$0.1\,Mpc gap between the two planes (see histogram on the right 
side of Fig.\,1 of \cite{2015ApJ...802L..25T}) and thus reveal that the double planar structure was 
in fact an artifact of small number statistics. How do the stellar populations in the new Centaurus dwarfs compare to Local Group look-alikes? 
\cite{2010A&A...516A..85C,  2011A&A...530A..59C, 2011A&A...530A..58C,2012A&A...541A.131C} conducted an 
extensive study of the resolved stellar content of dwarf galaxies in the Centaurus group. They 
investigated their star formation histories and metallicity content, and what effect the denser environment
has on shaping these properties. With the large number of new dwarf galaxies available from our study,
this work can be extended and pushed toward lower limits, allowing statistically more robust comparisons with the 
Local Group dwarf galaxy population.

%_______________________________________________________________________________

\begin{acknowledgements}

OM and BB are grateful to the Swiss National Science Foundation for financial support. 
HJ acknowledges the support of the Australian Research Council through Discovery projects DP120100475 and DP150100862. {The authors would like to thank Dmitry Makarov and the anonymous
referee for helpful comments that improved the paper.}
This project used data obtained with the Dark Energy Camera (DECam), which was constructed by the Dark Energy Survey (DES) 
collaborating institutions: Argonne National Lab, University of California Santa Cruz, University of Cambridge, Centro de Investigaciones Energeticas, Medioambientales y Tecnologicas-Madrid, University of Chicago, University College London, DES-Brazil consortium, University of Edinburgh, ETH-Zurich, Fermi National Accelerator Laboratory, University of Illinois at Urbana-Champaign, Institut de Ciencies de l'Espai, Institut de Fisica d'Altes Energies, Lawrence Berkeley National Lab, Ludwig-Maximilians Universitat, University of Michigan, National Optical Astronomy Observatory, University of Nottingham, Ohio State University, University of Pennsylvania, University of Portsmouth, SLAC National Lab, Stanford University, University of Sussex, and Texas A\&M University. Funding for DES, including DECam, has been provided by the U.S. Department of Energy, National Science Foundation, Ministry of Education and Science (Spain), Science and Technology Facilities Council (UK), Higher Education Funding Council (England), National Center for Supercomputing Applications, Kavli Institute for Cosmological Physics, Financiadora de Estudos e Projetos, Fundao Carlos Chagas Filho de Amparo a Pesquisa, Conselho Nacional de Desenvolvimento Cientfico e Tecnolgico and the Ministrio da Cincia e Tecnologia (Brazil), the German Research Foundation-sponsored cluster of excellence ``Origin and Structure of the Universe" and the DES collaborating institutions. This research was made possible through the use of the AAVSO Photometric All-Sky Survey (APASS), funded by the Robert Martin Ayers Sciences Fund.
\end{acknowledgements}

\bibliographystyle{aa}
\bibliography{bibliographie}

\end{document}